\documentclass[a4paper,12pt, epsfig]{article}
\pdfoutput=1
\usepackage{epsfig}
\usepackage{epstopdf}
\usepackage{graphicx}
\usepackage{ifthen}

\usepackage{amsmath}
\usepackage[psamsfonts]{amssymb}
\usepackage{euscript}

\usepackage{latexsym}
\usepackage[arrow,matrix,curve]{xy}

\newskip\humongous \humongous=0pt plus 1000pt minus 1000pt

\newif\ifdtup

\def\,{\hspace{-.1cm}}
\def\hsp{,\hspace{.7cm}}

\def\tf {\tilde{f}}
\def\fjl {\Phi\left(\frac{n}{q}j,\frac{n}{q} l\right)}
\def\F#1#2{\Phi\left(\frac{n}{q}{#1},\frac{n}{q}{#2}\right)}
\def\fc#1#2 {\frac{n}{q}#1\frac{n}{q}#2}

\def\e#1{\langle#1\rangle_0}
\def\ee#1{\left\langle#1\right\rangle}
\def\di#1{\left(\vcenter{\xymatrix{#1}}\right)}

\renewcommand{\theequation}{\arabic{section}.\arabic{equation}}
\renewcommand{\(}{\begin{equation}}
\renewcommand{\)}{end{equation} \vspace{-.05in}\linebreak}

\newcounter{saveeqn}
\newcounter{savealpheqn}

\newcommand{\alpheqn}{\setcounter{saveeqn}{\value{equation}}%
  \stepcounter{saveeqn}\setcounter{equation}{0}%
  \renewcommand{\theequation}{\mbox{\arabic{section}.\arabic{saveeqn}
\alph{equation}}}
  \renewcommand{\)}{\end{equation}}}
\def\part#1{\frac{\partial}{\partial{#1}}}%
\def\group#1{\refstepcounter{equation}\setcounter{saveeqn}
 {\value{equation}}%
  \label{#1}\setcounter{equation}{0}%
\renewcommand{\theequation}{\mbox{\arabic{section}.\arabic{saveeqn}
\alph{equation}}}
  \renewcommand{\)}{\end{equation}}}
\newcommand{\reseteqn}{\setcounter{equation}{\value{saveeqn}}%
  \renewcommand{\theequation}{\arabic{section}.\arabic{equation}}%
  \renewcommand{\)}{\end{equation}}}

\newcommand{\aalpheqn}{\setcounter{saveeqn}{\value{equation}}%
  \stepcounter{saveeqn}\setcounter{equation}{0}%
  \renewcommand{\theequation}{\mbox{
        \Alph{subsection}.\arabic{saveeqn}\alph{equation}}}
   \renewcommand{\)}{\end{equation}}}
\newcommand{\areseteqn}{\setcounter{equation}{\value{saveeqn}}%
  \renewcommand{\theequation}{\Alph{subsection}.\arabic{equation}}%
  \renewcommand{\)}{\end{equation}}}

\renewcommand{\thefootnote}{\alph{footnote}}
\renewcommand{\(}{\begin{equation}}
\renewcommand{\)}{\end{equation}}
\newcommand{\ba}{\begin{eqnarray}}
\newcommand{\ea}{\end{eqnarray}}

\newcommand{\bp}{\mathop{\vtop{\ialign{##\crcr
   $\hfil\displaystyle{}\hfil$\crcr\noalign{\kern-13pt\nointerlineskip}
   \BIG{(}\hskip0pt\crcr\noalign{\kern3pt}}}}}
\newcommand{\cbp}{\mathop{\vtop{\ialign{##\crcr
   $\hfil\displaystyle{}\hfil$\crcr\noalign{\kern-13pt\nointerlineskip}
   \BIG{)}\hskip0pt\crcr\noalign{\kern3pt}}}}}
\newcommand{\pa}{\mathop{\vtop{\ialign{##\crcr
    
$\hfil\displaystyle{\oplus}\hfil$\crcr\noalign{\kern+1pt\nointerlineskip 
}
   \hspace{.08in}$^{\alpha=0}$\hskip6pt\crcr\noalign{\kern3pt}}}}}
\renewcommand{\hsp}{,\hspace{.3in}}

\newcommand{\Z}{\ensuremath{\mathbb Z}}

\catcode`\@=11
\def\vereq#1#2{\lower3pt\vbox{\baselineskip1.5pt \lineskip1.5pt
\ialign{$\m@th#1\hfill##\hfil$\crcr#2\crcr\sim\crcr}}}
\catcode`\@=12

\renewcommand{\(}{\begin{equation}}
\renewcommand{\)}{\end{equation}}

\newcommand{\beas}{\begin{eqnarray*}}
\newcommand{\eeas}{\end{eqnarray*}}

\newcommand{\bquo}{\begin{quote}}
\newcommand{\enqu}{\end{quote}}

\newcommand{\C}{{\mathbb C}}
\newcommand{\cp}{{\mathrm{\mathbb CP}}}
\newcommand{\R}{{\mathbb R}}
\renewcommand{\Z}{{\mathbb Z}}

\newcommand{\beq}{\begin{equation}}
\newcommand{\eeq}{\end{equation}}
\newcommand{\bea}{\begin{eqnarray}}
\newcommand{\eea}{\end{eqnarray}}

\newskip\humongous \humongous=0pt plus 1000pt minus 1000pt

\newif\ifdtup

\catcode`\@=11

\@addtoreset{equation}{section}

\def\@normalsize{\@setsize\normalsize{15pt}\xiipt\@xiipt
\abovedisplayskip 14pt plus3pt minus3pt%
\belowdisplayskip \abovedisplayskip
\abovedisplayshortskip \z@ plus3pt%
\belowdisplayshortskip 7pt plus3.5pt minus0pt}

\def\small{\@setsize\small{13.6pt}\xipt\@xipt
\abovedisplayskip 13pt plus3pt minus3pt%
\belowdisplayskip \abovedisplayskip
\abovedisplayshortskip \z@ plus3pt%
\belowdisplayshortskip 7pt plus3.5pt minus0pt
\def\@listi{\parsep 4.5pt plus 2pt minus 1pt
      \itemsep \parsep
      \topsep 9pt plus 3pt minus 3pt}}

\relax

\catcode`@=12

\catcode`\@=11

\def\section{\@startsection{section}{1}{\z@}{3.5ex plus 1ex minus  .2ex}{2.3ex plus .2ex}{\large\bf}}

\def\thesection{\arabic{section}}
\def\thesubsection{\arabic{section}.\arabic{subsection}}

\def\appendix{\setcounter{section}{0}
 \def\thesection{Appendix \Alph{section}}
 \def\thesubsection{\Alph{section}.\arabic{subsection}}
 \def\theequation{\Alph{section}.\arabic{equation}}}
\renewcommand{\theequation}{\arabic{section}.\arabic{equation}}

\begin{document}
\def\thefootnote{\fnsymbol{footnote}}
\def\thetitle{Anchoring and Binning the Coordinate Bethe Ansatz}
\def\autone{Jarah Evslin}
\def\affa{Institute of Modern Physics, NanChangLu 509, Lanzhou 730000, China}
\def\affb{University of the Chinese Academy of Sciences, YuQuanLu 19A, Beijing 100049, China}

\begin{center}
{\large {\bf \thetitle}}

\bigskip

\bigskip

{\large \noindent  \autone{${}^{1,2}$} \footnote{jarah@impcas.ac.cn}} 

\vskip.7cm

1) \affa\\
2) \affb\\

\end{center}

\begin{abstract}
\noindent

\noindent
The Coordinate Bethe Ansatz (CBA) expresses, as a sum over permutations, the matrix element of an XXX Heisenberg spin chain Hamiltonian eigenstate with a state with fixed spins.  These matrix elements comprise the wave functions of the Hamiltonian eigenstates.  However, as the complexity of the sum grows rapidly with the length $N$ of the spin chain, the exact wave function in the continuum limit is too cumbersome to be exploited.  In this note we provide an approximation to the CBA whose complexity does not directly depend upon $N$.   This consists of two steps.  First, we add an anchor to the argument of the exponential in the CBA.  The anchor is a permutation-dependent integral multiple of $2\pi i$.  Once anchored, the distribution of these arguments simplifies, becoming approximately Gaussian.  The wave function is given by the Fourier transform of this distribution and so the calculation of the wave function reduces to the calculation of the moments of the distribution.  Second, we parametrize the permutation group as a map between integers and we bin these maps.  The calculation of the moments then reduces to a combinatorial exercise on the partitioning into bins.   As an example, we consider the matrix element between the classical and quantum ground states.



\end{abstract}

%
\setcounter{footnote}{0}
\renewcommand{\thefootnote}{\arabic{footnote}}




\section{Introduction}

\subsection{Motivation}

Man has always sought to understand the origin of the Yang-Mills mass gap.  In the instantaneous frame, it is a consequence of the ground state.  This ground state may be realized, in the Schrodinger picture, as a wave functional\footnote{Recall that these wave functionals associate a complex number to every field configuration on a time slice.} which satisfies the Schrodinger equation \cite{stuck}.  Despite decades of efforts, no such solution appears to be forthcoming.

On the other hand, Yang-Mills theory in 3+1 dimensions is quite similar to the $\cp^1$ nonlinear sigma model in $1+1$ dimensions.  Here also fractional instantons are somehow involved in the generation of a mass gap \cite{fateevmeron}.  Knowledge of the ground state and first excited state wave functionals of this model would unlock exciting doors, allowing a concrete understating of {\it{how}} the instantons generate the mass gap in the Minkowski theory, perhaps as a kind of infinite-dimensional generalization of the familiar story in quantum mechanics with a double well potential.  

Our motivation is based on an analogy, summarized in Table~\ref{antab}, between (i) The double well model in quantum mechanics, (ii) The $\cp^1$ nonlinear sigma model and (iii) Yang-Mills theory.   Consider the following states: (i) A position eigenstate corresponding to the point $x$, (ii) A wave functional which vanishes on all field configurations but one, which wraps a circle on $\cp^1$ at fixed latitude $\theta$ and (3) A single gauge-invariant wave functional supported on (the gauge orbit of) a gauge field configuration with each Chern-Simons number $k$.  While none of these states are Hamiltonian eigenstates, there is a potential for the variables $x$, $\theta$ and $k$ arising from the (i) the quantum mechanical potential itself, (ii) the kinetic term $\partial n\partial n$ and (iii) the Yang-Mills kinetic term.  The double-well potential has two degenerate minima by definition, while the sigma model potential has degenerate minima at $\theta=\pm\pi/2$ among the $|\theta\rangle$ states and Yang-Mills has degerate minima when $k$ is an integer.  In each case there are instantons of action $S$ representing tunneling between the minima.  As a result one expects that the wave functional is suppressed by roughly $e^{-S}$ deep inside of the barrier.

\begin{table}
\centering
\begin{tabular}{c|l|l|l}
&Double Well QM&$\cp^1$ model & Yang-Mills\\
\hline\hline
States Considered&Position Eigenstate $x$&Latitude $\theta$ circle&Chern-Simons no. $k$\\
\hline
Source of Potential&$V(x)$&$\partial_k n\partial_k n$&$F_{k\mu}F^{k\mu}$\\
\hline
Degenerate Vacua&$2$&$2$&$\infty$\\
\hline
Role of Instantons&Mass Gap&Mass Gap&Monopole Mass?\\
\hline
\end{tabular}
\caption{Three-way analogy motivating this work}
\label{antab}
\end{table}


In quantum mechanics, the mass gap may be seen as a consequence of a discrete choice in how the wave functions are connected across the barriers.  Is there a similar story in quantum field theory?  Does this cross-barrier bridge also render a monopole-operator tachyonic in Yang-Mills?  After all, it is known that in some $\mathcal{N}=2$ super Yang-Mills theories, when softly broken to $\mathcal{N}=1$, instantons do render some monopoles tachyonic, leading to confinement~\cite{sw}.   To answer these questions, we need at least to understand the basic features of the ground state and first excited wave functionals.  For example, does the first excited state wave functional have a node at the maximum of this potential, corresponding to $\theta=0$ in the sigma model or a half-integral $k$ in Yang-Mills?  How is the fractional instanton plasma, seen in the Euclidean space sigma model in Ref.~\cite{fateevmeron}, manifested in the vacuum state?

This sigma model is not only solvable but has already been solved \cite{zam}.  So, what are the wave functionals?  A map between the $\cp^1$ sigma model and the XXX Heisenberg spin chain was shown in \cite{haldane1,haldane2} at the level of low energy fluctuations and in \cite{affleck} in the full quantum theory.  The former applies to a spin chain of any spin $s$, with strong coupling at small spin while the latter, reviewed in \ref{mapapp}, strictly speaking yields an equivalence only at infinite $s$, although finite $s$ can be used as a definition for an $\cp^1$ sigma model whose target is a quantum deformed $\cp^1$.  The spectra of these spin chains are also well-known.  There are many formalisms for writing the wave functions corresponding to these states and so these states are also known.  With the XXX states known, and the map to the sigma model known, also the sigma model wave functionals are by definition known.  

So what are the sigma model wave functionals?   To actually take these spin chain solutions and map them to something intelligible on the sigma model side was Faddeev's challenge to his students in \cite{fc}.  The map is known in the coordinate basis of spins in the spin chain, and so to meet the challenge one needs the matrix elements of the spin chain Hamiltonian eigenvectors with the coordinate states, which have definite spins at each lattice site.  The challenge is indeed a challenge because, while many forms are by now known for the spin chain Hamiltonian eigenstates in the coordinate basis \cite{bethe,faddeev81,alcaraz,boos,katsura,gromovsep}, each grows in complexity either with the length $N$ of the chain or else with the distance of a coordinate state from a preferred spin state, such as the classical ground state.  

Our goal is to present a method for approximating the matrix elements which depends on the complexity of the state, but not directly on $N$.  The individual lattice sites are replaced by bins.  The intuition is that the states which survive to the continuum limit are those which are essentially homogeneous inside of each bin.  Homogeneous means that using the mapping to the $\cp^1$ model, each pair of adjacent lattice sites in the same bin corresponds to the same point in $\cp^1$.  Therefore the points in the sigma model correspond not to the original lattice sites, but rather to the bins.  To describe the sigma model ground state, one then needs to calculate the spin chain matrix element for each such configuration of bins.  This calculation is very different from the Coordinate Bethe Ansatz (CBA) because the complicated symmetric sum has been smoothed away.     The goal of the present note is to present a formalism which allows these bin states to be derived from the CBA.

\subsection{Outline}

After a review of the XXX spin chain in Sec.~\ref{xxxsez}, we begin in Sec.~\ref{ancorasez} with the first key ingredient in our construction, the anchor.  The CBA gives the matrix element $a$ between a given spin chain basis state and a given energy eigenstate as a sum of phases $e^{i\alpha(g)}$, one for each element $g$ of the permutation group $S_n$, where $n=N/2$ for the antiferromagnetic ground state, in which we will be primarily interested from now on.  Consider the continuum limit, corresponding to large $N$.  The sum may be replaced by a density function $\rho(\alpha)$ and so the matrix element becomes an integral
\beq
a=\sum_{g\in S_n} e^{i\alpha(g)}
\rightarrow\int_{-\infty}^{\infty} e^{i\alpha}\rho(\alpha)d\alpha.  \label{ft}
\eeq
In other words it is given by the Fourier transform of the density function $\rho(\alpha)$.  

If $\rho(\alpha)$ were a Gaussian distribution, this transform would be trivial.  If it were close to a Gaussian distribution, one could perform the Fourier transform perturbatively, using a moment expansion of $\rho(\alpha)$.  Unfortunately, we have observed numerically that $\rho(\alpha)$ is rich in fine structure.  In particular it contains a series of maxima with separations of order 2$\pi$, which dominate the moments, making the Gaussian approximation quite poor.

The anchor is a permutation-dependent integral multiple of $2\pi$ which we we will subtract from the arguments $\alpha$ of the phases.  We refer to the difference as the anchored argument $\alpha^\prime$.  Clearly subtracting the anchor does not affect the matrix elements, as these depend only upon $e^{i\alpha}$.   Our first main result is purely numerical.  We have observed, by calculating all values of $\alpha(g)$ on spin chains where $N\leq 22$, that the density of $\alpha^\prime$ is nearly free of substructure and the Gaussian approximation is quite good.   The standard deviation of the unanchored arguments $\alpha$ 
\beq
\sqrt{\int_{S_n}\alpha^2\rho(\alpha)-\left(\int_{S_n} \alpha \rho(\alpha)\right)^2} \label{sig}
\eeq
is of order $O(N)^{3/2}$.   Our second main result, which is shown analytically using the binning approximation described below, is that the standard deviation of the anchored $\alpha^\prime$ is only of order $O(N)$.  The anchored argument $\alpha^\prime$ therefore provides a more convenient starting place for a perturbative calculation of the Fourier transform (\ref{ft}) than the original argument $\alpha$.


The other key ingredient is introduced in Sec.~\ref{binsez}.  To calculate the moments of the density $\rho(\alpha^\prime)$, the elements of the group $S_n$ are realized as one to one maps from the integers $[1,n]$ to themselves.  We divide this interval into $q$ bins.  For each permutation $g$ one can determine how many elements of the $i$th bin map to the $j$th bin.  We will call this number $f_{ij}(g).$  For a given $g$, $f_{ij}(g)$ consists of $q^2$ (nonnegative) integers and so is an element of $\Z^{q^2}$. We rewrite the CBA in terms of the quantities $f_{ij}(g)$ as follows.   We make the binning approximation
\beq
a=\sum_{g\in S_n} e^{i\alpha^\prime(g)}
\sim\sum_{\vec{p}\in\Z^{q^2}}e^{i\alpha^\prime(\vec{p})}h(\vec{p})
\eeq
where $h(\vec{p})$ is the number of elements $g\in S_n$ such that $f_{ij}(g)=p_{ij}$ and $\alpha^\prime(\vec{p})$ is equal to $\alpha^\prime(g)$ where $g$ is a particular permutation such that $f_{ij}(g)=p_{ij}$.  Our third main result is our formula for $\alpha^\prime(\vec{p})$, or stated differently $\alpha^\prime$ as a function of the $f_{ij}$, in Eqs.~(\ref{redotti}) and (\ref{aeq}).  The moments of $\rho(\alpha^\prime)$ are then determined from the correlation functions of $f_{ij}$ which in turn depend on the functions $h(\vec{p})$.  We calculate $h(\vec{p})$ using standard combinatorial arguments.

Finally in Sec.~\ref{tornasez} we will, in the case of the matrix element between the classical and quantum ground states, apply the techniques introduced above to calculate the $O(N^2)$ contribution to the second moment of the anchored $\rho(\alpha^\prime)$.  We will see explicitly that its coefficient is small, but it does not vanish.   



\section{The Antiferromagnetic XXX Heisenberg Spin Chain} \label{xxxsez}

The $\cp^1$ sigma model is the continuum limit of a spin chain with an infinite spin at each lattice site.  Classically, the spin squared corresponds to the inverse coupling \cite{haldane1,haldane2} and so low spin corresponds to a high coupling.  In particular, at low spin one describes the sigma model at strong coupling and one does not expect a sensible description of individual instantons.  Therefore, it will be essential for us to eventually extend our analysis to higher spin.  However, in the present note we will restrict our attention to spin $s=1/2$.

\subsection{Finite Chain}

The spin $1/2$ Heisenberg spin chain consists of $N$ lattice sites.  At each lattice site lies a Hilbert space $\C^2$ with basis $\{|\,\uparrow\rangle,|\,\downarrow\rangle\}$.  The total Hilbert space is the $N$-fold tensor product\footnote{As was described in Ref.~\cite{vonneumann}, when $N=\infty$ this space decomposes into superselection selectors.  We will be interested in finite $N$ in the present note, however it is tempting to conjecture that the superselection sector of interest corresponds to the constant bin states that we will introduce in Sec.~\ref{binsez}.} of these $\C^2$.  At each lattice site $l$ lies an $\mathfrak{su}(2)$ Lie algebra with generators $\sigma^i_l$ satisfying
\beq
[\sigma^i_l,\sigma^j_m]=2i\delta_{lm}\epsilon^{ijk}\sigma^k_l. \label{comm}
\eeq
This algebra acts on the $\C^2$ Hilbert space at the site $l$, according to the usual 2-dimensional representation such that
\beq
\sigma_l^3 |\,\uparrow\rangle_l=|\,\uparrow\rangle_l\hsp \sigma_l^3 |\,\downarrow\rangle_l=-|\,\downarrow\rangle_l.
\eeq
The XXX spin chain corresponds to the Hamiltonian
\beq
H= J \sum_{i=1}^3 \sum_{l=1}^N (\sigma^i_l \sigma^i_{l+1}-\mathbf{1})
\eeq
where $\mathbf{1}$ is the identity.  We will let the constant $J$ be positive, corresponding to the antiferromagnetic spin chain.  Although the eigenvalues of $H$ depend on $J$, in this note we will only be interested in the eigenvectors, which are independent of $|J|$.  In particular, we will restrict our attention to the antiferromagnetic ground state $|\Omega\rangle$, which is the eigenstate of $H$ with minimal eigenvalue.  This state is the same for any positive value of $J$.

Any state can be decomposed into the basis consisting of the tensor product of the  $\{|\,\uparrow\rangle_l,|\,\downarrow\rangle_l\}$ bases at each lattice site.  An element of the basis is a string of $\uparrow$'s and $\downarrow$'s.  It is described by the set of positions $m(i)$ of the $i$th $\downarrow$ for all $i$.    Therefore an arbitrary state $|\Psi\rangle$ is fully characterized by the matrix elements
\beq
a\left(\{m(i)\}\right)=\langle\{m(i)\}|\Psi\rangle.
\eeq
The Hamiltonian commutes with rigid rotations, which are generated by the $\mathfrak{su}(2)$ Lie algebra with basis
\beq
\Sigma^i=\sum_{l=1}^N \sigma^i_l.
\eeq
Therefore it can be diagonalized simultaneously with $\Sigma^3$.  As a result, each Hamiltonian eigenstate can be taken to have a definite number $n$ of spin downs.

For all Hamiltonian eigenstates $\Psi$, the elements $a\left(\{m(i)\}\right)$ are given by the coordinate Bethe Ansatz \cite{bethe}
\beq
a\left(\{m(i)\}\right)=\sum_{g\in S_n}{\mathrm{exp}}\left(i\sum_{j=1}^n m(j) K(P(j))+\frac{i}{2}\sum_{j<k}\Phi(P(j),P(k))\right)
\eeq
where $P(j):[1,n]\rightarrow [1,n]$ is the permutation corresponding to $g\in S_n$.  The information about the state is contained in the functions $K\in[0,2\pi]$ and $\Phi\in[-\pi,\pi]$ which are related by
\beq
2{\rm{cot}}\left(\frac{1}{2}\Phi(i,j)\right)={\rm{cot}}\left(\frac{1}{2}K(i)\right)-{\rm{cot}}\left(\frac{1}{2}K(j)\right) \label{bethea}
\eeq
and by the Bethe equation
\beq
N K(i)=2\pi Q(i) + \sum_{j\neq i}^n \Phi(i,j) \label{betheeq}
\eeq
where $Q(i)$ is an integer.  In fact, a state is characterized by just the set of $\{Q(i)\}.$  The ground state for example corresponds to
\beq
N=2n\hsp Q(i)=2n-2i+1.
\eeq

The right hand side of Eq.~(\ref{comm}) contains an $\hbar$, which we have set to unity.  However in the classical limit it is instead set to zero, in which case the lowest energy state of $H$ becomes a classical ground state, such as
\beq
|0\rangle=|\,\uparrow\downarrow\uparrow\downarrow\cdot\cdot\cdot\rangle
\eeq
which corresponds to
\beq
n=\frac{N}{2}\hsp
m(i)=2i.
\eeq
In most of this paper we will restrict our attention to the matrix element between the classical ground state $|0\rangle$ and the quantum ground state $|\Omega\rangle$
\beq
a(m(i)=2i)=\langle 0|\Omega\rangle.
\eeq
The generalization of our results to other matrix elements with well-behaved continuum limits is essential for our goals.  While we suspect that this will be a straightforward generalization of the calculations below, we leave these extension to future work.

\subsection{The Thermodynamic Limit}

One may automatically solve Eq.~(\ref{bethea}) by introducing spectral parameters $\lambda(i)$, related to $K(i)$ and $\Phi(i)$ by
\beq
e^{iK(j)}=\left(\frac{\lambda(j)+\frac{i}{2}}{\lambda(j)-\frac{i}{2}}\right)\hsp
e^{i\Phi(j,k)}=\left(\frac{\lambda(j)-\lambda(k)+i}{\lambda(j)-\lambda(k)-i}\right) \label{lambdadef}
\eeq
so that
\beq
K(j)=-i\ {\mathrm{ln}}\left(\frac{\lambda(j)+\frac{i}{2}}{\lambda(j)-\frac{i}{2}}\right)=\pi-2\ {\mathrm{ArcTan}}(2\lambda(j)) \label{kl}
\eeq
and
\beq
\Phi(j,k)=-i\  {\mathrm{ln}}\left(\frac{\lambda(j)-\lambda(k)+i}{\lambda(j)-\lambda(k)-i}\right)=2\ {\mathrm{ArcCot}}(\lambda(j)-\lambda(k)). \label{fl}
\eeq
We recall that $K\in[0,2\pi]$ and $\Phi\in[-\pi,\pi]$ and so in Eqs.~(\ref{kl}) and (\ref{fl}) the ranges of both ArcTan and ArcCot must be taken to be $[-\pi/2,\pi/2]$.  Using (\ref{lambdadef}) Bethe's equation (\ref{betheeq}) can be rewritten as a condition on the spectral parameters
\beq
\left(\frac{\lambda(j)+\frac{i}{2}}{\lambda(j)-\frac{i}{2}}\right)^N=\prod_{k\neq j} \left(\frac{\lambda(j)-\lambda(k)+i}{\lambda(j)-\lambda(k)-i}\right).
\eeq

It will prove more convenient to rewrite Bethe's equation using (\ref{kl}) and (\ref{fl}) to obtain
\beq
{\mathrm{ArcTan}}(2\lambda(j))=\pi\left(\frac{1}{4}-\frac{Q(j)}{N}\right)+\frac{1}{N}\sum_{k\neq j}^n\left(\frac{\pi}{2}-{\mathrm{ArcCot}}(\lambda(j)-\lambda(k))\right).
\eeq
We have kept ArcCot$(\lambda(j)-\lambda(k))\in[-\pi/2,\pi/2]$ and ArcTan$(2\lambda(j))\in[-\pi/2,\pi/2]$ .  We would like to replace the $\pi/2-$ArcCot above with $\mathrm{ArcTan}$, where ArcTan$\in[-\pi/2,\pi/2]$.   However, using our conventions $\pi/2-$ArcCot$(\lambda(j)-\lambda(k))\in[0,\pi]$.  Therefore, to compensate for the difference in the principal values of $\pi/2-$ArcCot and ArcTan, we will need to subtract $\pi$ whenever ArcCot$(\lambda(j)-\lambda(k))$ is negative, which occurs when $\lambda(j)<\lambda(k)$.  We will choose the $\lambda(j)$ to be monotonically increasing in $j$, and so we will need to subtract $\pi$ for each $k$ such that $k>j$.  In other words, to bring ArcTan into the fundamental domain we must subtract $\pi(n-j)$ from the sum, which must be added to the $Q$ term.  Restricting attention to the ground state $|\Omega\rangle$, $N=2n$, $Q(j)=2n-2j+1$, the spectral parameters are the solutions of
\bea
{\mathrm{ArcTan}}(2\lambda(j))&=&\pi\left(\frac{1}{4}-\frac{Q(j)}{2n}+\frac{n-j}{2n}\right)+\frac{1}{2n}\sum_{k\neq j}^n{\mathrm{ArcTan}}(\lambda(j)-\lambda(k))\nonumber\\
&=&\pi\left(-\frac{1}{4}+\frac{j}{2n}\right)+\frac{1}{2n}\sum_{k\neq j}^n{\mathrm{ArcTan}}(\lambda(j)-\lambda(k)). \label{atan}
\eea

To pass to the continuum limit, one replaces the lattice site index $j\in[1,n]$ with
\beq
x(j)=-\frac{1}{4}+\frac{j}{2n}\in\left[-\frac{1}{4},\frac{1}{4}\right]. \label{x}
\eeq
Sometimes it is convenient to replace $j$ by $j-1/2$ in this expression to make it symmetric in $x\rightarrow -x$, however this will only affect subdominant contributions in $1/n$ and will not affect our main results here.  Now all functions $f(j)$ can be replaced by interpolating functions $\tilde{f}(x)$,  by demanding
\beq
\tilde{f}(x(j))=f(j). \label{tilde}
\eeq
By abuse of notation, we will drop the tildes and write simply $f$ for both the original discrete function and its continuous interpolation.  The interpolation is not uniquely defined, however if one imposes (\ref{atan}) then the choice of interpolation is irrelevant, since the equation only restricts the values at integral points where (\ref{tilde}) fully determines $f(x)$.  

To fix $f(x)$ at all $x\in\left[-\frac{1}{4},\frac{1}{4}\right]$, one replaces the sum in Eq.~(\ref{atan}) with an integral
\beq
\frac{1}{2n}\sum_{k\neq j}^n{\mathrm{ArcTan}}(\lambda(j)-\lambda(k))\rightarrow \int_{-1/4}^{1/4}dy\ {\mathrm{ArcTan}}(\lambda(x)-\lambda(y)) \label{sumint}
\eeq
so that the spectral function $\lambda(x)$ is determined by
\beq
{\mathrm{ArcTan}}(2\lambda(x))=\pi x+\int_{-1/4}^{1/4}dy\ {\mathrm{ArcTan}}(\lambda(x)-\lambda(y)). \label{vaceq}
\eeq
The replacement (\ref{sumint}) is not an equality.  It changes the equation.  The solutions $\lambda(x)$ will not be solutions of the original equation, even at the lattice sites $x(j)$.  It is expected that this correction is subdominant in the $1/n$ expansion. However, these subleading corrections to the $\lambda(x)$ may in principle provide leading contributions to the matrix elements.

 
 \subsection{The Ground State}
One can now solve (\ref{vaceq}) to find the above functions of $x$ for the quantum ground state $|\Omega\rangle$.  First, let us define the density  
\beq
\rho(x)=\frac{1}{\partial\lambda(x)/\partial x}
\eeq
which is unrelated to the density of phases $\rho(\alpha)$ introduced above.
The derivative of Eq.~(\ref{vaceq}) with respect to $x$ is
\beq
\frac{2}{1+4\lambda(x)^2}\frac{1}{\rho(x)}=\pi+\int_{-1/4}^{1/4}\frac{dy}{1+(\lambda(x)-\lambda(y))^2}\frac{1}{\rho(x)}.
\eeq
Now multiply through by $\rho(x)$.  The function $\lambda:[-1/4,1/4]\rightarrow[-\infty,\infty]$ is a bijection and so we can pull back any function $f(x)$ to obtain $f(\lambda)$.  Let $\lambda=\lambda(x)$ and $\mu=\lambda(y)$.  This allows us to rewrite the entire equation using functions of $\lambda$ and $\mu$, 
\beq
\frac{2}{1+4\lambda^2}=\pi\rho(\lambda)+\int_{-\infty}^{\infty}\frac{\rho(\mu)d\mu}{1+(\lambda-\mu)^2} \label{lmeq}
\eeq
where the integration measure was converted using
\beq
dy=\rho(\mu)d\mu.
\eeq

The equation (\ref{lmeq}) is usually solved using Fourier transforms.  We will review the argument here, as we need to go a few steps beyond the textbook treatment to obtain all functions of $x$ explicitly.   The Fourier transform of the left hand side, omitting the factor of two for now, is
\beq
\int_{-\infty}^\infty \frac{e^{i\lambda \alpha}d\lambda}{1+4\lambda^2}.
\eeq
The integrand has simple poles at $\lambda=\pm i/2$.  If $\alpha>0$ ($\alpha<0$) then the integrand vanishes exponentially for a large semicircular contour on the upper (lower) half of the complex plane.  The corresponding contour encircles the pole at $+i/2$ ($-i/2$), where the residue is $-ie^{-\alpha/2}/4$ ($ie^{\alpha/2}/4$).  The contour is counterclockwise (clockwise) and so the residue theorem yields
\beq
\int_{-\infty}^\infty \frac{e^{i\lambda \alpha}d\lambda}{1+4\lambda^2}=\frac{\pi}{2}e^{-|\alpha|/2}. \label{ftid}
\eeq
Defining the Fourier transform of the density by
\beq
\tilde{\rho}(\alpha)=\frac{1}{2\pi}\int_{-\infty}^{\infty}e^{i\alpha\lambda}\rho(\lambda)d\lambda
\eeq
the Fourier transform allows Eq.~(\ref{lmeq}) to be rewritten
\beq
\mathrm{LHS}=\frac{2}{1+4\lambda^2}=\frac{1}{2}\int_{-\infty}^{\infty}e^{-i\lambda\alpha}e^{-|\alpha|/2}d\alpha \label{LHS}
\eeq
and the right hand side
\bea
=\mathrm{RHS}&=&\pi\int_{-\infty}^{\infty}d\alpha e^{-i\lambda\alpha}\tilde{\rho}(\alpha)+\int_{-\infty}^{\infty}d\alpha\tilde{\rho}(\alpha)\int_{-\infty}^{\infty}\frac{e^{-i\mu\alpha}d\mu}{1+(\lambda-\mu)^2}\nonumber\\
&=&\int_{-\infty}^{\infty}d\alpha \tilde{\rho}(\alpha)\left[\pi e^{-i\lambda\alpha}+\pi e^{-|\alpha|}e^{-i\lambda\alpha)}\right] \label{RHS}
\eea
where the integral over $\mu$ was performed as in Eq.~(\ref{ftid}).
Taking the Fourier transform of this equation yields
\beq
\frac{e^{-|\alpha|/2}}{2}=\pi\tilde{\rho}(\alpha)\left[1+e^{-|\alpha|}\right]
\eeq
and so the Fourier transformed density is
\beq
\tilde{\rho}(\alpha)=\frac{1}{2\pi}\frac{1}{e^{|\alpha|/2}+e^{-|\alpha|/2}}. \label{fd}
\eeq

To obtain the density, one need only Fourier transform Eq.~(\ref{fd}).  First note that on the real line it is equal to the analytic function given by simply removing the absolute values.  With a small perturbation which can later be removed, this function shrinks exponentially on either the positive or negative semicircle of the complex plane.  Let us choose the positive semicircle.  This contour encircles the poles at
\beq
\alpha=\pi i (2k+1)\hsp k\in \Z
\eeq
where the residues are $-i (-1)^k e^{\pi(2k+1)\lambda}/(2\pi)$.  Therefore the density is
\beq
\rho(\lambda)=\int_{-\infty}^{\infty}e^{-i\alpha\lambda}\tilde{\rho}(\alpha)d\alpha
=\sum_{k=0}^{\infty}(-1)^k e^{\pi(2k+1)\lambda}=\frac{e^{\pi\lambda}}{1+e^{2\pi\lambda}}=\frac{1}{2{\mathrm{cosh}}(\pi\lambda)}.
\eeq
Thus
\beq
\frac{d\lambda}{dx}=\frac{1}{\rho(\lambda)}=2{\mathrm{cosh}}(\pi\lambda). \label{ldiffeq}
\eeq
This equation is the starting point for studies of the thermodynamics of this model.  

We will need explicit expressions for the various functions of $x$.  To find these, we must solve Eq.~(\ref{ldiffeq}).  Multiplying through by $dx/2$cosh$(\pi\lambda)$ and integrating one obtains
\bea
x+\frac{1}{4}&=&\int_{-1/4}^{x}dx=\int_{-\infty}^{\lambda(x)}\frac{d\lambda}{e^{\pi\lambda}+e^{-\pi\lambda}}=\int_{-\infty}^{\lambda(x)}d\lambda\sum_{k=0}^{\infty}(-1)^k e^{\pi(2k+1)\lambda}\nonumber\\
&=&\sum_{k=0}^{\infty}(-1)^k\frac{e^{\pi(2k+1)\lambda}}{\pi(2k+1)}\bigg\rvert_{-\infty}^{\lambda(x)}=\sum_{k=0}^{\infty}(-1)^k\frac{e^{\pi(2k+1)\lambda(x)}}{\pi(2k+1)}=\frac{-i}{2\pi}{\mathrm{ln}}\left(\frac{1+ie^{\pi\lambda(x)}}{1-ie^{\pi\lambda(x)}}\right)\nonumber\\
&=&\frac{1}{\pi}{\mathrm{ArcTan}}\left(e^{\pi\lambda(x)}\right)
\eea
which is easily inverted to obtain
\beq
\lambda(x)=\frac{1}{\pi}{\mathrm{ln}}\left({\mathrm{Tan}}\left[\pi\left(x+\frac{1}{4}\right)\right]\right). \label{lamfin}
\eeq
Substituting this into Eqs.~(\ref{kl}) and (\ref{fl}) gives the needed results
\bea
K(x)&=&\pi-2{\mathrm{ArcTan}}\left(\frac{2}{\pi}{\mathrm{ln}}\left({\mathrm{Tan}}\left[\pi\left(x+\frac{1}{4}\right)\right]\right)\right)\\
\Phi(x,y)&=&2\ {\mathrm{ArcCot}}\left(\frac{1}{\pi}
{\mathrm{ln}}\left(
\frac{{\mathrm{Tan}}\left[\pi\left(x+\frac{1}{4}\right)\right]}{{\mathrm{Tan}}\left[\pi\left(y+\frac{1}{4}\right)\right]}
\right)
\right). \label{kffin}
\eea
Similarly one finds
\beq
\rho(x)=\frac{{\mathrm{cos}}(2\pi x)}{2}.
\eeq

Note that the function $\lambda(x)$ given in Eq.~(\ref{lamfin}) is an exact solution of the continuum equation (\ref{vaceq}) but not of the exact discrete equation (\ref{atan}).  At large $n$ with $j$ constant, the $j$th equation in Eq.~(\ref{atan}) is violated by $c_j/n$ where $c_j$ is independent of $n$ to leading order.  The left hand side is always larger.  Numerically we have found $c_1=0.45$, $c_{10}=0.18$, $c_{100}=0.009$, $c_{1000}=0.004$ and so on.  Shifting an individual $\lambda$ to adjust for this shift yields a change of order $1/n$.  However it is not obvious that when all $\lambda$ are consistently adjusted together, the correction will vanish at large $n$ at fixed $x$.  

In fact in Subsec.~\ref{tornasez} we will see that the $O(N^2)$ contribution to the variance of the anchored $\alpha$ is the difference between two terms which differ by about 1\%.  In principle, it is possible that such a small difference is an artefact of the continuum approximation and would vanish if we solved the original discrete system.  To test this, we have used $({\ref{atan}})$ to solve for the left hand side, which we then substituted into the right hand side and so on iteratively 200 times with chain lengths of thousands and we found that convergence appears to arrive after of order 100 recursions, with a total change in $\lambda$ of less than about $1\%$ at every site, and much less than $1\%$ far from the boundaries.

\section{The Anchor} \label{ancorasez}

\begin{figure} 
\begin{center}
\includegraphics[width=2.5in,height=1.7in]{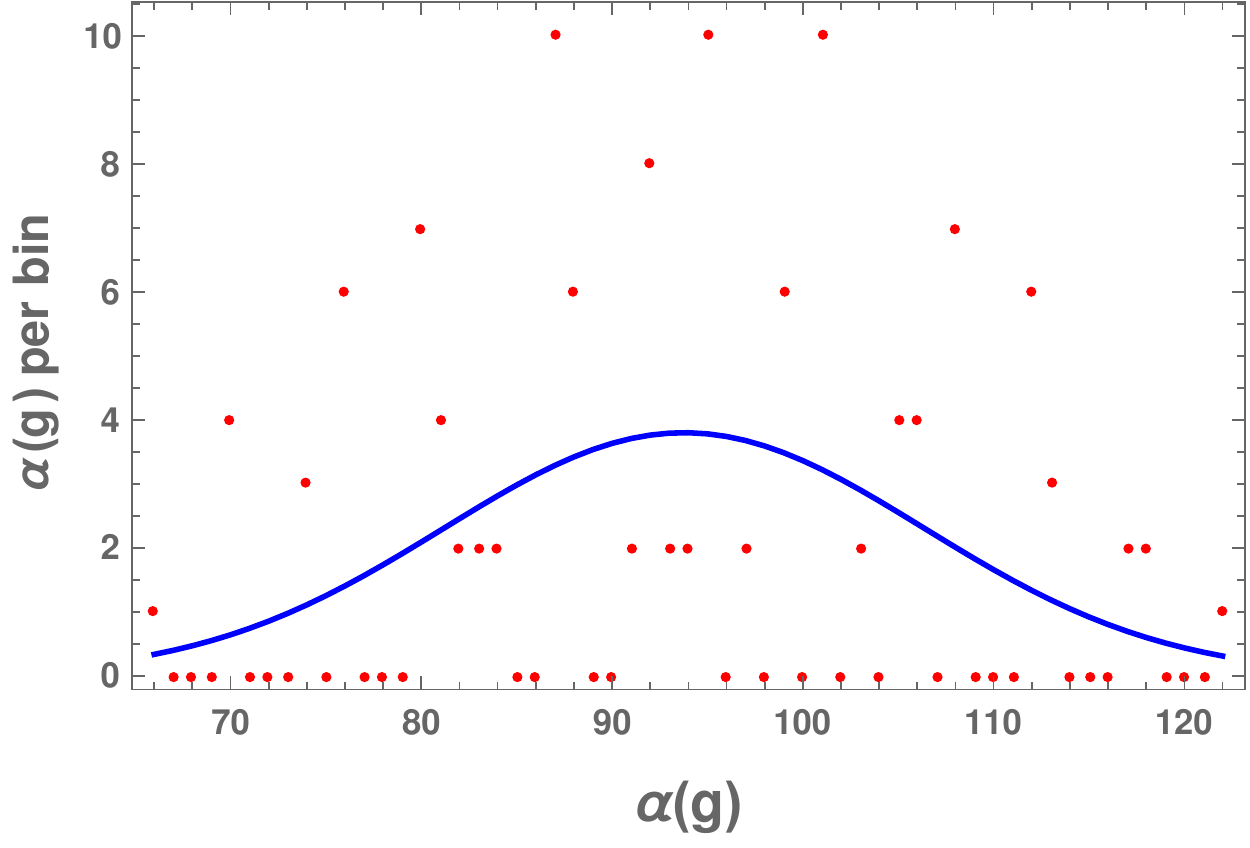}
\includegraphics[width=2.5in,height=1.7in]{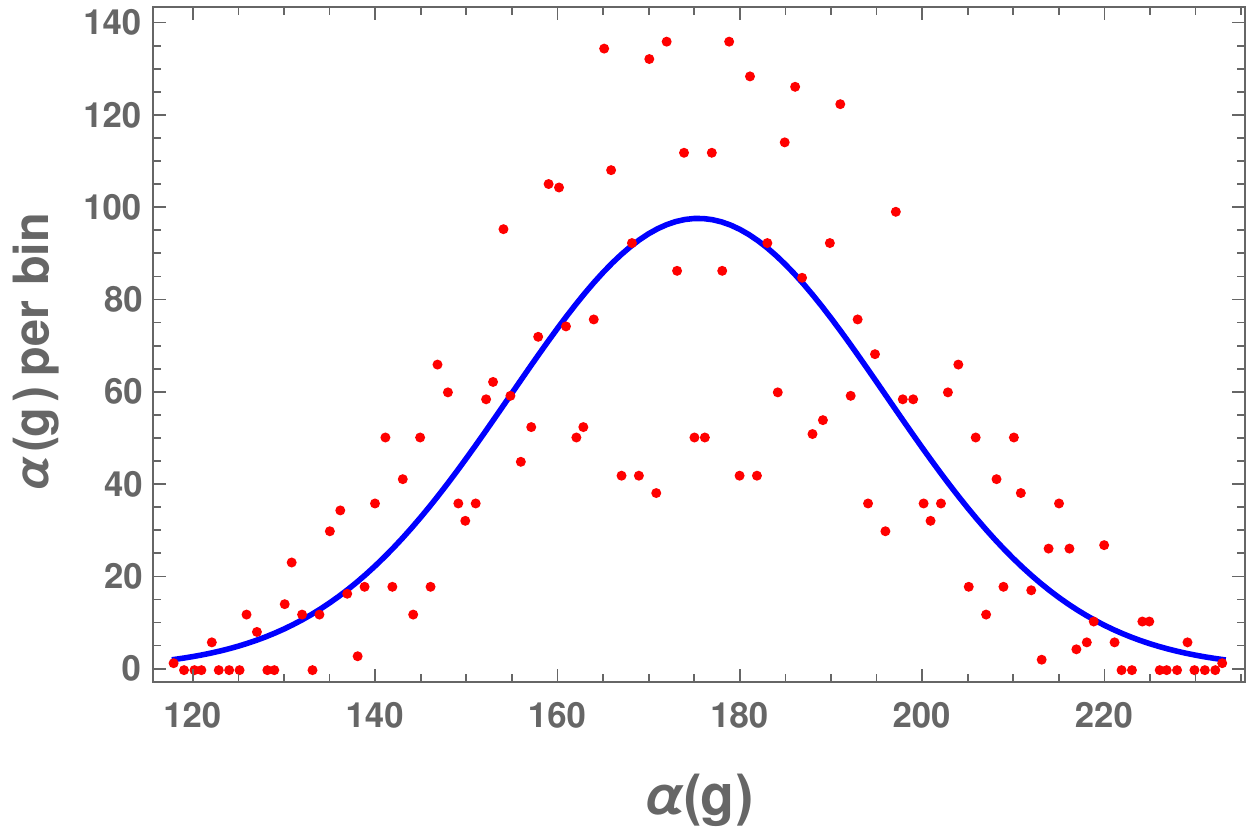}
\includegraphics[width=3.0in,height=2.3in]{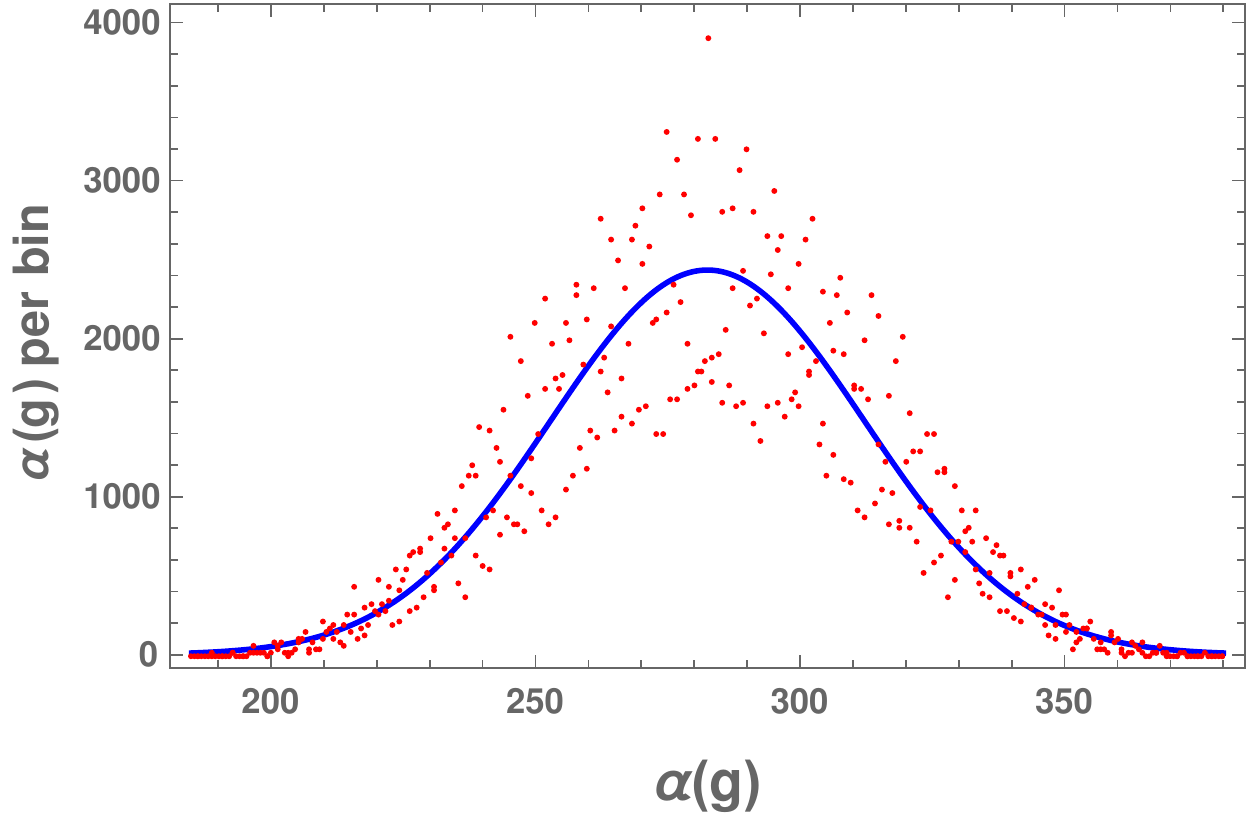}
\caption{Histograms of the distributions of $\alpha(g)$ at $N=10$ (top-left), $N=14$ (top-right) and $N=18$ (bottom) computed numerically for $\langle 0|\Omega\rangle$.  The distribution is dominated by structure, making the best fit Gaussian approximation, show in blue, quite poor everywhere.  The bin size is $1$ for $N=10$ and $N=14$ and $0.5$ for $N=18$.}
\label{noafig}
\end{center}
\end{figure}

Recall that the coordinate Bethe Ansatz expresses the matrix elements in the form
\beq
a=\sum_{g\in S_n} e^{i\alpha(g)} \label{betheb}
\eeq
where the phase
\beq
\alpha(g)=\alpha_1(g)+\alpha_2(g)\hsp
\alpha_1(g)=\sum_{j=1}^n m(j) K(P(j))\hsp
\alpha_2(g)=\frac{1}{2}\sum_{j<k}^n\Phi(P(j),P(k)) \label{adef}
\eeq
depends on the permutation $g\in S_n$.  At large $n$, the sum (\ref{betheb}) becomes an integral (\ref{ft}) with measure given by the density $\rho(\alpha)$.  Eq.~(\ref{ft}) states that the matrix elements are given by the Fourier transform of $\rho(\alpha)$.  This function is shown in Fig.~\ref{noafig} in the case of the matrix element $\langle 0|\Omega\rangle$ at $N=10,\ 14$\ and $18$.  The numerical precision is very high, so the scatter seen here is intrinsic to the function.  Therefore we see that $\rho(\alpha)$ is unfortunately rich in substructure.  In fact, this substructure lies at sufficiently large scales so as to contribute to the Fourier transform, and so its evaluation is a difficult task.  

The role of the anchor $\alpha_3(g)$ is to shift
\beq
\alpha(g)\rightarrow\alpha^\prime(g)=\alpha(g)-\alpha_3(g)
\eeq
so as to cancel out the substructure.  The anchor $\alpha_3(g)$ will be an integral multiple of $2\pi$ and so the shift will not affect $e^{i\alpha}$.  Therefore the substitution of $\alpha$ with $\alpha^\prime$ leaves the matrix elements invariant.  To see how the anchor works, and to motivate it, we will first consider the substructure created by two actions of the cyclic group $\Z_n$.

\subsection{Type I Cyclic Permutations}

One can define a free action of the cyclic group $\Z_n$ on the permutation group $S_n$ as follows.  Let the generator $1\in\Z_n$ act on $g\in S_n$ by
\beq
1:S_n\rightarrow S_n:g\mapsto g^\prime:P(j)\mapsto P^\prime(j)=P(j+1{\mathrm{\ mod\ }}n).
\eeq
Restrict our attention to matrix elements with the classical ground state $|0\rangle$, which corresponds to $m(j)=2j$.  In this case, and only in this case, we will now show that $\Z_n$ is an exact symmetry of the phases
\beq
e^{i\alpha(g)}=e^{i\alpha({g^\prime})}.
\eeq

Indeed, $\alpha(g^\prime)$ is easily calculated
\bea
\alpha(g^\prime)&=&\sum_{j=1}^n m(j) K(P(j+1{\mathrm{\ mod\ }}n))+\frac{1}{2}\sum_{j<k}^n\Phi(P(j+1{\mathrm{\ mod\ }}n),P(k+1{\mathrm{\ mod\ }}n))\nonumber\\
&=&\sum_{j=2}^{n} m(j-1) K(P(j))+m(n)K(P(1))\nonumber\\
&&+\frac{1}{2}\sum_{j=2}^{n-1}\sum_{k=j+1}^{n}\Phi(P(j),P(k))+\frac{1}{2}\sum_{j=2}^{n}\Phi(P(j),P(1)).
\eea
Now, fixing $m(j)=2j$ we find
\beq
\alpha(g^\prime)-\alpha(g)=-2\sum_{j=2}^{n} K(P(j))+2(n-1)K(P(1))-\frac{1}{2}\sum_{j=2}^{n}\Phi(P(1),P(j))+\frac{1}{2}\sum_{j=2}^{n}\Phi(P(j),P(1)).
\eeq
Using the antisymmetry of $\Phi$ this simplifies to
\beq
\alpha(g^\prime)-\alpha(g)=-2\sum_{j=1}^{n} K(P(j))+2nK(P(1))-\sum_{j=2}^{n}\Phi(P(1),P(j)).
\eeq
$K$ is symmetrically distributed about $\pi$ and so the first term on the right hand side is just $-2n\pi$.  Bethe's equation (\ref{betheeq}) on the other hand gives the sum of the second and third terms to be $2\pi Q(P(1))$.  Putting this all together we obtain
\beq
\alpha(g^\prime)-\alpha(g)=2\pi(-n+2n-2P(1)+1)=2\pi(n-2P(1)+1). \label{t1diff}
\eeq

This is an integer multiple of $2\pi$.  Thus we have shown that these cyclic permutations leave each summand in the matrix elements invariant, and yet they affect the arguments $\alpha(g)$ and so complicate the distribution $\rho(\alpha)$.  Clearly, to calculate this distribution, it would be desirable to remove these spurious shifts.  How can this be done?

Consider a second action of the generator of the cyclic group.  Now
\beq
\alpha(g^{\prime\prime})-\alpha(g^\prime)=2\pi(n-2P^\prime(1)+1)=2\pi(n-2P(2)+1)
\eeq
and so
\beq
\alpha(g^{\prime\prime})-\alpha(g)=2\pi(2n-2P(2)-2P(1)+2).
\eeq
In general the element $k$ of the cyclic group shifts the arguments by
\beq
\alpha(g^{(k)})-\alpha(g)=2\pi\sum_{i=1}^k(n-2P(i)+1).
\eeq

How can we modify $\alpha(g)$ to prevent these spurious shifts?  Recall that $P:[1,n]\rightarrow [1,n]$ is a bijection and so it is invertible and the inverse transforms under the cyclic action by
\beq
(P^\prime)^{-1}(j)=P^{-1}(j)-1.
\eeq
Choose any integer $k\in[1,n]$ and define
\beq
\alpha^I_3(g)=-2\pi\sum_{i=1}^{P^{-1}(k)}(n-2P(i)+1).
\eeq
How does this transform?
\bea
\alpha^I_3(g^\prime)&=&-2\pi\sum_{i=1}^{(P^{\prime})^{-1}(k)}(n-2P^\prime(i)+1)\\
&=&-2\pi\sum_{i=1}^{P^{-1}(k)-1}(n-2P(i+1)+1)=-2\pi\sum_{i=2}^{P^{-1}(k)}(n-2P(i)+1)\nonumber
\eea
and so the difference is
\beq
\alpha^I_3(g^\prime)-\alpha^I_3(g)=2\pi(n-P(1)+1)=\alpha(g^\prime)-\alpha(g).
\eeq

\begin{figure} 
\begin{center}
\includegraphics[width=2.5in,height=1.7in]{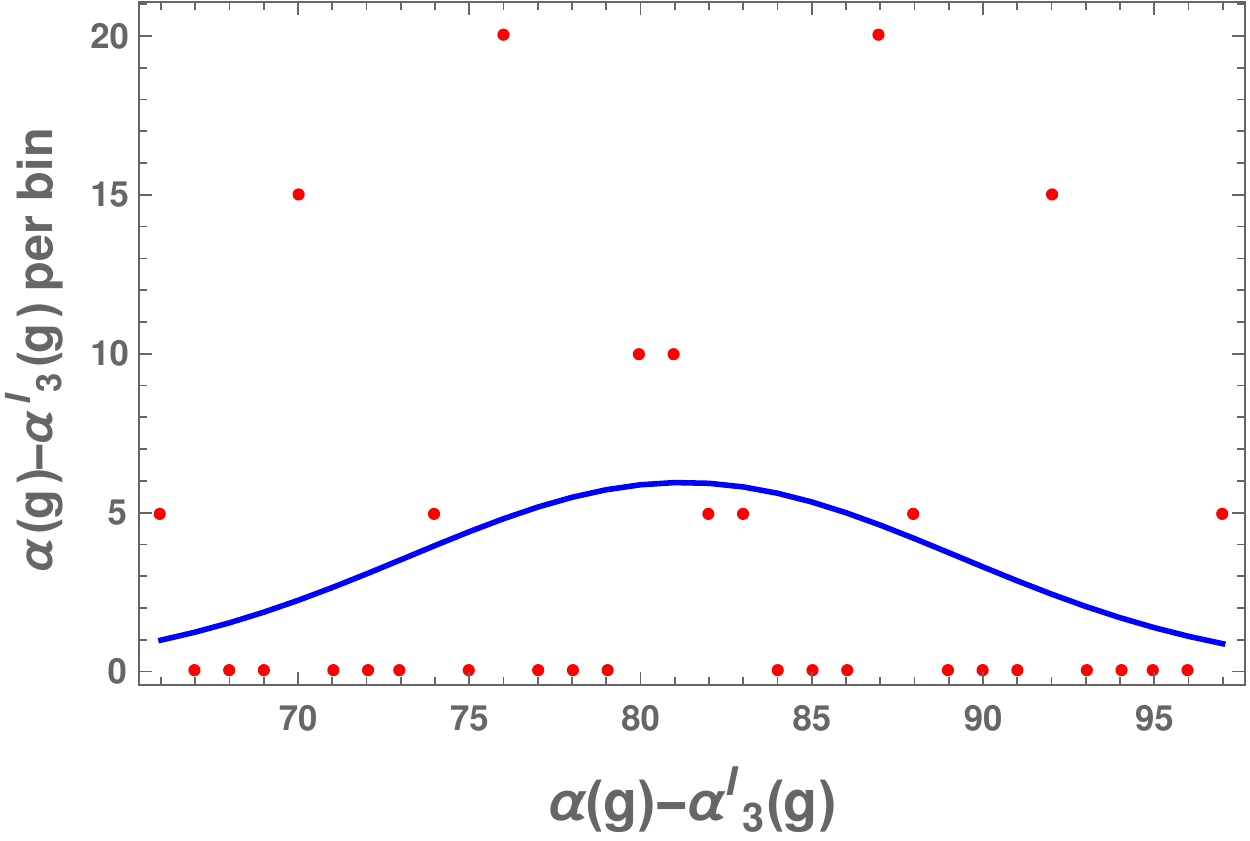}
\includegraphics[width=2.5in,height=1.7in]{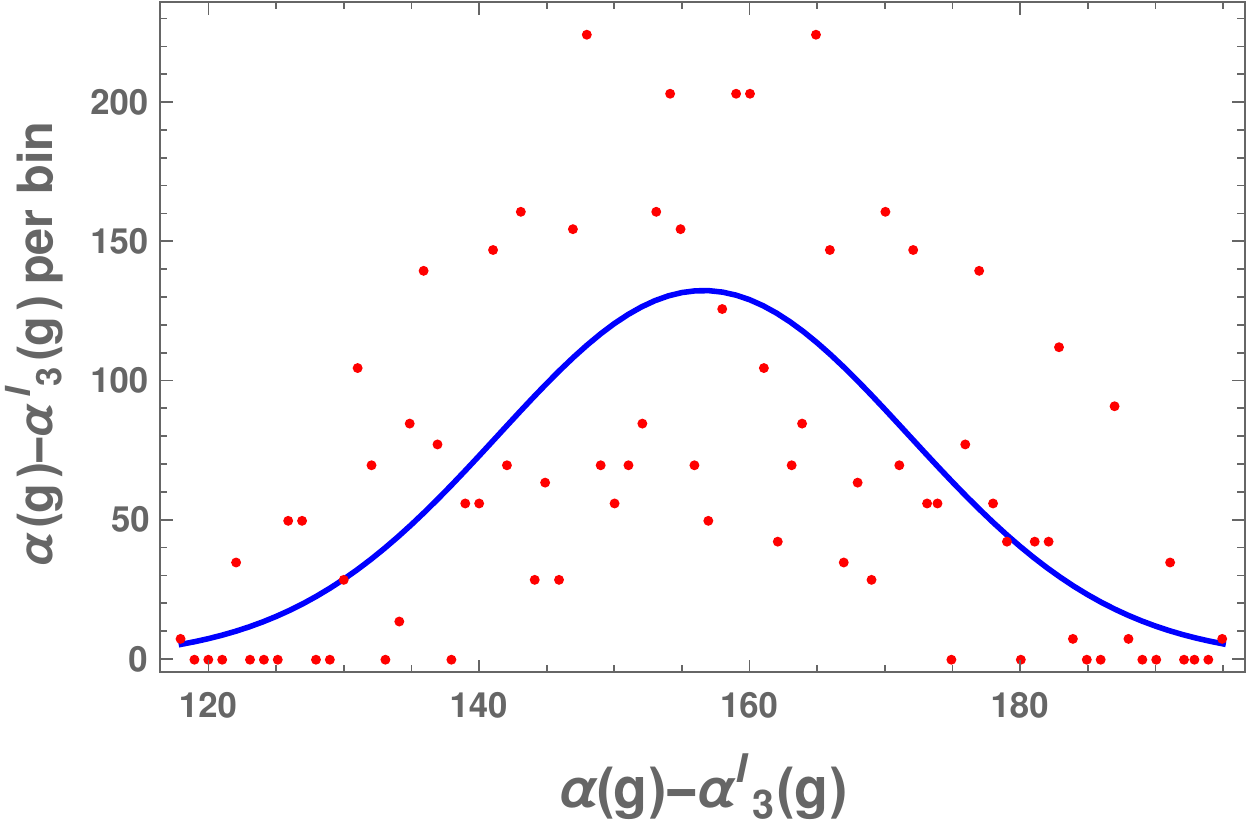}
\includegraphics[width=3.0in,height=2.3in]{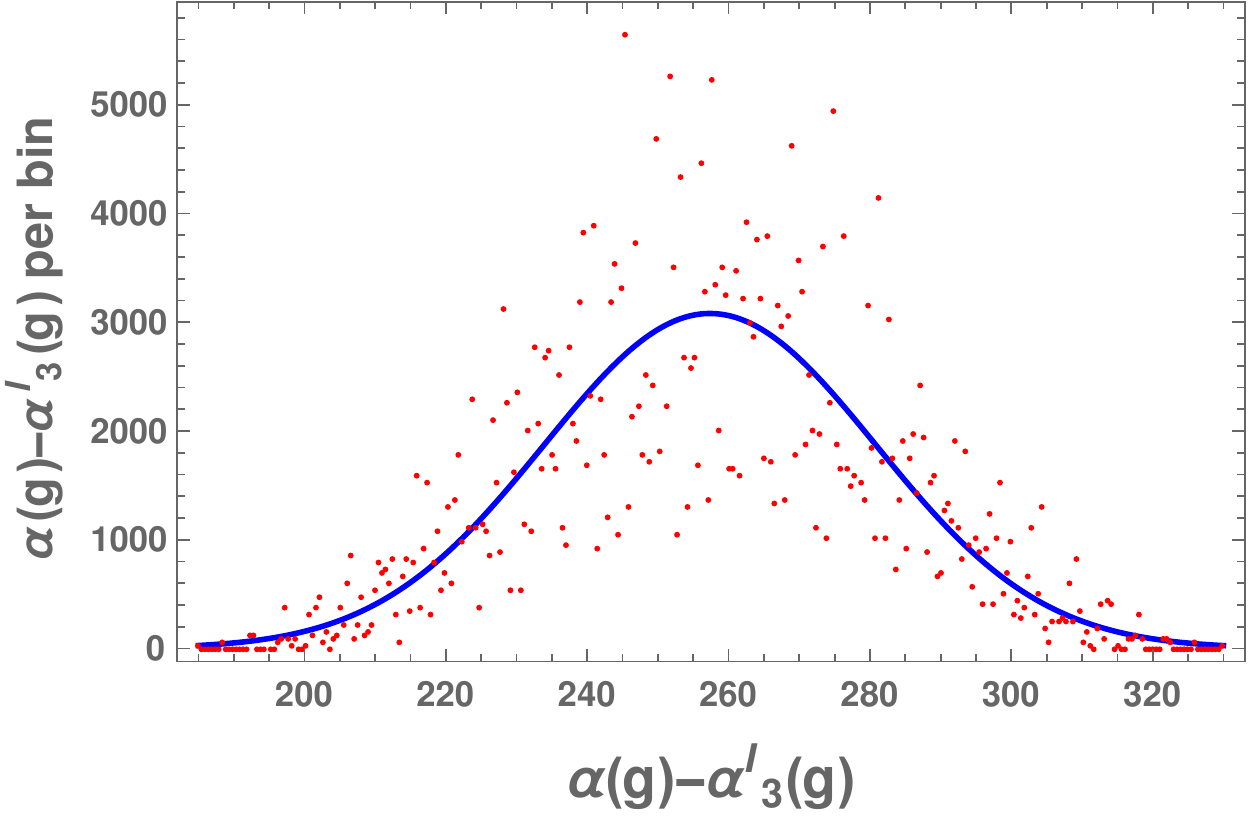}
\caption{Histograms of the distributions of $\alpha(g)-\alpha_3^I(g)$ at $N=10$ (top-left), $N=14$ (top-right) and $N=18$ (bottom) computed numerically for $\langle 0|\Omega\rangle$.   As compared with $\alpha(g)$ in Fig.~\ref{noafig}, the standard deviations have dropped from 12.6, 20.6 and 29.7 to 8.0, 15.2 and 23.5 respectively.  However the Gaussian approximation is still quite poor.}
\label{anc1fig}
\end{center}
\end{figure}

As $\alpha^I_3$ and $\alpha$ transform identically under the cyclic permutations, their difference $\alpha(g)-\alpha^I_3(g)$ is invariant.  Thus $\alpha^I_3(g)$ so defined is an anchor which fixes these cyclic permutations.  However it is not the only such anchor.  One may add to it any other integral multiple of $2\pi$ which is invariant under these cyclic transformations and so obtain another such anchor.  Below we will see that there is another cyclic action which is not fixed, and so this choice of $\alpha^I_3(g)$ is not optimal.  In Fig.~\ref{anc1fig} we see that this $\alpha^I_3$ does reduce the scatter in the distribution of the phase arguments.  At $N\sim 15$ the reduction in the variance is about a factor of 2 but at higher $N$ it is smaller as each orbit of the $\Z_n$ action is quite small in $S_n$.   It also leaves considerable substructure and so is not sufficient for the calculation of matrix elements in a moment expansion.

\subsection{Type II Cyclic Permutations} \label{tipo2}

The symmetric group $S_n$ admits another free $\Z_n$ action, whose generator acts by
\beq
1:g\rightarrow g^\prime:P^\prime(j)=P(j)+1{\mathrm{\ mod\ }}n. \label{cycgen}
\eeq
This action does {\it{not}} leave the phases invariant.  But, for $x$ not too close to the boundaries, it leaves the phases $e^{i\alpha(g)}$ reasonably invariant while dramatically shifting the arguments $\alpha(g)$.  Repeating the calculation as above, with this action, one obtains
\bea
\alpha(g^\prime)-\alpha(g)&=&\sum_{j\neq P^{-1}(n)}^n m(j) (K(P(j)+1)-K(P(j)))+m(P^{-1}(n))(K(1)-K(n))\nonumber\\
&&+\frac{1}{2}\sum_{j<l;j,l\neq P^{-1}(n)}^n\left(\Phi(P(j)+1,P(l)+1)-\Phi(P(j),P(l))\right)\nonumber\\
&&+\frac{1}{2}\sum_{j=1}^{P^{-1}(n)-1}\left(-\Phi(1,P(j)+1)-\Phi(P(j),n)\right)\nonumber\\
&&+\frac{1}{2}\sum_{j=P^{-1}(n)+1}^{n}\left(\Phi(1,P(j)+1)+\Phi(P(j),n)\right).
\eea
This time the calculation is more difficult.  Again consider the classical ground state $m(j)=2j$.  Now if $P$ is a cyclic permutation
\beq
P(j)=j+k{\mathrm{\ mod\ }}n.
\eeq
then
\beq
P^\prime(j)=P(j)+1{\mathrm{\ mod\ }}n=j+k+1{\mathrm{\ mod\ }}n = P(j+1{\mathrm{\ mod\ }}n)
\eeq
and so type II cyclic permutations are, in this case, identical to type I cyclic permutations.  Therefore as before 
\beq
\alpha(g^\prime)-\alpha(g)=\Delta:=2\pi(n-2P(1)+1)=-2\pi\left(n-2P^{-1}(n)-1\right). \label{deleq}
\eeq
The trivial rewriting in the last step will allow this result to approximately generalize to other permutations $P$ as we will now explain.

\begin{figure} 
\begin{center}
\includegraphics[width=2.5in,height=1.7in]{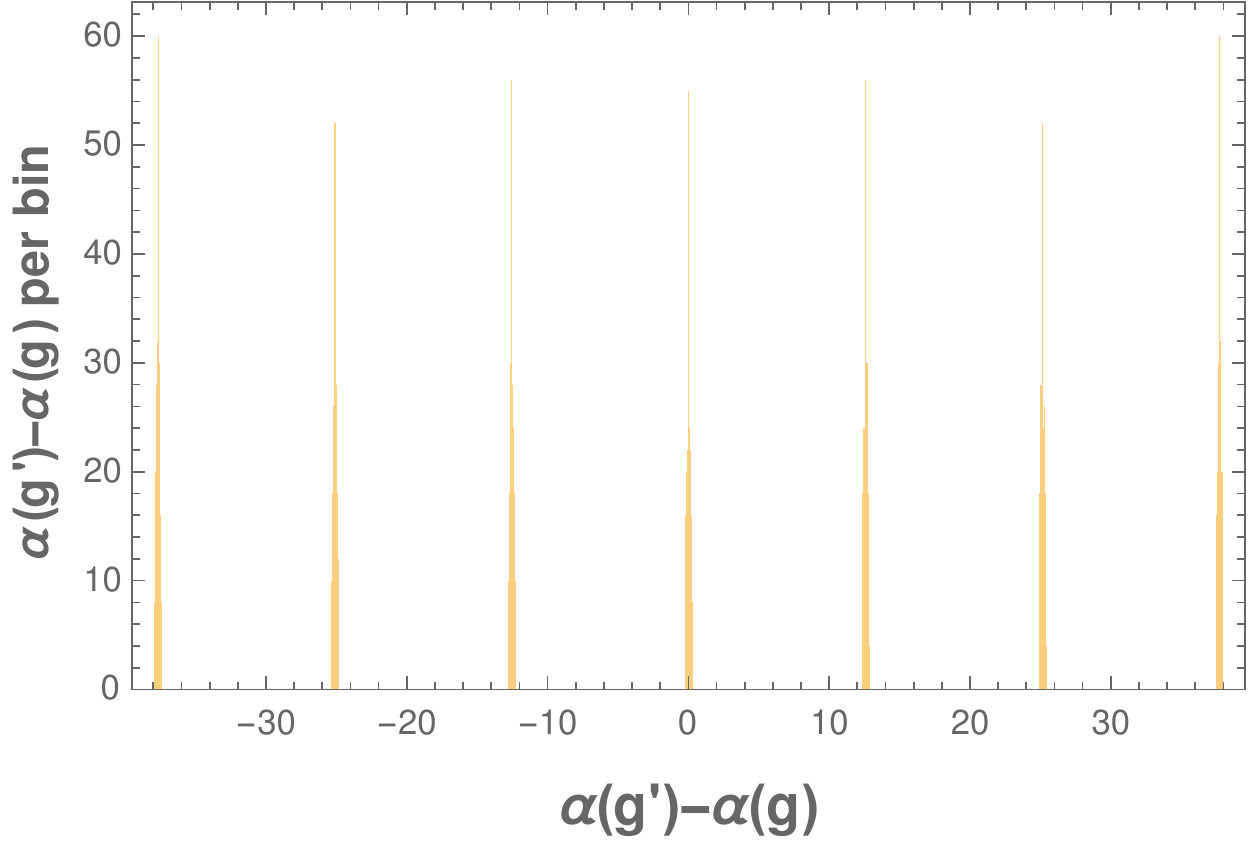}
\includegraphics[width=2.5in,height=1.7in]{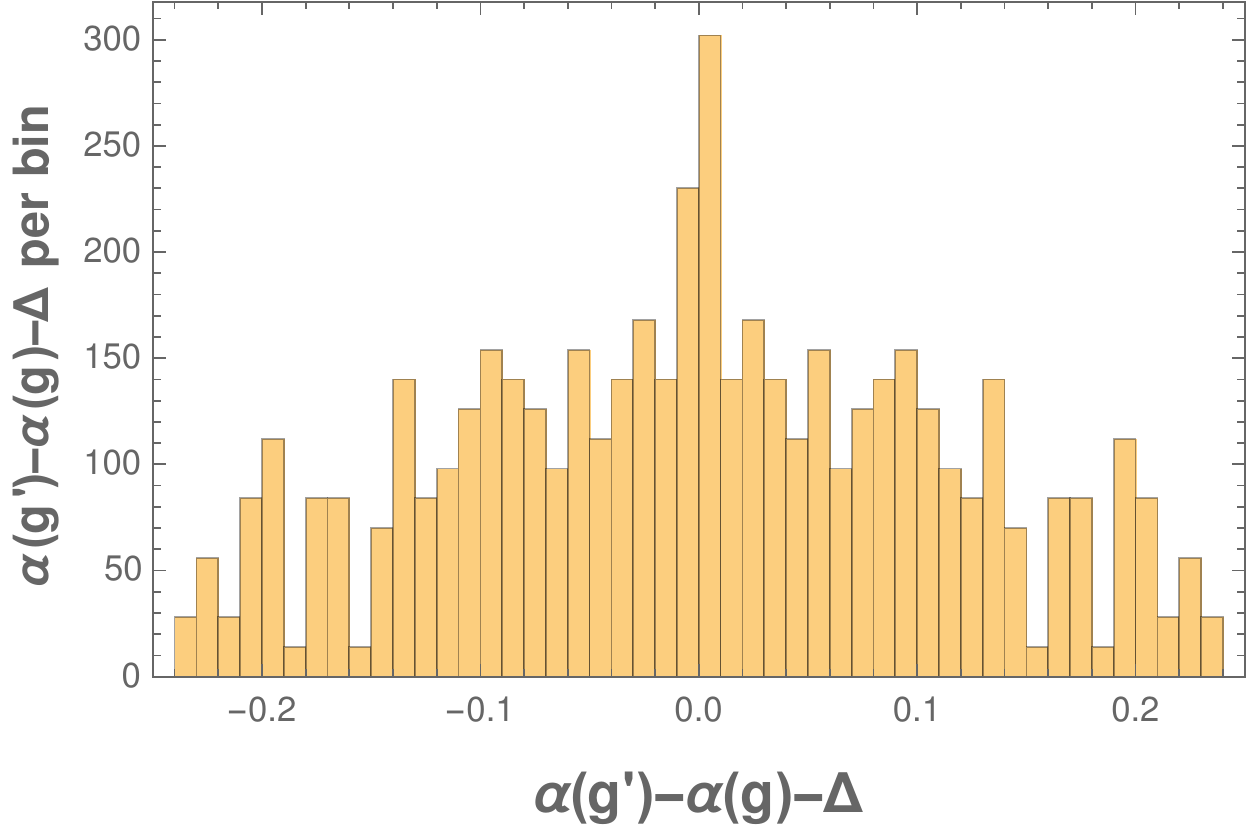}
\includegraphics[width=2.5in,height=1.7in]{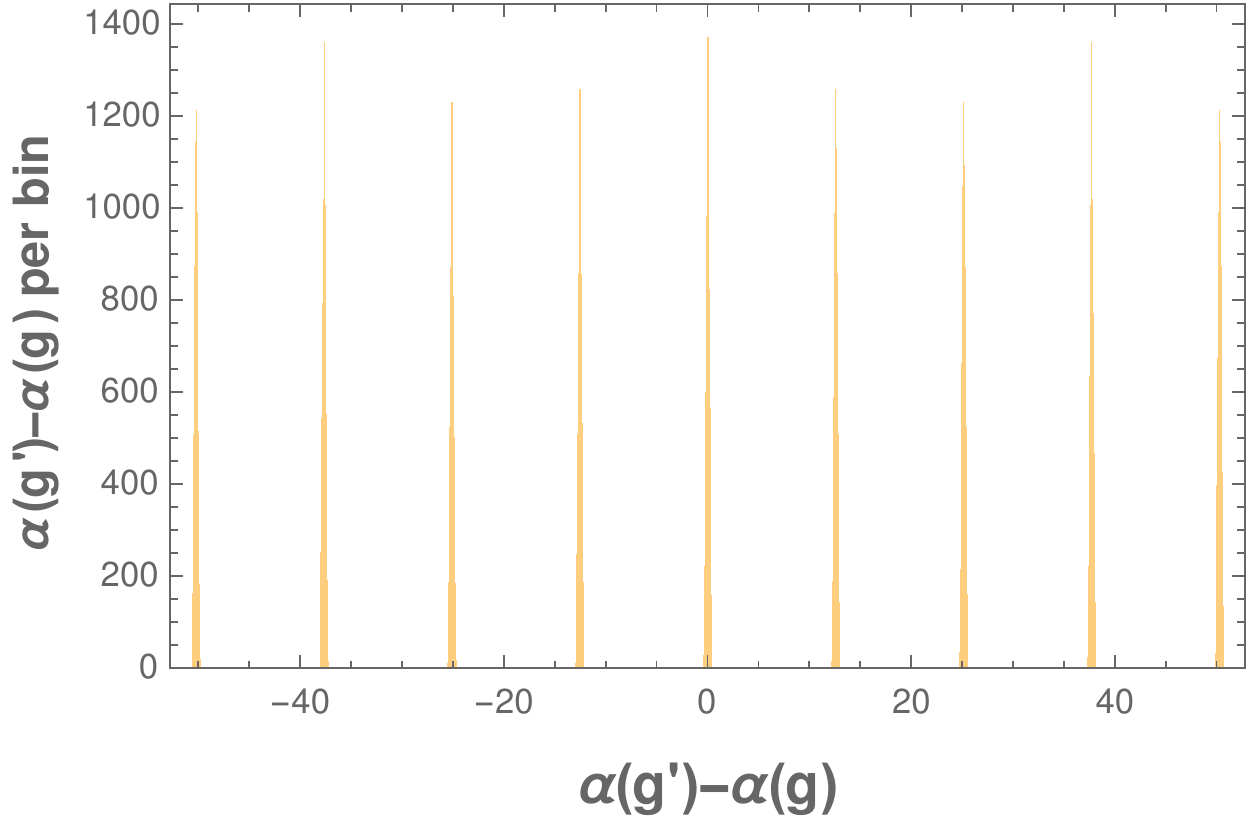}
\includegraphics[width=2.5in,height=1.7in]{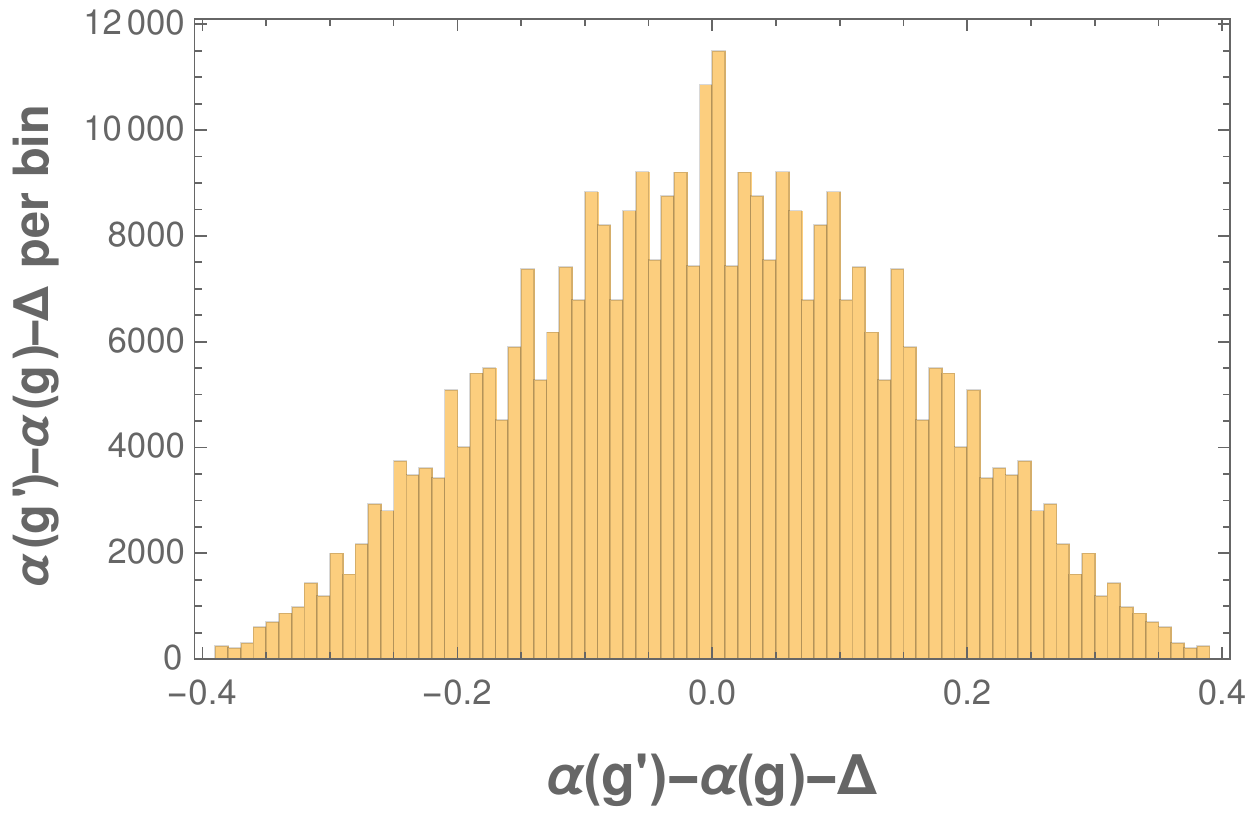}
\caption{Histograms of the distributions of $\alpha(g^\prime)-\alpha_(g)$ (left) and $\alpha(g^\prime)-\alpha_(g)-\Delta$ (right) at $N=14$ (top) and $N=18$ (bottom) computed numerically for $\langle 0|\Omega\rangle$. The bin width is $0.01$.  One sees that $\alpha(g^\prime)-\alpha(g)$ is nearly an integral multiple of $2\pi$, in fact $4\pi$ as $n$ is odd.  On the other hand, $\alpha(g^\prime)-\alpha_(g)-\Delta$ is quite small.}
\label{deltafig}
\end{center}
\end{figure}

Any two permutations $P$ in the symmetric group $S_n$ are related by a series of basic permutations in which pairs of adjacent numbers are permuted.  In particular, any $P$ is related to a cyclic permutation, for which (\ref{deleq}) holds, by some series of basic permutations.  We have checked numerically, for $N\sim 15$, that $\alpha(g^\prime)-\alpha(g)-\Delta$ varies by less than about $0.05$ under each basic permutation.  In this sense, $\Delta$, as defined in (\ref{deleq}) is a reasonable approximation for $\alpha(g^\prime)-\alpha(g)$ for any permutation $P$, even those which are not cyclic.  On the other hand,  under some basic permutations $\alpha(g^\prime)-\alpha(g)$ jumps by an integer multiple of $2\pi$.   Therefore, $\Delta$ contains all of the $2\pi$ jumps resulting from the unit type II cyclic permutation (\ref{cycgen}).  The distributions of $\alpha(g^\prime)-\alpha(g)$ and $\alpha(g^\prime)-\alpha(g)-\Delta$  are shown in Fig.~\ref{deltafig}.  It is evident here that the second has a much smaller scatter.

\begin{figure} 
\begin{center}
\includegraphics[width=2.5in,height=1.7in]{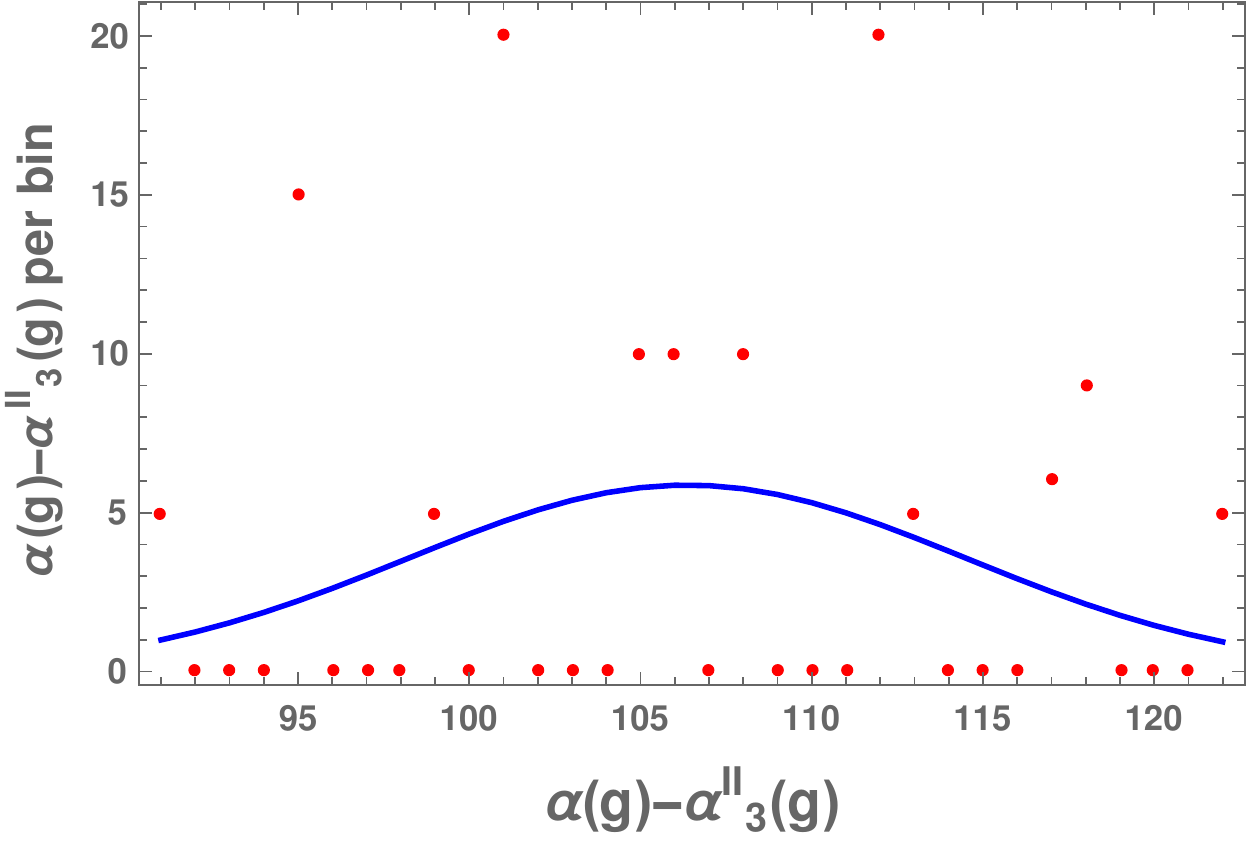}
\includegraphics[width=2.5in,height=1.7in]{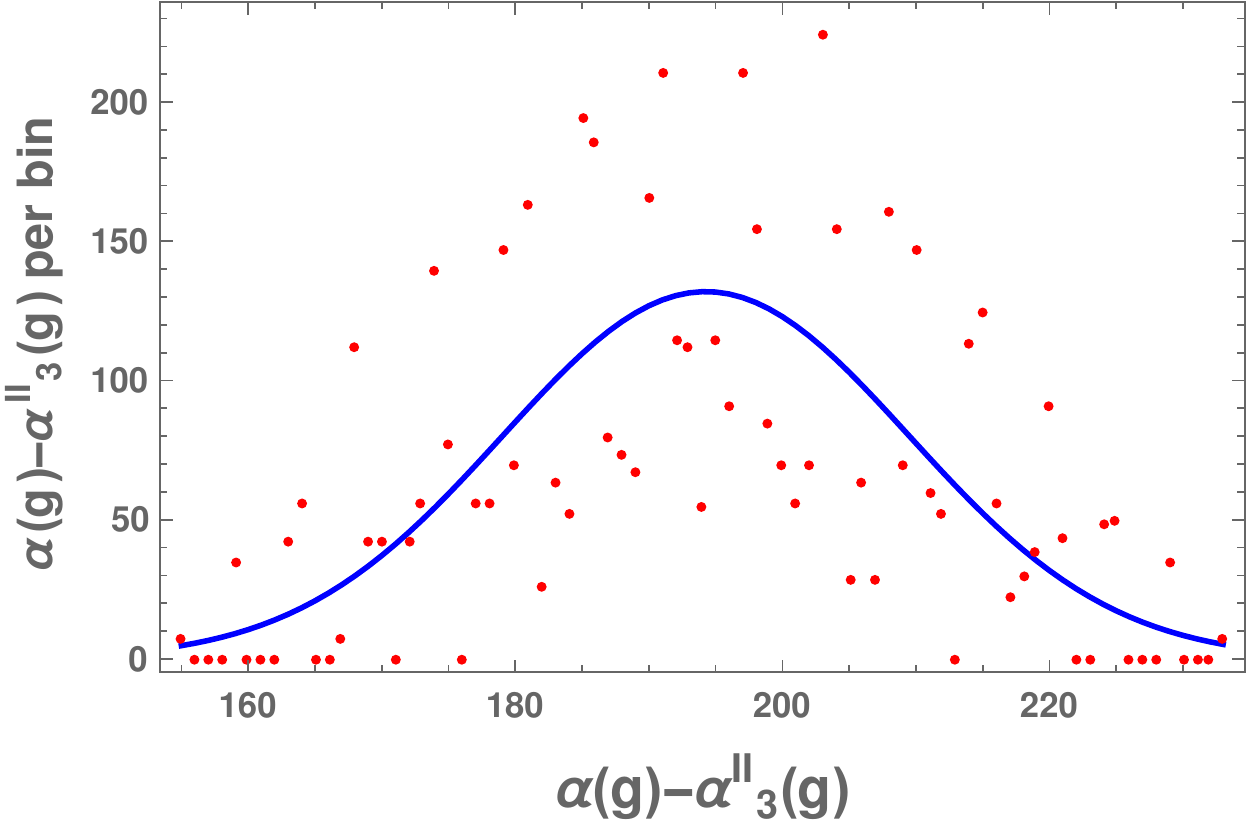}
\includegraphics[width=3.0in,height=2.3in]{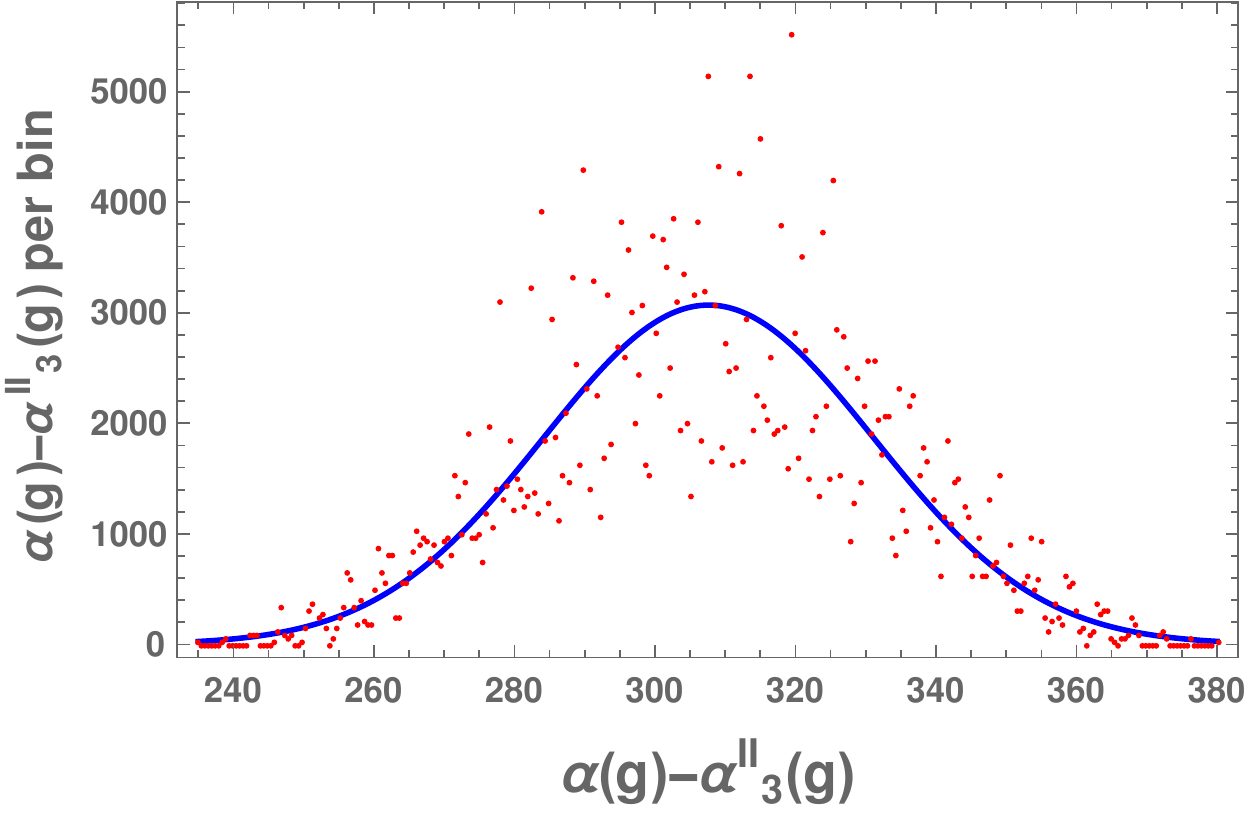}
\caption{Histograms of the distributions of $\alpha(g)-\alpha_3^{II}(g)$ at $N=10$ (top-left), $N=14$ (top-right) and $N=18$ (bottom) computed numerically for $\langle 0|\Omega\rangle$.   As compared with $\alpha(g)$ in Fig.~\ref{noafig}, the standard deviations have dropped from 12.6, 20.6 and 29.7 to 8.2, 15.2 and 23.6 respectively.  However the Gaussian approximation is still quite poor.}
\label{anc2fig}
\end{center}
\end{figure}


Again it is not difficult to construct an anchor which reproduces this transformation law for an arbitrary type II cyclic permutation
\beq
\alpha_3^{II}(g)=2\pi\sum_{i=1}^{P(k)-1}(n-2P^{-1}(i)+1)
\eeq
where $k$ is arbitrary.  However this anchor does not leave the type I cyclic permutations invariant, and as can be seen in Fig.~\ref{anc2fig} it causes a reduction in the scatter of $\alpha$ which is comparable to that of $\alpha_3^I$. 

\subsection{A Universal Anchor}

\begin{figure} 
\begin{center}
\includegraphics[width=2.5in,height=1.7in]{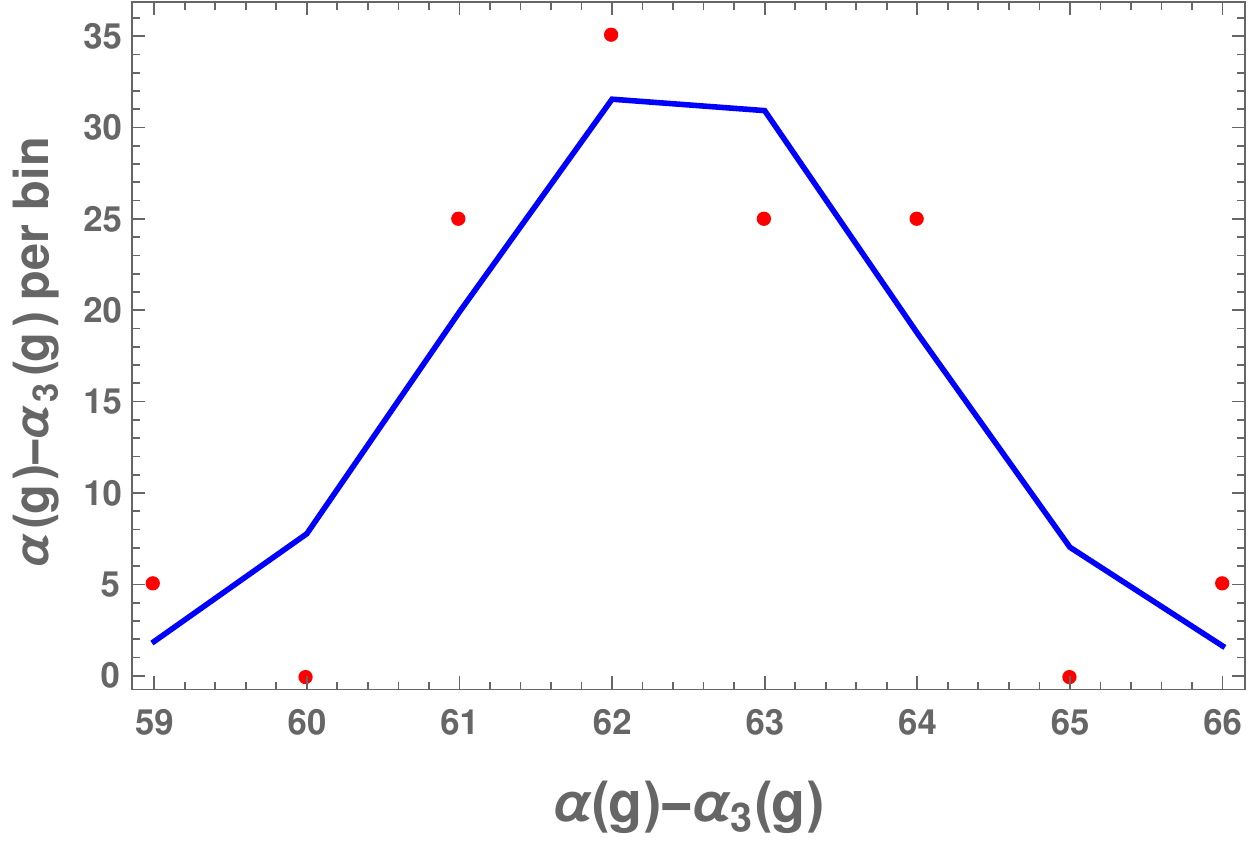}
\includegraphics[width=2.5in,height=1.7in]{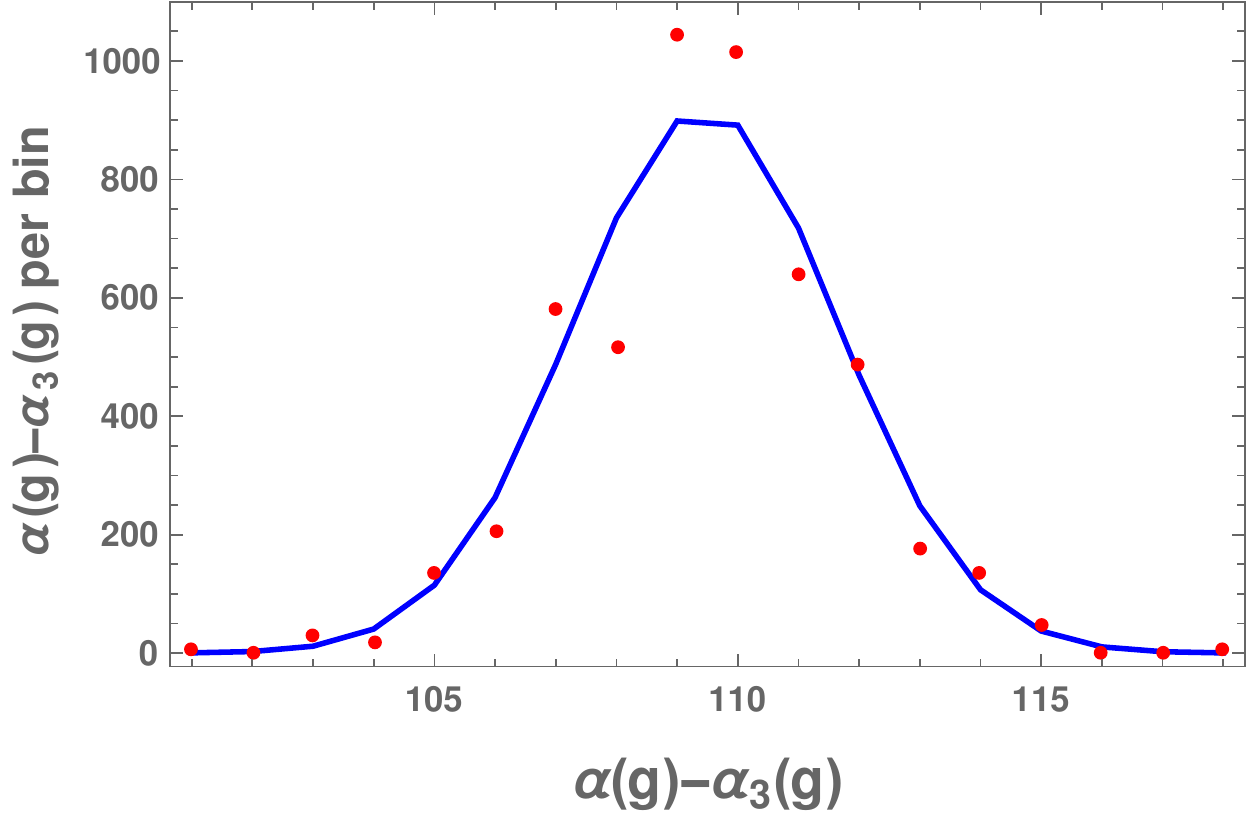}
\includegraphics[width=3.0in,height=2.3in]{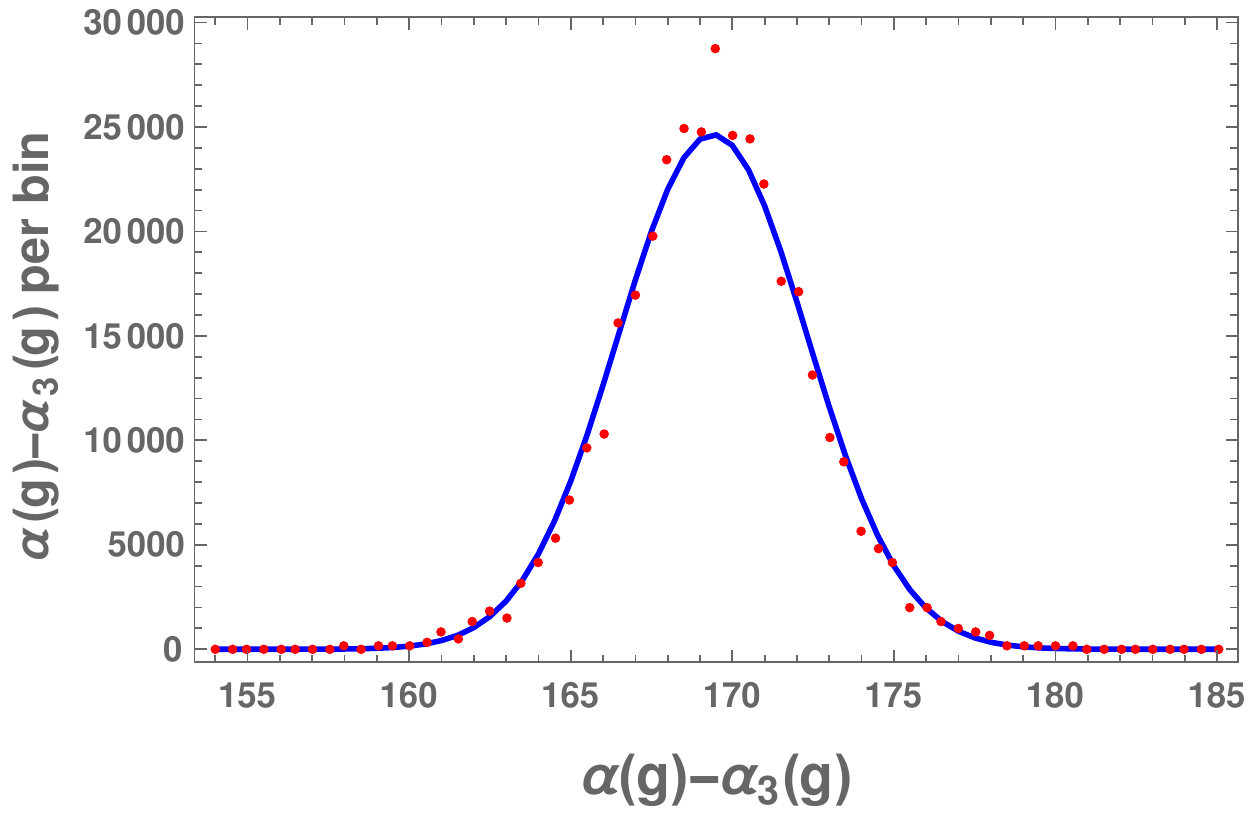}
\caption{Histograms of the distributions of $\alpha(g)-\alpha_3(g)$ at $N=10$ (top-left), $N=14$ (top-right) and $N=18$ (bottom) computed numerically for $\langle 0|\Omega\rangle$.   As compared with $\alpha(g)$ in Fig.~\ref{noafig}, the standard deviations have dropped from 12.6, 20.6 and 29.7 to 1.4, 2.2 and 3.2 respectively.  Finally the Gaussian approximation is reasonable.}
\label{anc3fig}
\end{center}
\end{figure}

We propose that the argument $\alpha(g)$ in the Bethe Ansatz (\ref{betheb}) and (\ref{adef}) be replaced by the anchored argument
\beq
\alpha^\prime(g)= \alpha_1(g)+\alpha_2(g)-\alpha_3(g)
\eeq
where the anchor $\alpha_3(g)$ is defined by
\beq
\alpha_3(g)=2\pi \sum_{j<k} \theta(P(j)-P(k)) \label{ancora}
\eeq
where the Heaviside step function is
\beq
\theta(x)=\left\{\begin{tabular}{ll}$1$ \ \ if $x>0$\\$0$\ \ \   otherwise.\\
\end{tabular} \right. 
\eeq
The trivial permutation $P(j)=j$ gives $\alpha_3(g)=0$.  More generally, this counts the number of pairs of sites whose order is flipped by $g$.

We will see that the anchored argument $\alpha^\prime(g)$ has a number of nice properties, not shared by $\alpha(g)$.  In this subsection we will see that it is invariant under type I permutations.  $\alpha_3(g)$ jumps by $\Delta$ under type II permutations, and so while $\alpha^\prime(g)$ is not invariant under these permutations, its shift is relatively modest.  Of course we do not want invariance under type II permutations, as these affect $e^{i\alpha}$.   Numerically, one can see in Fig.~\ref{anc3fig} that the density of $\alpha^\prime$ has far less pronounced substructure than $\alpha$ and is much better approximated by a Gaussian distribution.  Later, using the binning approximation, we will show analytically that the variance of $\alpha$ is of order $O(N^{3/2})$ but that of $\alpha^\prime$ is only $O(N)$.  Altogether these observations lead us to believe that $\alpha^\prime$ will be a more convenient variable than $\alpha$ for the evaluation of the Fourier transform in Eq.~(\ref{ft}).

How does the anchor work?  For example, begin with the identity permutation $P(j)=j$.  Now consider the type I cyclic permutation, it yields
\beq
P^\prime(j)=P(j+1)=j+1\ {\mathrm{mod}\ }n. \label{pp}
\eeq
In this case $P(n)=1$ and so a single entry has moved from the right to the left of all other $n-1$ entries, all of which were smaller.  Thus the sum gains contributions from all elements with $j=1$
\beq
\alpha_3(g^\prime)=\alpha_3(g^\prime)-\alpha_3(g)=2\pi \sum_{j<k}^n\theta(P(j)-P(k)) =2\pi \sum_{k=2}^n 1 = 2\pi(n-1)=2\pi(n-2P(1)+1).
\eeq
And so we see that $\alpha_3$ transforms just like $\alpha$ under this cyclic permutation of type I.  In fact, the transformation (\ref{pp}) is not only the generator of type I cyclic permutations, but also type II cyclic permutations, which coincide in this example because $g$ is just a shift.  Now
\beq
P^{-1}(n)=n\hsp
\Delta=2\pi (n-1)=\alpha_3(g^\prime)-\alpha_3(g)
\eeq
and so the anchor compensates for the type II permutation as well, as it must since this is also a type I permutation.

What about general elements of $S_n$?  Beginning with an arbitrary element $g\in S_n$, a type I permutation yields
\beq
P^\prime(j)=P(j+1\ {\textrm{mod}}\ n)
\eeq
and so our anchor transforms to
\bea
\alpha_3(g^\prime)&=&2\pi \sum_{j=1}^{n-1}\sum_{k=j+1}^n \theta(P^\prime(j)-P^\prime(k))\\
&=&2\pi \sum_{j=1}^{n-1}\sum_{k=j+1}^n\theta(P(j+1)-P(k+1\ {\textrm{mod}}\ n))\nonumber\\
&=&2\pi \sum_{j=1}^{n-2}\sum_{k=j+1}^{n-1}\theta(P(j+1)-P(k+1))+2\pi \sum_{j=1}^{n-1}\theta(P(j+1)-P(1))\nonumber\\
&=&2\pi \sum_{j=2}^{n-1}\sum_{k=j+1}^{n}\theta(P(j)-P(k))+2\pi \sum_{j=2}^{n}\theta(P(j)-P(1))\nonumber\\
\eea
yielding a difference of 
\bea
\alpha_3(g^\prime)-\alpha_3(g)&=&2\pi \sum_{j=2}^{n}\theta(P(j)-P(1))-2\pi \sum_{k=2}^{n}\theta(P(1)-P(k))\\
&=&2\pi(n-P(1))-2\pi(P(1)-1)=2\pi(n-2P(1)+1)\nonumber
\eea
which equals $\alpha(g^\prime)-\alpha(g)$ calculated in Eq.~(\ref{t1diff}).  Therefore $\alpha(g)-\alpha_3(g)$ is invariant under type I permutations.

What about type II permutations? Now
\beq
P^\prime(j)=P(j)+1\ {\textrm{mod}}\ n
\eeq
and so
\bea
\alpha_3(g^\prime)&=&2\pi \sum_{j=1}^{n-1}\sum_{k=j+1}^n \theta(P^\prime(j)-P^\prime(k))\\
&=&2\pi \sum_{j=1}^{n-1}\sum_{k=j+1}^n\theta((P(j)+1 \ {\textrm{mod}}\ n)-(P(k)+1\ {\textrm{mod}}\ n)))\nonumber\\
&=&2\pi \sum_{j=1,j\neq P^{-1}(n)\ }^{n-1}\sum_{k=j+1,k\neq P^{-1}(n)}^n\theta((P(j)-P(k))))\nonumber\\
&&+\sum_{k=P^{-1}(n)+1}^n\theta(1-P(k)-1)+\sum_{j=1}^{P^{-1}(n)-1}\theta(P(j)+1-1)\nonumber\\
&=&2\pi \sum_{j=1,j\neq P^{-1}(n)\ }^{n-1}\sum_{k=j+1,k\neq P^{-1}(n)}^n\theta(P(j)-P(k))+2\pi(P^{-1}(n)-1).\nonumber\\
\eea
The difference is then
\bea
\alpha_3(g^\prime)-\alpha_3(g)&=&-2\pi \sum_{k=P^{-1}(n)+1}^n\theta(n-P(k))-2\pi \sum_{j=1}^{P^{-1}(n)-1}\theta(P(j)-n)+2\pi(P^{-1}(n)-1)\nonumber\\
&=&-2\pi(n-P^{-1}(n))+2\pi(P^{-1}(n)-1)\nonumber\\
&=&-2\pi(n-2P^{-1}(n)+1)=\Delta
\eea
which agrees with the approximation to the shift in $\alpha(g)$ found in Subsec.~\ref{tipo2}.  Therefore $\alpha(g)-\alpha_3(g)$ is approximately invariant under both kinds of cyclic permutations.

\begin{figure} 
\begin{center}
\includegraphics[width=2.5in,height=1.7in]{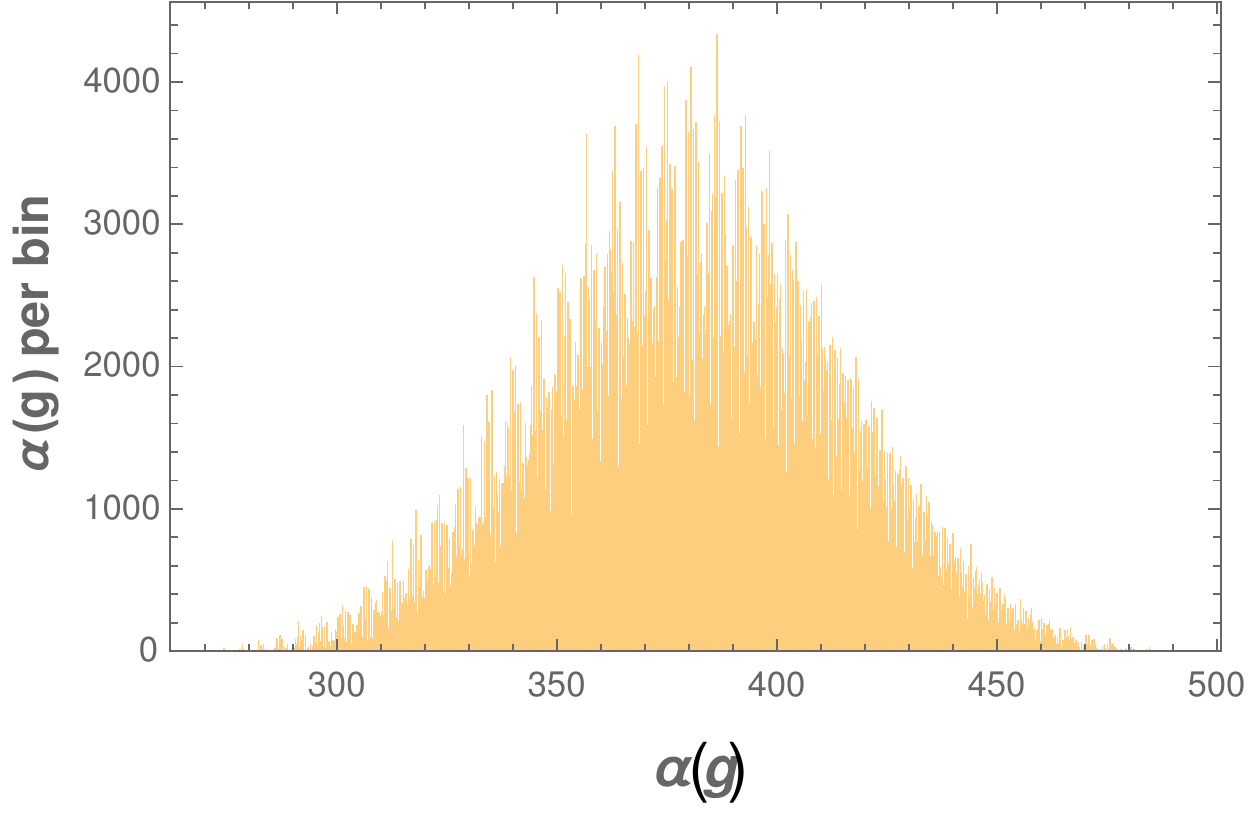}
\includegraphics[width=2.5in,height=1.7in]{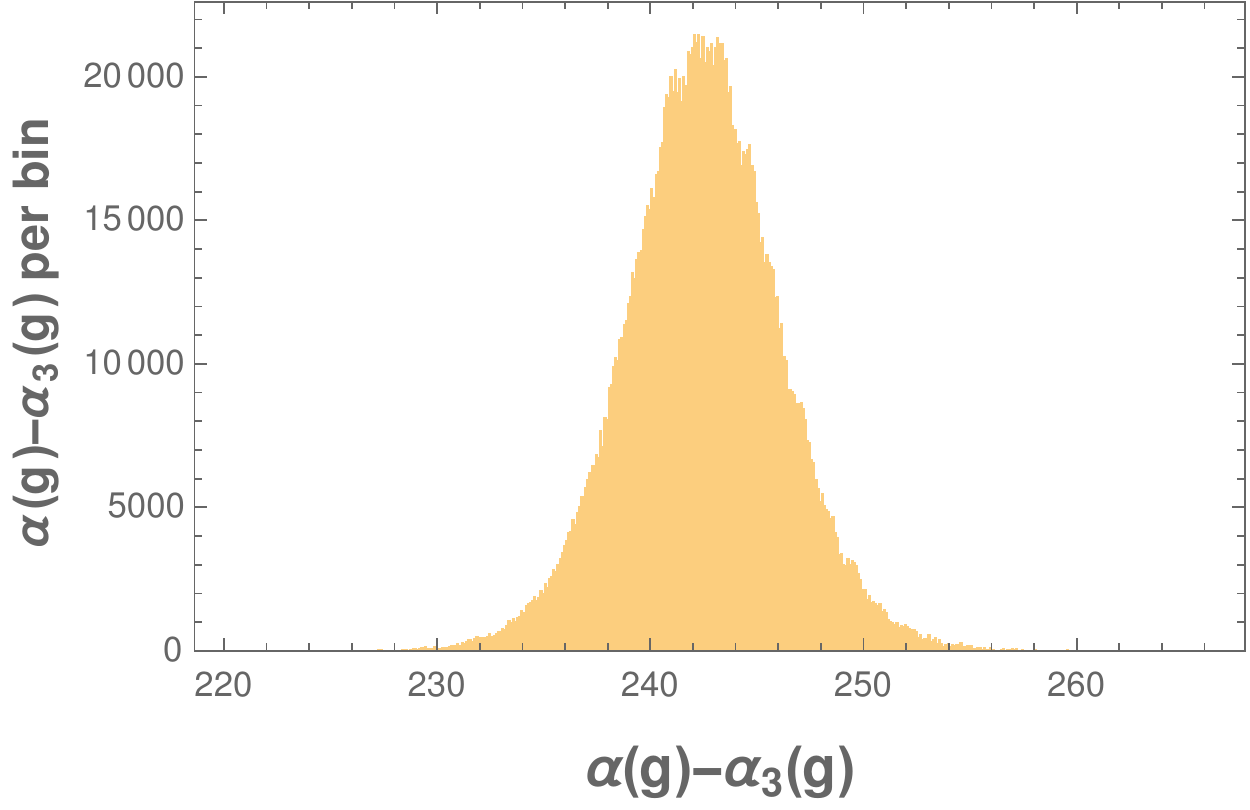}
\includegraphics[width=3.0in,height=2.3in]{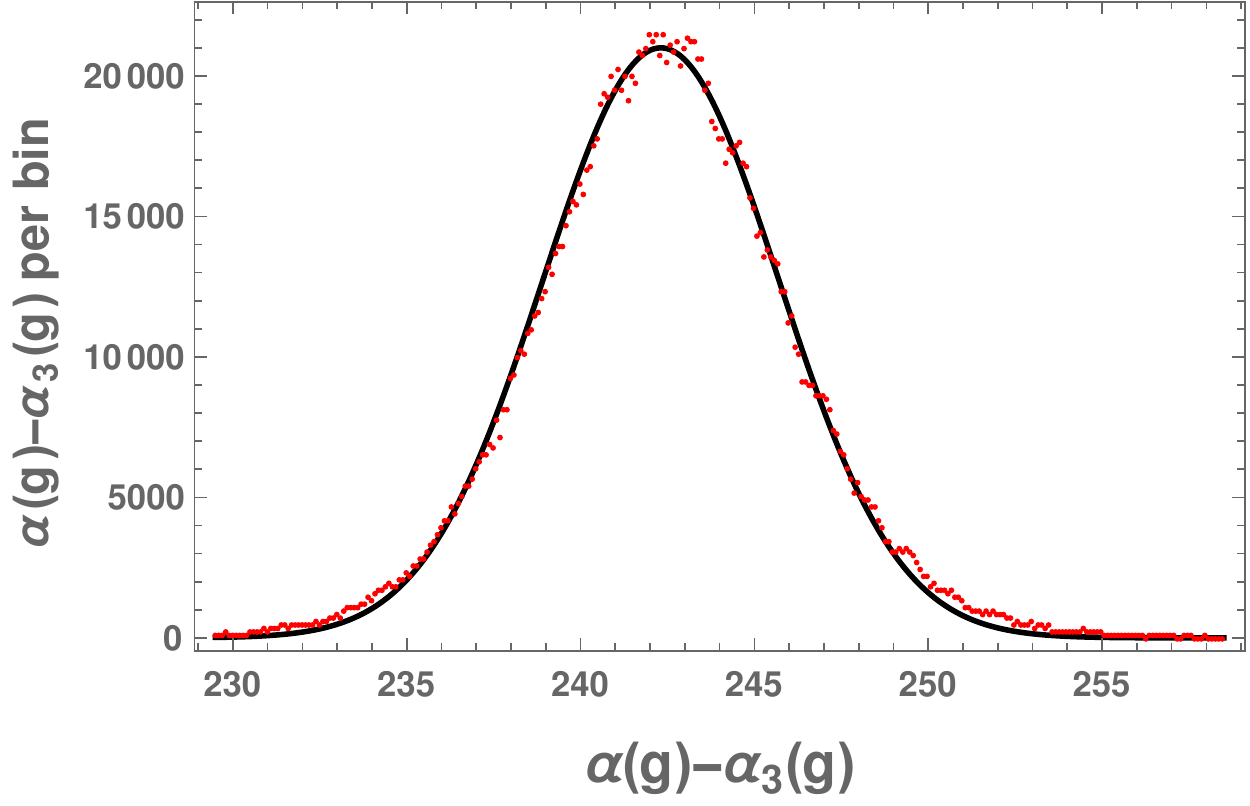}
\caption{Histograms of the distributions of $\alpha(g)$ (top-left) and $\alpha(g)-\alpha_3(g)$ (others) computed numerically at $N=22$ for $\langle 0|\Omega\rangle$.  The matrix element is dominated by structure at scales near $2\pi$.  $\alpha(g)$ has rich substructure at this scale, which dominates the matrix element.   $\alpha(g)-\alpha_3(g)$ is much thinner, with no evidence of substructure at this scale.  In the bottom panel one sees that $\alpha(g)-\alpha_3(g)$ closely fits a Gaussian of deviation $3.4$ (black curve), although there is slight leptokurtosis.  The bin width is $0.1$ and cyclic permutations of type I are used to fix $P(1)=1$.}
\label{n11fig}
\end{center}
\end{figure}

We need more.  We need $\alpha(g)-\alpha_3(g)$ to be free of substructure, so that its moments yield a well-behaved expansion about a Gaussian.   In the case of the matrix element of the classical and quantum ground states $\langle 0|\Omega\rangle$ at $N=22$, so that $n=11$, these properties are demonstrated numerically in Fig.~\ref{n11fig}.  One sees that the full width half maximum of $\alpha$ is about 70, which is about $2n^{3/2}$ as expected.  On the other hand $\alpha(g)-\alpha_3(g)$ is much thinner, with a full width half maximum of only about $6$.  We see in the bottom panel that a Gaussian provides a reasonable fit to the anchored $\alpha(g)-\alpha_3(g)$.  If this has any substructure, it lies at scales far beneath $2\pi$ where it has little effect on Eq.~(\ref{ft}) and so the matrix elements.

This is our first main result.  With the anchor (\ref{ancora}) the distribution of phases $\rho(\alpha)$ in the CBA becomes approximately a Gaussian and so the calculation of the matrix elements in Eq.~(\ref{ft}) requires only that one determine its moments.    In the rest of this note, we will describe a method for the calculation of these moments. 

\subsection{Other matrix elements}

\begin{figure} 
\begin{center}
\includegraphics[width=2.5in,height=1.7in]{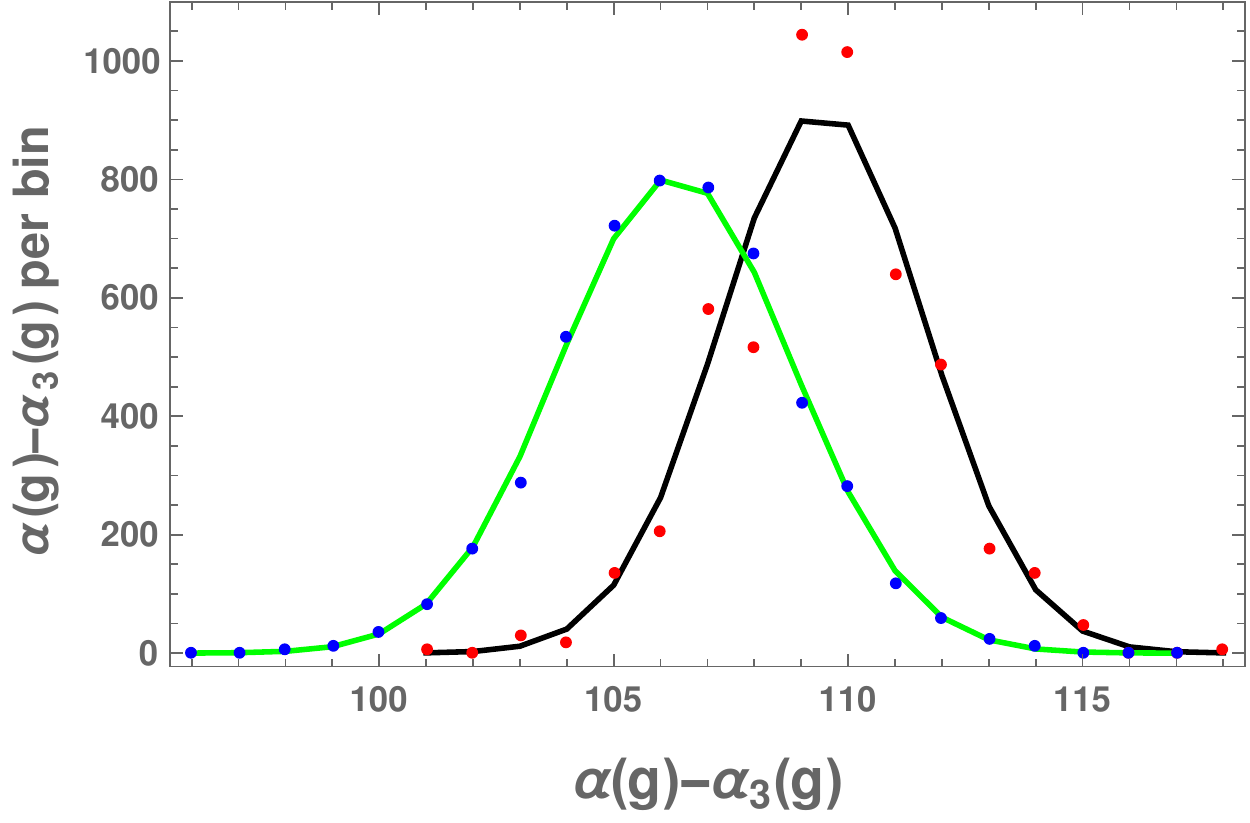}
\includegraphics[width=2.5in,height=1.7in]{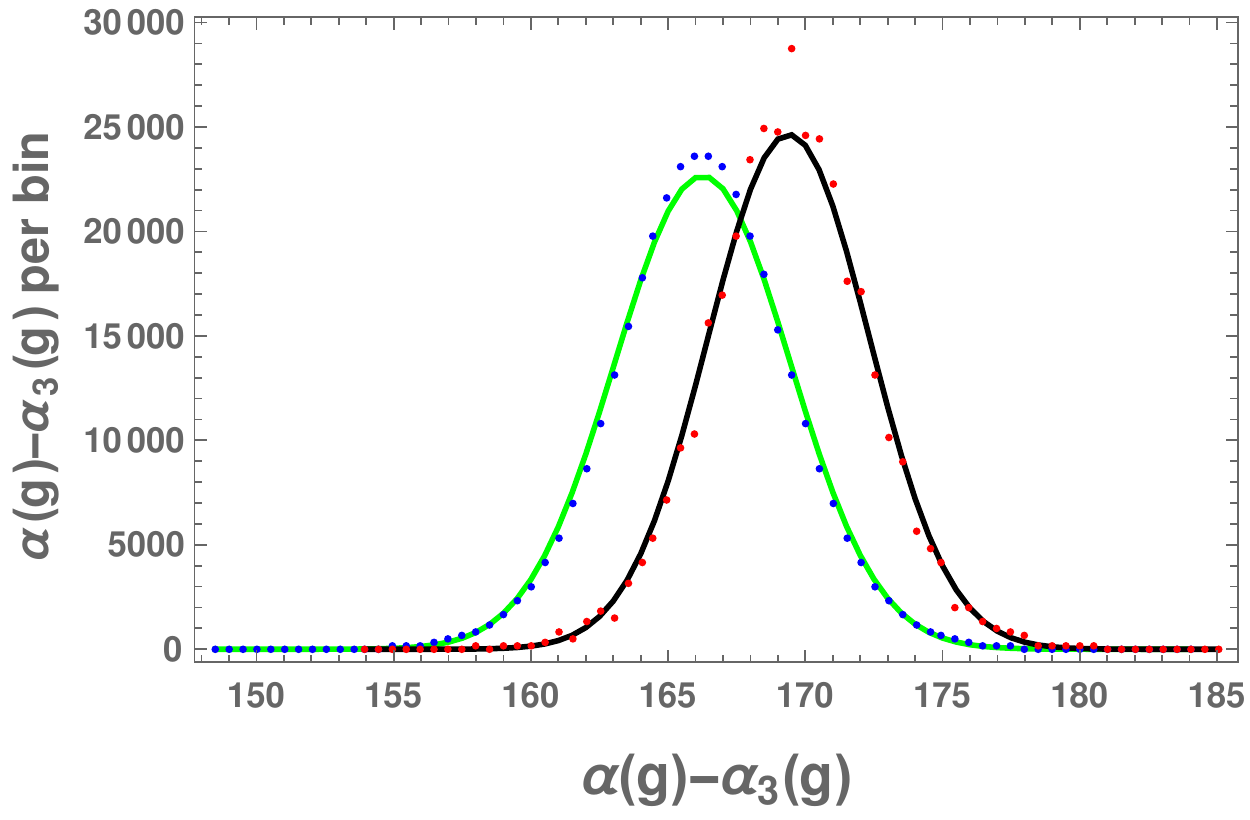}
\caption{Histograms of the distributions of $\alpha(g)-\alpha_3(g)$ computed numerically at $N=14$ (left) and $N=18$ (right) for $\langle 0|\Omega\rangle$ (red with a black Gaussian fit) and $\langle 1|\Omega\rangle$ (blue with a green Gaussian fit).  }
\label{quasifig}
\end{center}
\end{figure}

Of course we are not only interested in the matrix element $\langle 0|\Omega\rangle$.  Our anchor was motivated by the fact that the type I shift symmetry leaves $e^{i\alpha(g)}$ invariant in the case of the classical ground state $|0\rangle$.  This is not true for other states.  So how well does the anchor perform when $m(i)\neq 2i$, corresponding to other left hand sides of the matrix element?   We have only investigated this question numerically.

\begin{figure} 
\begin{center}
\includegraphics[width=3.0in,height=2.3in]{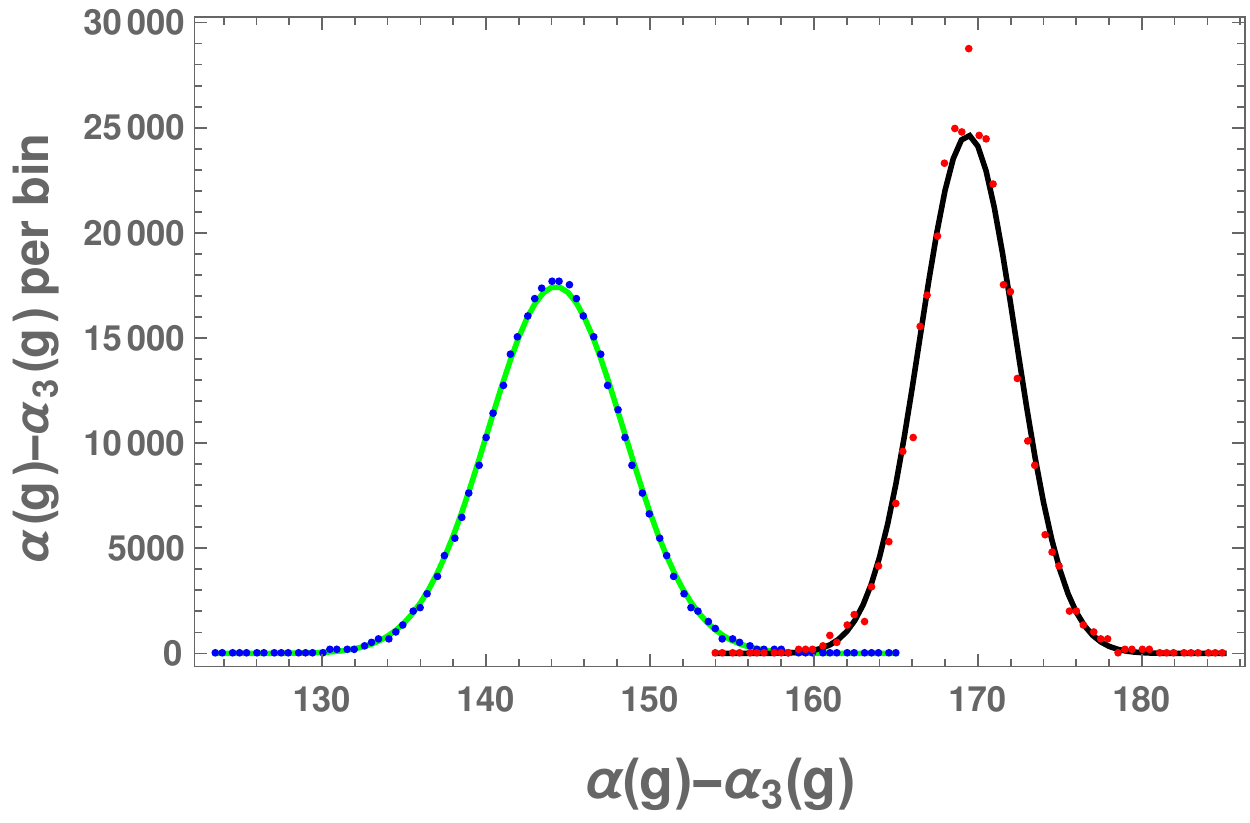}
\caption{Histograms of the distributions of $\alpha(g)-\alpha_3(g)$ computed numerically at $N=18$ for the classical ground state $\langle 0|\Omega\rangle$ (red with a black Gaussian fit) and the generic state $\langle 2|\Omega\rangle$ (blue with a green Gaussian fit).  The Gaussian approximation is quite good in both cases, although the variances differ.}
\label{randfig}
\end{center}
\end{figure}

\begin{figure} 
\begin{center}
\includegraphics[width=3.0in,height=2.3in]{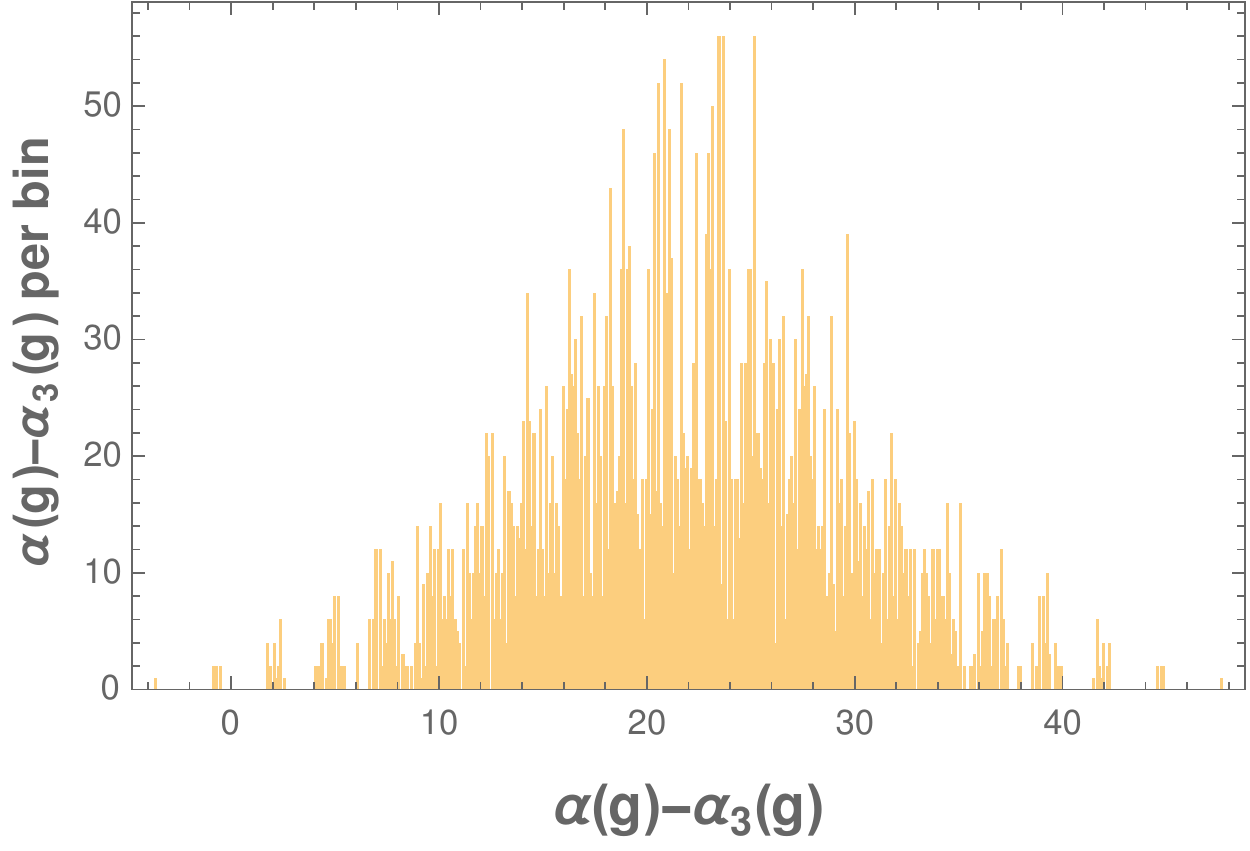}
\caption{Histogram of the distribution of $\alpha(g)-\alpha_3(g)$ computed numerically at $N=14$ \ for $\langle 3|\Omega\rangle$.  There appears to be a periodic substructure.  If this persists at large $N$ it will be an obstruction to calculating this matrix element.  However we believe that matrix elements of states so far from the classical ground state will be exponentially suppressed.}
\label{mattofig}
\end{center}
\end{figure}

First let us consider a small change, leaving all $m(i)=2i$ except for $m(2)=3$.  Let us call this state $|1\rangle$.  In Fig.~\ref{quasifig} we see that this shift in $m(2)$ leads to a shift in $\alpha-\alpha_3$, but the shape and variance are not noticeably affected.  What about matrix elements with states that are further from the classical vacuum?  Consider two more states
\beq
|2\rangle:=|1,2,4,7,10,11,14,15,18\rangle \hsp
|3\rangle:=|1,2,3,4,5,6,7\rangle
\eeq
at $N=18$ and $N=14$ respectively.  In Fig.~\ref{randfig} we compare the distribution of $\alpha(g)$ in the case of $\langle 0|\Omega\rangle$ with that of $\langle 2|\Omega\rangle$.  The values of $m(i)$ in the state $|2\rangle$ were chosen at random, so that it may represent a generic state.  One sees that for this state the shape of $\rho(\alpha^\prime)$ is still quite similar to the ground state and the increase in the variance is modest.  On the other hand, the state $|3\rangle$ was chosen to be as far as possible from the classical ground state.  In Fig.~\ref{mattofig} we see that in this case the density function has noticeable, periodic substructure which will no doubt affect the Fourier transform and will be difficult to capture in the moment expansion.  The variance is also considerably larger than in the case of the other states, although still far smaller than that of the unanchored $\rho(\alpha)$.  The viability of our strategy for calculating the matrix elements requires that the contributions of such states to physical observables be suppressed at large $N$.  We note that this is the $n=N/2$ state with the highest energy.

\section{Binning} \label{binsez}

Exact calculations of matrix elements have been a major industry for decades.  However as we are interested in the continuum field theory, our goal is somewhat different.  It is more difficult, because we will need a method which calculates matrix elements for states which differ at arbitrarily many lattice sites from any given reference state.  This distance is in general infinite, and so if our proposal requires a computation time which is polynomial in this distance then we are lost.   That said, we do not need a closed form answer.  It is sufficient to present a method for the calculation of any matrix element, so long as the time required for a given precision, as measured in units accessible to the continuum field theory, does not increase with $N$ but only with some suitable measure of the complexity of the state.  Our task is also easier because we are not interested in all states.  We are only interested in those states which survive the continuum limit.  In particular, nearby lattice sites should have similar behaviors, in the sense that they map nearby pairs of lattice sites to the same target space point via the map in Ref.~\cite{affleck}, which is reviewed in~\ref{mapapp}.  

\subsection{The Binning}

This motivates the following approach.  Let $n/q$ be an integer.  We will divide the interval $[1,n]$ into $q$ bins
\beq
{\mathcal{S}}_i=\left[\frac{n}{q}(i-1)+1,\frac{n}{q}i\right]\hsp i\in [1,q].
\eeq
Recall that an element $g\in S_n$ is completely characterized by a bijection $P:[1,n]\rightarrow[1,n]$.   Let
\beq
f_{ij}(g)=\sum_{k\in {\mathcal{S}}_i}\sum_{l\in {\mathcal{S}}_j}\delta_{P(k),l}=|P({\mathcal{S}}_i)\cap {\mathcal{S}}_j|
\eeq
where $|\mathcal{S}|$ is the cardinality of the set $\mathcal{S}$.  In other words, $f_{ij}(g)$ is the number of entries of ${\mathcal{S}}_i$ which $P$ maps into ${\mathcal{S}}_j$.  Clearly $f_{ij}(g)$ contains only some of the information in $P$, while $P$ is fully equivalent to $g$.  We will rely upon

\noindent
{\it{The Binning Postulate:}}\ For the calculation of a given quantity $X$ to any precision $\epsilon>0$, there exists a sufficiently high $q(\epsilon)$ such that, if $X$ is calculated replacing all $g$ with the same $\{f_{ij}\}$ by the same $g_{f}$ then the introduced error in $X$ will be bounded by $\epsilon$.

It may be that the binning postulate is false, or that it is true only at some leading orders in $N$.  Certainly it is false for many quantities $X$.   It is our hope that the binning postulate is true, however, for all $X$ accessible in the continuum field theory.  This requires that, in the continuum limit, the homogeneous bins (bins with nearly constant N\'eel order parameter) dominate the matrix elements.  In other words, we conjecture that {\bf{each point in the continuum field theory corresponds to a bin on the spin chain}}, and so none of the bins' internal structure survives in the continuum field theory.  

\begin{figure} 
\begin{center}
\includegraphics[width=2.5in,height=1.7in]{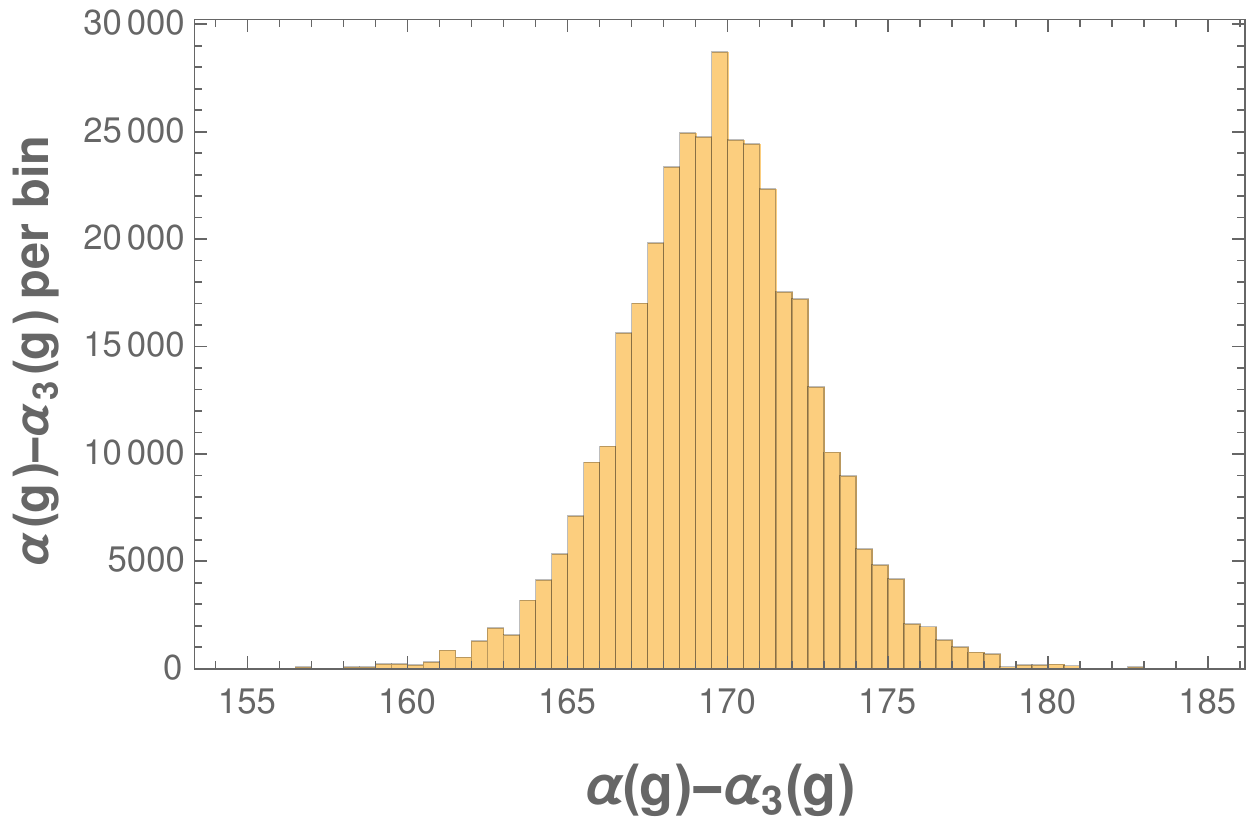}
\includegraphics[width=2.5in,height=1.7in]{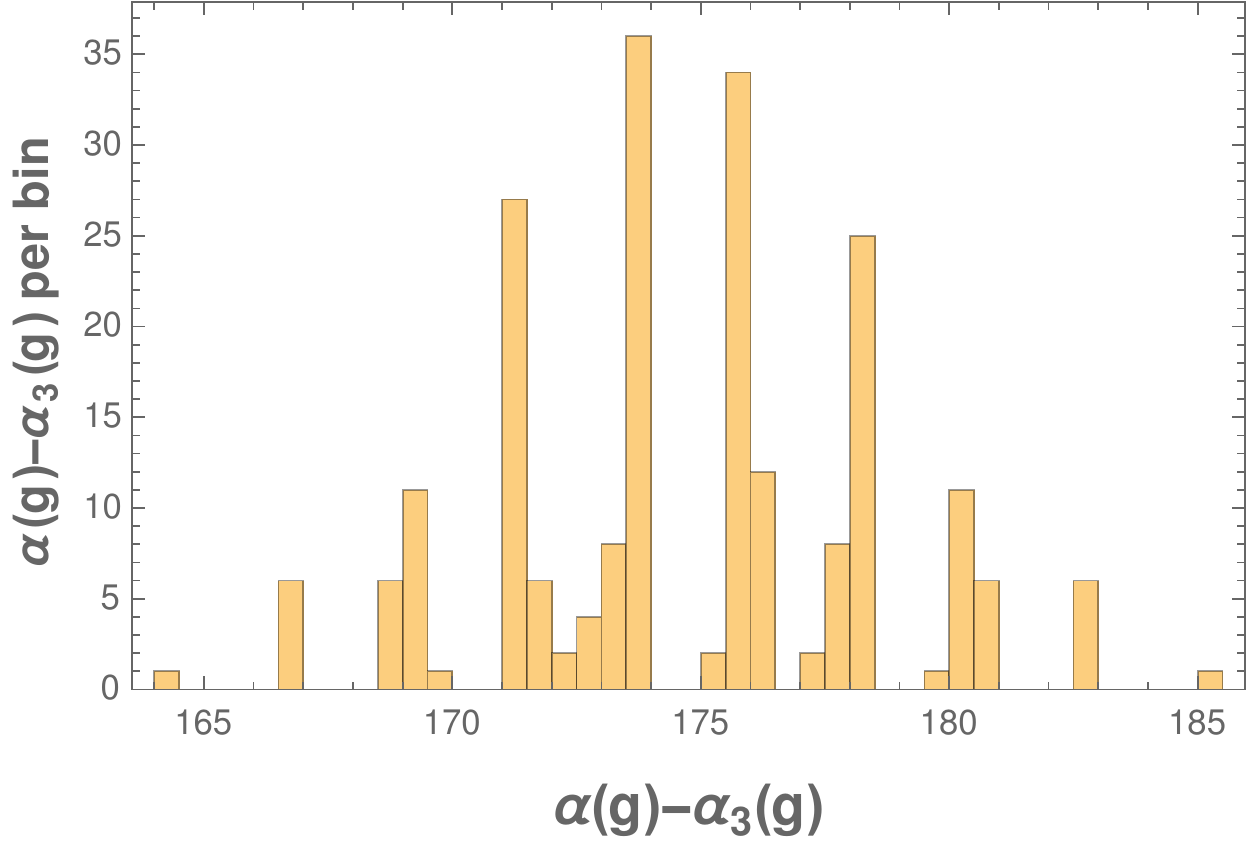}
\caption{Histograms of the distributions of $\alpha^\prime(g)=\alpha(g)-\alpha_3(g)$ at $N=18$.  In the left panel, all values of $g$ are considered.  In the right panel, only those values with $f_{ij}(g)=3\delta_{ij}$ are considered,  yielding a standard deviation of 3.8.  Therefore the intrabin scatter of $\alpha^\prime(g)$ is comparable to the full scatter in this case, and so by no means negligible.}
\label{binfig}
\end{center}
\end{figure}

At least at the small values of $N$ accessible to brute force numerical calculations, there is no evidence that the binning postulate holds for $\alpha$ itself.   As shown in Fig.~\ref{binfig} the intrabin and interbin variances of $\alpha^\prime$ at $N=18$ are comparable.   Whether it holds at large $N$ may depend on the relation between $q$ and $N$ assumed in this limit.  Needless to say, understanding this issue is critical to the success of our intended program and it remains possible that an inevitable failure of the binning postulate will obstruct our approach.

With these strong conjectures in hand our strategy is clear.  We will recast our problem in terms of $f$, assuming that with a suitable choice of $g_f$ the intrabin contributions to various quantities vanish in the $q\rightarrow\infty$ limit.  

We have checked this in some cases as follows.  The expressions below often contain nested sums over bins with inequalities, such as $\sum_{i=1}^q\sum_{j=i+1}^q$.  The summand in which two bins are equal, such as $i=j$, is not clearly defined by our procedure.  For example, in terms involving $\alpha_3$ or $\Phi$ it depends on the permutations of elements inside of a bin.  This is intrabin information which is present in $g$ but not in $f_{ij}$.  We have tried different prescriptions for these diagonal summands, such as  $\sum_{i=1}^q\sum_{j=i}^q$ and also a one half weight for the diagonal summand $i=j$,  in several expressions throughout the paper.  In each case this led to a correction which is suppressed by a factor of $1/q$ with respect to the leading term.   For example, the $1/q^2$ in Eq.~(\ref{cost}) can be made to disappear by adopting a half weight.   However, in the calculation of the $O(n^2)$ contribution to the variance of $\alpha$ we have assumed a symmetric form (\ref{xeq}) of the anchor $\alpha_3$, which fixes the convention for the diagonal summand and we found that this convention greatly simplifies the computation.

Now our binning approximation is
\bea
\alpha_1(g)&=&\sum_j^n m(j) K(P(j))\sim \alpha_1(f)=\sum_{i,j=1}^q m\left(\frac{n}{q} i \right) f_{ij}(g) K\left(\frac{n}{q}j\right).\label{af}\\
\alpha_2(g)&=&\frac{1}{2}\sum_{j<k}^n\Phi(P(j),P(k))\sim\alpha_2(f)=\frac{1}{2}\sum_{i<k}^q\sum_{j,l=1}^q f_{ij}(g) f_{kl}(g) \Phi\left(\frac{n}{q}j,\frac{n}{q} l\right)\nonumber\\
\alpha_3(g)&=&2\pi \sum_{j<k} \theta(P(j)-P(k))\sim\alpha_3(f)=2\pi\sum_{i<k}^q\sum_{j>l}^q  f_{ij}(g) f_{kl}(g).\nonumber\\
\alpha(f)&=&\alpha_1(f)+\alpha_2(f)-\alpha_3(f).\nonumber
\eea
Here and from now on, we drop the prime on the anchored argument $\alpha$ as we will no longer need the unanchored $\alpha$.  These expressions are the definitions of our binned $\alpha(f)$, and so no large $n$ or $q$ limit needs to be taken.  However, even in the case of quantities for which the binning postulate holds, we expect in general that calculations of these quantities using $\alpha(f)$ will differ from those using the exact $\alpha(g)$ at subleading orders in an expansion in either $n/q$ or in $q$.

Our strategy will be as follows.  The matrix elements of interest can be expressed in terms of moments of $\rho(\alpha)$ where $\alpha$ is a function of $g\in S_n$.  Therefore the moments are averages over the group $S_n$.  The binning approximation lets us replace $\alpha(g)$ with $\alpha(f)$.  The moments of $\alpha(f)$ are averages over the space of values of $f$, and no longer over the full group $S_n$.  Now equation (\ref{af}) gives $\alpha(f)$ explicitly, and so allows one to express the moments of $\alpha$ in terms of those of $f$, which, as we will see below, can in turn be calculated using standard combinatoric arguments. 


\subsection{Simplifications at First Order}

This can be somewhat simplified.  First note that each of the $n/q$ elements of ${\mathcal{S}}_i$ is mapped to some ${\mathcal{S}}_j$ by $P$.  This yields the sum rule
\beq
\sum_{j=1}^q f_{ij}(g)=\frac{n}{q}.
\eeq
Similarly all $n/q$ elements of ${\mathcal{S}}_j$ are in some $P({\mathcal{S}}_i)$ yielding the second sum rule
\beq
\sum_{i=1}^q f_{ij}(g)=\frac{n}{q}.
\eeq
These sum rules hold individually for every $g\in S_n$.  

Let us define the expectation value of $f_{ij}$ by
\beq
\langle f_{ij} \rangle=\frac{1}{n!}\sum_{g\in S_n} f_{ij}(g).
\eeq
Higher correlators are defined similarly.  It is quite clear that $\langle f_{ij} \rangle$ is independent of $i$ and $j$.  Therefore the expectation value of either sum rule yields
\beq
\langle f_{ij}\rangle=\frac{1}{q} \left\langle \sum_{i=1}^q f_{ij}\right\rangle=\frac{1}{q}\left\langle\frac{n}{q}\right\rangle=\frac{n}{q^2}.
\eeq
This quantity will appear so often that we will name it
\beq
\beta=\frac{n}{q^2}.
\eeq

Many quantities are more simply expressed in terms of the reduced
\beq
\tf_{ij}(g)=\frac{f_{ij}(g)}{\beta}-1. \label{redotti}
\eeq
From the corresponding properties of $f_{ij}$ one finds
\beq
\sum_{i=1}^q\tf_{ij}(g)=\sum_{j=1}^q\tf_{ij}(g)=0
\hsp\langle \tf_{ij}\rangle=0. \label{ftsomma}
\eeq
These sum rules hold exactly for any value of $q$ and $n$, so long as $n/q$ is an integer.

We can now use (\ref{af}) to express the Bethe phases in terms of $\tf$.  The first is
\beq
\alpha_1(f)=\beta\sum_{i,j=1}^q  m\left(\frac{n}{q} i \right) (1+\tf_{ij}) K\left(\frac{n}{q}j\right).
\eeq
Let us fix our reference state to be the classical ground state $|0\rangle$ and so $m(j)=2j$.   Then this becomes
\bea
\alpha_1(f)&=&\beta\sum_{i,j=1}^q  2i\frac{n}{q} (1+\tf_{ij}) K\left(\frac{n}{q}j\right).\label{a1f}\\
&=&2\beta\frac{n}{q}\left(\sum_{i=1}^q i\right)\left(\sum_{j=1}^q K\left(\frac{n}{q}j\right)\right)+\frac{2n^2}{q^3}\sum_{i,j=1}^q  i \tf_{ij} K\left(\frac{n}{q}j\right)\nonumber\\
&=&n^2\pi\left(1+\frac{1}{q}\right)+\frac{2n^2}{q^3}\sum_{i,j=1}^q  i \tf_{ij} K\left(\frac{n}{q}j\right)\nonumber
\eea
where we used the fact that $K(i)$ is symmetric about $\pi$.  The $1/q$ correction to the first term is an artefact of our treatment of interbin effects, and could be changed if we changed our prescription for these by, for example, adding terms $\alpha_4$ to consider cases in which $i=k$ but nonetheless a given element of ${\mathcal{S}}_i$ is less than one of ${\mathcal{S}}_k={\mathcal{S}}_i$ and so should be included in the sum.  The binning postulate states that such corrections should not appear in continuum field theory observables.

Next we will treat $\alpha_2(f)$
\beq
\alpha_2(f)=\frac{1}{2}\beta^2 \sum_{i<k}^q\sum_{j,l=1}^q (1+\tf_{ij})(1+\tf_{kl}) \Phi\left(\frac{n}{q}j,\frac{n}{q} l\right). \label{a2f}
\eeq
Note that the term with no $\tf$ vanishes because
\beq
\sum_{j,l=1}^q\Phi\left(\frac{n}{q}j,\frac{n}{q} l\right)\sim \frac{q^2}{n^2}\sum_{j,l=1}^n\Phi\left(j,l\right)=0.
\eeq
This expression is exact only at $q=n$ and also in the large $q$ limit for any $n$.  The deviation from zero at subleading orders in the $q$ expansion is an artefact of the binning approximation, which should not contribute to physical quantities, and so we will neglect it from now on. 

It may appear that the term linear in $\tf$ in Eq.~(\ref{a2f}) vanishes as a result of the sum rule, but it does not as $i<k$ and so it is not summed over all bins.  However $j$ and $l$ are summed over all bins, and so we can apply the binned version of the Bethe equation (\ref{betheeq}), which in the case of the ground state $|\Omega\rangle$ is
\beq
2q K(j)=2\pi \left(2q-2j+\frac{q}{n}\right) + \sum_{l\neq j}^q \fjl . \label{bethebin}
\eeq
Now we are ready to evaluate the terms linear in $\tf$.  It turns out that they are equal, so we will show the evaluation of the $\tf_{ij}$ term
\bea
\frac{1}{2}\beta^2 \sum_{i<k}^q\sum_{j,l=1}^q \tf_{ij} \fjl&=&\frac{1}{2}\beta^2\sum_{i}^q\left(\sum_{k=i+1}^q\right)\sum_{j=1}^q\tf_{ij}\left(2qK(j)+2\pi\left(-2q+2j-\frac{q}{n}\right)\right)\nonumber\\
&=&\beta^2\sum_{i}^q\left(q-i \right)\sum_{j=1}^q\tf_{ij}\left(qK(j)+2\pi\left(-q+j-\frac{q}{2n}\right)\right)
\eea
which can be cleaned using the sum rule
\beq
\frac{1}{2}\beta^2 \sum_{i<k}^q\sum_{j,l=1}^q \tf_{ij} \fjl=-\frac{n^2}{q^3}\sum_{i,j}^qi\tf_{ij}K(j) -2\pi\beta^2\sum_{i,j}^qi\tf_{ij}j.
\eeq
As the $\tilde{f}_{kl}$ term is equal to the $\tf_{ij}$ term, we have found
\bea
\alpha_2(f)&=&-4\pi\beta^2\sum_{i,j}^qi\tf_{ij}j-2\frac{n^2}{q^3}\sum_{i,j}^qi\tf_{ij}K(j)\nonumber\\&&+\frac{1}{2}\beta^2 \sum_{i<k}^q\sum_{j,l=1}^q \tf_{ij}\tf_{kl} \fjl. \label{a2f}
\eea
Here we see our first major cancellation.  The second term of $\alpha_2(f)$ exactly cancels the second term in $\alpha_1(f)$ as written in Eq.~(\ref{a1f}).  Thus the function $K$ disappears from the phase factor $\alpha(f)$, and only a constant remains of $\alpha_1(f)$.

Finally we turn to $\alpha_3(f)$
\beq
\alpha_3(f)=2\pi\beta^2\sum_{i<k}^q\sum_{j>l}^q (1+\tf_{ij})(1+\tf_{kl}).
\eeq
The term with no $\tf$ is easily evaluated
\beq
2\pi\beta^2\sum_{i<k}^q\sum_{j>l}^q 1=2\pi\beta^2\left(\frac{q(q-1)}{2}\right)^2=\frac{\pi}{2}n^2\left(1-\frac{1}{q}\right)^2.
\eeq
This cancels half of the remaining constant term in $\alpha_1(f)$ in Eq.~(\ref{a1f}).  These constant terms then yield
\beq
\langle \alpha_1+\alpha_2-\alpha_3\rangle=\frac{n^2}{2}\pi\left(1+\frac{1}{q^2}\right). \label{cost}
\eeq
In the case $q=n$, corresponding to no binning\footnote{Later, when we calculate correlation functions of $f$'s, we will need to assume that $q\lesssim \sqrt{n}$, but that is not necessary here.}, the expectation values for $\langle \alpha_1+\alpha_2\rangle$ and for this full anchored combination are visible in Fig.~\ref{n11fig} and one indeed sees that the later is a bit more than half of the former.  Why a bit more?  Should not $1/q^2$ be an artifact?  When $n\rightarrow\infty$, $n/q^2$ should either tend to a constant or else go to zero more slowly than $1/q$.  And so one expects that a $1/q^2$ correction will be a $1/n$ correction.  Such a $1/n$ correction is expected, as we have made a rather arbitrary choice in definition of $\alpha_3(g)$ in Eq.~(\ref{ancora}).  We have not included contributions from the terms $j=k$.  If we include these contributions, then the anchor is increased by $2\pi n$ and so the expectation value decreases by $2\pi n$.  In this case the expectation value of the anchored phase is slightly less than half of the unanchored phase.  
The expectation value of $\alpha$ contributes a phase to the matrix elements, and so needless to say we need to be concerned about an $O(n)$ change in its expectation value.  The fact that such subleading effects, as subtle as the choice of whether to include the $j=k$ term in the anchor, may have such a large effect on our results means that care will be needed, in particular in such zero point effects which can leak into the next order in $n$.  

The Bethe phase $\alpha$ can be simplified yet further.  We have seen that it contains terms which are constant, linear and quadratic in $\tf$.  The constant terms where summed in Eq.~(\ref{cost}).  The two linear terms are equal, and so to evaluate their sum we will simply multiply the $\tf_{ij}$ term by $2$
\beq
4\pi\beta^2\sum_{i<k}^q\sum_{j>l}^q\tf_{ij}=4\pi\beta^2\sum_{i,j=1}^q(q-i)j\tf_{ij}=-4\pi\beta^2\sum_{i,j=1}^qi\tf_{ij}j. \label{morto}
\eeq
This is equal to the first term in $\alpha_2(f)$ as written in Eq.~(\ref{a2f}), leading to our second major cancelation.  Putting all remaining terms together we have found our master formula for the anchored phase
\beq
\alpha(f)=\frac{n^2}{2}\pi\left(1+\frac{1}{q^2}\right)+\frac{1}{2}\beta^2 \sum_{i<k}^q\sum_{j,l=1}^q \tf_{ij}\tf_{kl} \fjl-2\pi\beta^2\sum_{i<k}^q\sum_{j>l}^q \tf_{ij}\tf_{kl}. \label{aeq}
\eeq

We will soon see that $\langle \tf\tf\rangle\sim 1/n$ at leading order in $n$ and so we may already try to estimate the fluctuations of the anchored phase in the large $n$ and $q$ limit.  The first term is a constant and so does not contribute.  The second two have $\beta^2=n^2/q^4$.  The $q^4$ cancels with the sums, up to a factor of order unity.  Now the variance depends on the square of this, and so it will be of order
$O(n^4)$.  On the other hand the four point function of $\tilde{f}$ in the Gaussian approximation would give $O(1/n^2)$, and so we find a variance of $O(n^2)$ and so a standard deviation of $O(n)$.  

The canceled term in Eq.~(\ref{morto}) has a larger variance.  Consider the square of this term.  The term contains $\beta^2$ and so its square contains $\beta^4$, yielding $n^4$ as above.  Again, as above, the $1/q^8$ in the $\beta^4$ is canceled by eight sums over bins.  The difference is that this term only contains a single power of $\tf$, and so its square only contains a two-point function of $\tf$, yielding $1/n$.  And so the variance is of order $O(n^3)$.  In Subsec.~\ref{ncubosez} we will see that this leading order term is nonvanishing.  Thus we arrive at our second main result:  Anchoring reduces the variance of the argument $\alpha$ from $O(n^3)$ to $O(n^2)$.


\subsection{Bin Statistics from Partitions: One Point} \label{punto}

Finally we are ready to calculate correlation functions of $f$.  These are averages of products of $f_{ij}(g)$ over the symmetric group $S_n$.  To calculate them, one must count how many members $g\in S_n$ give each value for a given polynomial in $f_{ij}(g)$.  Let us warm up by considering a single $f_{ij}$.  How many elements of $g$ satisfy $f_{ij}(g)=p$? 

 Let us call this number
\beq
h_{ij}(p)={\mathrm{Number\ of\ }}\ g\in S_n {\mathrm{\ such\ that}}\ f_{ij}(g)=p.
\eeq
As the symmetric group $S_n$ has $n!$ elements, the probability that a given $g$ satisfies $f_{ij}(g)=p$ is then
\beq
\tilde{h}_{ij}(p)=\frac{h_{ij}(p)}{n!}.
\eeq
Recall that $P$ must map each integer in $[1,n]$ to a distinct integer in $[1,n]$.  If $p$ elements of ${\mathcal{S}}_i$ are to map to ${\mathcal{S}}_j$, one needs to choose which $p$ elements of ${\mathcal{S}}_j$ are in $P({\mathcal{S}}_i)$.  Recalling that each bin has $n/q$ elements, the number of choices is ${n/q} \choose p$.   One must also choose the $n/q-p$ elements of the complement of ${\mathcal{S}}_j$ which are in $P({\mathcal{S}}_i)$.  The corresponding number of choices is ${n-n/q} \choose {n/q-p}$.  Finally, one may permute the elements of ${\mathcal{S}}_i$ and its complement, yielding factors of $(n/q)!$ and $(n-n/q)!$ respectively.  The result is
\beq
h_{ij}(p)={{n/q} \choose p}{{n-n/q} \choose {n/q-p}}\left(n/q\right)!\left(n-n/q\right)!.  
\eeq

These later factors are independent of $p$ and so will not be important in future calculations, as they only contribute to the overall normalization which is fixed by the fact that
\beq
\sum_{p=0}^\infty h_{ij}(p)=n!.
\eeq
So let us separate all of the $p$-independent terms into a constant $c$
\bea
\tilde{h}_{ij}(p)&=&\frac{h_{ij}(p)}{n!}=\frac{c}{p!}\left(\frac{(n/q)!}{(n/q-p)!}\right)^2\left(\frac{(n-2n/q)!}{(n-2n/q+p)!}\right)=\frac{c}{p!}\frac{\left(\left(n/q\right)^{\underline{p}}\right)^2}{\left(n-2n/q+1\right)^{\overline{p}}}\nonumber\\
c&=&\frac{\left(\left(n-n/q\right)!\right)^2}{n!\left(n-2n/q\right)!}
\eea
where we have defined the falling and rising factorials
\beq
m^{\underline{n}}=\frac{m!}{(m-n)!}\hsp m^{\overline{n}}=\frac{(m+n-1)!}{(m-1)!}.
\eeq
Curiously, $\tilde{h}_{ij}(p)/c$ is the $p$th term in the Gauss series for the hypergeometric function ${}_2F_1(n/q,n/q,-n+2n/q;-1)$.  

So far these expressions are exact for all $n$ and $q$.  We will be interested in the limit where $n/q\rightarrow \infty$ while $p$, which is of order $\beta=n/q^2$, will be finite or slowly tend to $0$.  In this limit the rising and falling factorials are of the form $x!/(x+r)!$ with $r<<\sqrt{x}$.   Indeed, $r$ will be finite and $\sqrt{x}$ infinite.  When $x=n/q$ and $r=p\sim n/q^2$ this inequality implies $n\lesssim q^3$.  

To find a suitable approximation for the ratios of factorials in this limit, we combine the expansion
\beq
\left(1+\frac{a}{n}\right)^n=e^a\left(1-\frac{a^2}{2n}+\frac{a^3}{3n^2}+\frac{a^4}{2n^2}-\frac{a^4}{4n^3}-\frac{a^5}{6n^3}-\frac{a^6}{48n^3}+O\left(\frac{1}{n^4}\right)\right)
\eeq 
with Stirling's approximation
\beq
n!=\sqrt{2\pi n}n^n e^{-n}\left(1+\frac{1}{12n}+\frac{1}{288n^2}+O\left(\frac{1}{n^3}\right)\right)
\eeq
and the binomial expansion to obtain our main tool
\bea
x^{\underline{-r}}&=&\frac{1}{(x+1)_{\overline{r}}}=\frac{x!}{(x+r)!}\\
&\sim& x^{-r}\left(1+\frac{-r-r^2}{2x}+\frac{2r+9r^2+10r^3+3r^4}{24x^2}+\frac{-6r^2-17r^3-17r^4-7r^5-r^6}{48 x^3}\right).\nonumber \label{tool}
\eea

With this tool in hand, we can approximate $h$.  If we let $q\sim O(n^{1/2})$ and expand to order $O(n^{-1})$, for example, we find
\beq
\tilde{h}_{ij}(p)=\frac{c}{p!}\beta^p\left[1+\frac{(1+2\beta)p-p^2}{n/q}+\frac{\left(12\beta-3-\frac{1}{\beta}\right)p+\left(12\beta+9+\frac{6}{\beta}\right)p^2+\left(-12-\frac{8}{\beta}\right)p^3+\frac{3}{\beta}p^4}{6n}\right]. \label{h}
\eeq
Note that the leading term is a Poisson distribution times $e^\beta c$.  Therefore the expectation value of any function of $f$ can be given in terms of Poisson correlators
\beq
\langle p\rangle_0=\beta\hsp \langle p^2\rangle_0=\beta^2+\beta\hsp \langle p^3\rangle_0=\beta^3+3\beta^2+\beta\hsp \langle p^4\rangle_0=\beta^4+6\beta^3+7\beta^2+\beta,\ ...\ . \label{poisstab}
\eeq
In particular, by setting the expectation value of $1$ to be equal to $1$, we can fix $c$ at any desired order.  In this case the relation between Eq.~(\ref{h}) and the Poisson distribution yields
\bea
1=\langle 1\rangle&=&e^\beta c \left[\langle 1\rangle_0+\frac{(1+2\beta)\langle p\rangle_0-\langle p^2\rangle_0}{n/q}\right.\\
&&+\left.\frac{\left(12\beta-3-\frac{1}{\beta}\right)\langle p\rangle_0+\left(12\beta+9+\frac{6}{\beta}\right)\e{p^2}+\left(-12-\frac{8}{\beta}\right)\e{p^3}+\frac{3}{\beta}\e{p^4}}{6n}\right].\nonumber
\eea
Then inserting the Poisson expectation values from Eq.~(\ref{poisstab}) one finds
\beq
1=e^\beta c \left[1+\frac{\beta^2}{n/q}+\frac{3\beta^3+7\beta^2-3\beta}{6n}\right]
\eeq
and so obtains $c$ at $O(n^{-1})$
\beq
c=\frac{e^{-\beta}}{1+\frac{\beta^2}{n/q}+\frac{3\beta^3+7\beta^2-3\beta}{6n}}.
\eeq

Any other correlator can be found similarly, using (\ref{h}) to relate the desired correlator to a combination of Poisson correlators.  For example,
\bea
\langle f_{ij}\rangle&=&\langle p\rangle=e^\beta c \left[\langle p\rangle_0+\frac{(1+2\beta)\langle p^2\rangle_0-\langle p^3\rangle_0}{n/q}\right.\nonumber\\
&&+\left.\frac{\left(12\beta-3-\frac{1}{\beta}\right)\langle p^2\rangle_0+\left(12\beta+9+\frac{6}{\beta}\right)\e{p^3}+\left(-12-\frac{8}{\beta}\right)\e{p^4}+\frac{3}{\beta}\e{p^5}}{6n}\right]\nonumber\\
&=&\beta.
\eea
This spectacular order by order cancellation is in fact required by the sum rule, as was argued above, and so provides a consistency check of our approximations.

Higher orders in $n$ have useful information for correlators of distinct $f_{ij}$.  However, for our purposes in this Subsection, for correlators at a single $\{i,j\}$ it suffices to use the leading term, given by the Poisson distribution.  At this order
\beq
\ee{f_{ij}^n}=\e{p^n}.
\eeq
We can then find arbitrary correlators of $\tf$ at the same point.  For example
\beq
\ee{\tf^2}=\ee{\left(\frac{f}{\beta}-1\right)^2}=\frac{\ee{f^2}}{\beta^2}-\frac{2\ee{f}\ee{1}}{\beta}+1=\frac{\e{p^2}}{\beta^2}-\frac{2\e{p}}{\beta}+1=\frac{1}{\beta}.
\eeq
This is reasonable.  It means that so long as $\beta>>1$, $\tf$ will stay away from its minimal value of $-1$, where $f$ vanishes, and so is reasonably well approximated by a Gaussian.  As $\alpha$ is quadratic in $\tf$, to determine its variance we will need four point functions of $\tf$.  If all $\tf$ are at the same point, the leading order contribution is
\bea
\ee{\tf^4}&=&\ee{\left(\frac{f}{\beta}-1\right)^4}=\frac{\ee{f^4}}{\beta^4}-\frac{4\ee{f^3}\ee{1}}{\beta}+\frac{6\ee{f^2}\ee{1}}{\beta}-\frac{4\ee{f}\ee{1}}{\beta}+1\nonumber\\
&=&\frac{3}{\beta^2}+\frac{1}{\beta^3}=3\ee{\tf}^2+\frac{1}{\beta^3}. \label{f4}
\eea
The first term is usual disconnected contribution to the four point function, in which the $\tf$s are paired into 3 possible pairs of pairs and their two point correlations are used.  These give a result of order $1/n^2$ which, combined with the $n^4$ in $\alpha^2$ in Eq.~(\ref{aeq}) yields $\beta^2$ and so a variance of order $O(n^2)$.  

\subsection{Bin Statistics from Partitions: Multiple Points}

In general we will need correlators of $\tf_{ij}$ with different indices.  There are two ways to generalize the above calculation to multiple indices.  The first is to use the sum rule to extrapolate new correlators from old correlators.  This is sufficient to derive all of the Gaussian terms, as these simply come from the two point function, and the sum rule together with one two point function yields all two point functions.  So the sum rule approach will be sufficient for our application in Sec.~\ref{tornasez}, which concerns the calculation of the $O(n^2)$ terms.  However, once that order is understood, the reader may wish to calculate the subleading terms.  These come from the essentially Poisson terms like the last term in Eq.~(\ref{f4}) and many, but not all, of these can be derived from the previous case using sum rules.  

Let us begin with the sum rule approach.  Once we know that in the large $n$ and $q$ limit, with $\beta$ unconstrained
\beq
\ee{\tf_{ij}\tf_{ij}}=\frac{1}{\beta}
\eeq
the sum rule (\ref{ftsomma}) implies that
\beq
\ee{\tf_{ij}\tf_{il}}=\ee{\tf_{ij}\tf_{kj}}=-\frac{1}{\beta (q-1)}\sim-\frac{1}{\beta q}
\eeq
for all $i\neq k$ and $j\neq l$.  In the last expression we have used the large $q$ limit.   A repeated application of the same sum rule yields
\beq
\ee{\tf_{ij}\tf_{kl}}=\frac{1}{\beta q^2}=\frac{1}{n}
\eeq
for $i\neq k$ and $j\neq l$.  
We will denote these correlations using the following diagrams
\bea
&&\ee{\tf_{ij}}=\left(\xymatrix{  \bullet}\right)
\hsp
\ee{\tf_{ij}^2}=\left(\xymatrix{\bullet^2}\right)\\
&&
\ee{\tf_{ij}\tf_{il}}=\left(\xymatrix{\bullet\ar[r]&\bullet\ar[l]}\right)
\hsp
\ee{\tf_{ij}\tf_{kl}}=\left(\vcenter{\xymatrix{&\bullet\ar@{-->}[dl]\\\bullet\ar@{-->}[ur]&}}\right) .\nonumber
\eea
Here the rows are the $\{j,l\}$ indices which are contracted with $\Phi$ in our master formula (\ref{aeq}), while the columns are the $\{i,k\}$ indices which are ordered.  Recall that $g\in S_n$ is represented as a map $P:[1,n]\rightarrow [1,n]$ and so the rows correspond to the bins in the image and the columns to the bins in the domain of the map.  Inverting $g$ corresponds to a transpose of the diagram, but this does not affect the statistics as it is an automorphism of $S_n$ and so each diagram will be equal to its transpose.  Similarly, rows can be freely interchanged, as can columns, without changing the value.   Solid lines connect entries directly related by the sum rule, and so introduce factors of $-1/q$ whereas dashed lines connect entries which are connected by two sum rules.  At this leading order in $n$ the dashed lines introduce factors of $1/q^2$. 

The $O(n^2)$ approximation to the four point functions then follow from simply summing together the three pairs of products of two point functions.  For example, if $i\neq k$ and $j\neq l$ then at leading order
\beq
\ee{\tf_{ij}^2\tf_{kl}^2}=\ee{\tf_{ij}^2}\ee{\tf_{kl}^2}=\frac{1}{\beta^2}
\eeq
while
\beq
\ee{\tf_{ij}^2\tf_{k_1l}\tf_{k_2l}}=\ee{\tf_{ij}^2}\ee{\tf_{k_1l}\tf_{k_2l}}=\frac{1}{\beta}\left(-\frac{1}{\beta q}\right)=-\frac{1}{\beta^2 q}
\eeq
corresponding to the diagrams
\beq
\ee{\tf_{ij}^2}\ee{\tf_{kl}^2}=\left(\vcenter{\xymatrix{&\bullet^2\\\bullet^2}}\right)
\hsp
\ee{\tf_{ij}^2}\ee{\tf_{k_1l}\tf_{k_2l}}=\left(\vcenter{\xymatrix{&\bullet\ar[r]&\bullet\ar[l]\\\bullet^2}}\right).
\eeq
There are contributions from other combinations of pairings of the points, but these are subdominant in $q$.

In general to calculate correlators at distinct points, the sum rules are not sufficient.  However the above partition argument can be generalized.  For concreteness, let us consider a correlator corresponding to a diagram with 2 rows and 2 columns.  This means that we will be interested in two domain bins $\{i,k\}$ and two image bins $\{j,l\}$.  We will need to calculate the joint probability distributions of
\beq
f_{ij}(g)=p_1\hsp f_{il}(g)=p_2\hsp f_{kj}(g)=p_3\hsp f_{kl}(g)=p_4.  \label{joint}
\eeq
The joint probability $\tilde{h}$ is just the number of elements $g$ satisfying (\ref{joint}) divided by $n!$.  

It can be calculated as in the $1\times 1$ case treated above.  First, one needs to choose $p_1$ elements of ${\mathcal{S}}_j$ to be in $P({\mathcal{S}}_i)$.  There are $n/q \choose p_1$ such choices.  Similarly there are $n/q\choose p_2$ choices for the intersection of $P({\mathcal{S}}_i)$ and ${\mathcal{S}}_l$.  This leaves $(n/q-p_1-p_2)$ elements of ${\mathcal{S}}_i$ which must map into the complement of ${\mathcal{S}}_j$ and ${\mathcal{S}}_l$, which has $(n-2n/q)$ elements, yielding $n-2n/q \choose n/q-p_1-p_2$ possibilities.  Now we have counted the possible images of ${\mathcal{S}}_i$, we must do the same for ${\mathcal{S}}_k$.  Recall that $p_3$ elements of ${\mathcal{S}}_k$ are mapped into ${\mathcal{S}}_j$.  However, $p_1$ elements of ${\mathcal{S}}_j$ are already full, and so only $n/q-p_1$ slots are available.  Thus the number of possible images of this map is $n/q-p_1 \choose p_3$.  Similarly the choice of images of ${\mathcal{S}}_k$ in ${\mathcal{S}}_l$ yields a factor of $n/q-p_2 \choose p_4$.  Now $n/q-p_3-p_4$ elements rest in ${\mathcal{S}}_k$ which must be mapped into the remaining $n-3n/q+p_1+p_2$ elements in the complement of ${\mathcal{S}}_j$ and ${\mathcal{S}}_l$, yielding a factor of $n-3n/q+p_1+p_2 \choose n/q-p_3-p_4$.  Finally, once one has chosen which slots are occupied, one multiplies by the various permutations of the domains, yielding $(n/q)!^2(n-2n/q)!$.  As always, this last factor is independent of the $p_i$ and so can be absorbed into a normalization constant to be fixed later.  Expanding these 6 choose functions into factorials and absorbing all terms independent of the $p_i$ into the constant $c$, one obtains
\bea
\tilde{h}_{ij}(p_i)&=&\frac{c}{p_1!p_2!p_3!p_4!}\left(\frac{(n/q)!}{(n/q-p_1-p_2)!}\right)\left(\frac{(n/q)!}{(n/q-p_3-p_4)!}\right)\label{box}\\
&&\left(\frac{(n/q)!}{(n/q-p_1-p_3)!}\right)\left(\frac{(n/q)!}{(n/q-p_2-p_4)!}\right)\left(\frac{(n-4n/q)!}{(n-4n/q+p_1+p_2+p_3+p_4)!}\right)\nonumber\\
&=&\frac{c}{p_1!p_2!p_3!p_4!}\frac{(n/q)^{\underline{p_1+p_2}}(n/q)^{\underline{p_3+p_4}}(n/q)^{\underline{p_1+p_3}}(n/q)^{\underline{p_2+p_4}}}{(n-4n/q+1)^{\overline{p_1+p_2+p_3+p_4}}}.
\eea
Again this expression is exact for all $n$ and $q$.  One sees that the terms with isolated $p_i$'s cancel, only those with entire rows $\{p_1+p_3,p_2+p_4\}$\ or columns $\{p_1+p_2,p_3+p_4\}$\ remain.  

The first four ratios enforce the correlations caused by the sum rules corresponding to each of the two rows and each of the two columns, while the last enforces the sum rule on the entire matrix.  This may be expanded using our main tool (\ref{tool}) and any correlation function may then be calculated as a sum of the corresponding Poisson correlation functions as above.  In particular the $1/q$ terms in general always yield factors of $-1/q$ associated with any two elements of the same row and column.  However, since each ratio of factorials only appears once, no diagram may contain two such lines in the same row and column.  Triplets instead appear in the $1/q^2$ terms, and quadruplets at $1/q^3$.  Similarly the last term in the second line of Eq.~(\ref{box}) yields dashed diagonal lines with factors of order $1/q^2$, although the coefficient is more complicated than at leading order.  While at $O(n^2)$ we have seen that the diagrams reduce to pairs of two point functions, at $O(n)$ it appears that only connected diagrams contribute to the four point function.  This may be expected since the $1/n^3$ term in Eq.~(\ref{f4}), which is at the correct order, only appears when the irreducible correlation of four points is considered.

The generalization to $j$ domain bins (columns) and $k$ image bins (rows) is clear.  There are $jk$ choices of maps and so a factor of $p_1!...p_{jk}!$ in the denominator.  The numerator consists of $j+k$ descending factorials, each $(n/q)$ with an argument equal to the sum of the $p$'s in the corresponding row or column.  The denominator is a single ascending factorial $(n-jk n/q+1)^{\overline{\sum_{i=1}^{jk}p_{i}}}$.



\section{Testing the Anchor} \label{tornasez}

In the large $q$ limit, what is the variance of $\alpha$?  

\subsection{The Variance of $\alpha_1$ at $O(n^3)$} \label{ncubosez}

Let us warm up with $\alpha_1$ as given in Eq.~(\ref{a1f}).  There are two terms.  First, a constant term, which doesn't contribute. We will drop it. Next is
\beq
x=\frac{2n^2}{q^3}\sum_{i,j=1}^q  i \tf_{ij} K\left(\frac{n}{q}j\right).
\eeq
As $\langle\tilde{f}_{ij}\rangle=0$, $\langle x\rangle=0$ and so the variance is
\beq
\ee{x^2}=\frac{4n^4}{q^6}\sum_{i,j=1}^q \sum_{k,l=1}^q  i\ k \ee{\tf_{ij}\tf_{kl}} K\left(\frac{n}{q}j\right)K\left(\frac{n}{q}l\right).
\eeq
This is the sum of four terms depending on whether $i=k$ and whether $j=l$, each summand corresponding to a diagram.  

When $i\neq k$ and $j\neq l$ one uses
\beq
\di{&\bullet\ar@{-->}[dl]\\\bullet\ar@{-->}[ur]&}=\ee{\tf_{ij}\tf_{kl}}=\frac{1}{n}
\eeq
to obtain the contribution
\bea
x_1&=&\frac{4n^4}{q^6}\left(\sum_{i=1}^q i\right)\left(\sum_{k=1}^q k\right)\frac{1}{n}\left(\sum_{j=1}^q K\left(\frac{n}{q}j\right)\right)\left(\sum_{l=1}^q K\left(\frac{n}{q}l\right)\right)\\
&=&\frac{4n^3}{q^6}\left(\frac{q^2}{2}\right)^2\left(\pi q\right)^2=\pi^2 n^3.
\eea
When $i=k$ and $j\neq l$, the matrix element
\beq
\di{\bullet\ar[d]\\\bullet\ar[u]}=\ee{\tf_{ij}\tf_{il}}=-\frac{q}{n}
\eeq
yields 
\bea
x_2&=&\frac{4n^4}{q^6}\left(\sum_{i=1}^q i^2\right)\left(-\frac{q}{n}\right)\left(\sum_{j=1}^q K\left(\frac{n}{q}j\right)\right)\left(\sum_{l=1}^q K\left(\frac{n}{q}l\right)\right)\\
&=&-\frac{4n^3}{q^5}\left(\frac{q^3}{3}\right)\left(\pi q\right)^2=-\frac{4\pi^2}{3} n^3.
\eea
Next one considers $j=l$ but $i\neq k$, with matrix element
\beq
\di{\bullet\ar[r]&\bullet\ar[l]}=\ee{\tf_{ij}\tf_{kj}}=-\frac{q}{n}
\eeq
to find
\bea
x_3&=&\frac{4n^4}{q^6}\left(\sum_{i=1}^q i\right)\left(\sum_{k=1}^q k\right)\left(-\frac{q}{n}\right)\left(\sum_{j=1}^q K\left(\frac{n}{q}j\right)^2\right)\\
&=&-\frac{4n^3}{q^5}\left(\frac{q^2}{2}\right)^2\left(\sum_{j=1}^q K\left(\frac{n}{q}j\right)^2\right)=-n^3\ee{K^2}
\eea
where we have defined the average
\beq
\ee{K^2}=\frac{1}{q}\sum_{j=1}^q K\left(\frac{n}{q}j\right)^2 .
\eeq
Finally the case $i=k$, $j=l$
\beq
\di{\bullet^2}=\ee{\tf_{ij}^2}=\frac{q^2}{n}
\eeq
provides the last contribution
\bea
x_4&=&\frac{4n^4}{q^6}\left(\sum_{i=1}^q i^2\right)\left(\frac{q^2}{n}\right)\left(\sum_{j=1}^q K\left(\frac{n}{q}j\right)^2\right)\\
&=&\frac{4n^3}{q^4}\left(\frac{q^3}{3}\right)\left(\sum_{j=1}^q K\left(\frac{n}{q}j\right)^2\right)=\frac{4}{3}n^3\ee{K^2} . \nonumber
\eea

Summing these contributions one finds the variance of $\alpha_1$
\beq
\ee{x^2}=\sum_{i=1}^4 x_i=\left(\frac{\ee{K^2}-\pi^2}{3}\right)n^3.
\eeq
Recall that the average value of $K$ is $\pi$, and $K$ is not constant, so $\ee{K^2}>\pi^2$ and therefore the $O(n^3)$ contribution does not vanish.

What about the unanchored $\alpha=\alpha_1+\alpha_2$?  Recall that the $x$ term in $\alpha_1$ is canceled by a term in $\alpha_2$, and so could the $O(n^3)$ contribution to the unanchored $\alpha_1+\alpha_2$ vanish?  The $\Phi$ term enters at $O(n^2)$ so it may seem promising.  The trouble is the first term in (\ref{a2f}).  It is identical to the $\alpha_1$ term considered here except with $K(nj/q)$ replaced by $-2\pi j/q$.  The $j\neq l$ cases then give $-\pi^2$, as $\pi$ is the average value of $2\pi j/q$.  The $j=l$ cases give $4\pi^2/3$, as the average of $j^2/q^2$ is $1/3$.  These again appear in the numerator and do not cancel.  Thus, just the same calculation as above shows that the $O(n^3)$ terms in the variance do not cancel without the anchor.

\subsection{The Variance of $\alpha$ at $O(n^2)$}

Once the anchor is included, one arrives at our master formula for $\alpha$ in Eq.~(\ref{aeq}).  Here all terms that could potentially give $n^3$ contributions on dimensional grounds are gone.  The constant term does not contribute to the variance and so we will drop it.  We will also shift $\alpha_3$ to make it antisymmetric in $j$ and $l$, thus eliminating the zero point which created a nonzero expectation value for $\ee{\alpha_3}$.  We do not know if such a shift is necessary for the binning postulate.  However it does not affect the previous arguments concerning the role of the anchor.  The variance in $\alpha$ will therefore be equal to that of
\bea
x&=&\frac{1}{2}\beta^2 \sum_{i<k}^q\sum_{j,l=1}^q \tf_{ij}\tf_{kl} \fjl+2\pi\beta^2\sum_{i<k}^q\sum_{j,l}^q \tf_{ij}\tf_{kl} \left(\frac{1}{2}-\theta(j-l)\right)\nonumber\\
&=&\frac{1}{2}\beta^2 \sum_{i<k}^q\sum_{j,l=1}^q \tf_{ij}\tf_{kl} \left(\fjl+2\pi-4\pi\theta(j-l)\right). \label{xeq}
\eea
The variance is just
\bea
\ee{x^2}&=&\frac{\beta^4}{4}\sum_{i_1<k_1}^q\sum_{i_2<k_2}^q\sum_{j_1,j_2,l_1,l_2=1}^q\ee{\tf_{i_1j_1}\tf_{i_2j_2}\tf_{k_1l_1}\tf_{k_2l_2}}\\
&&\times\left(\F{j_1}{l_1}+2\pi-4\pi\theta(j_1-l_1)\right)\left(\F{j_2}{l_2}+2\pi-4\pi\theta(j_2-l_2)\right) \nonumber.
\eea

We are interested in the $O(n^2)$ contribution, which arises entirely from the Gaussian correlations, corresponding to disconnected pairs of 2 point functions.  When more than one pairing is available, the sum over pairings may increase the diagram by a factor of 2 or 3 however this requires fixing one of the indices, which costs a factor of $q$ and so diagrams with equal contributions from multiple pairings will always be subleading in $1/q$.  Thus we need only consider diagrams with only a single choice of dominant pairing.  In addition, diagrams with more than three rows or columns will lead to vanishing contributions, as the sums of both indices of $\Phi$ vanish and also the sums of the anchor terms vanish due to the zero point shift corresponding to the $2\pi$ in (\ref{xeq}).  Thus in all we will only need to sum five diagrams.

We begin with easiest, corresponding to
\beq
\left(\vcenter{\xymatrix{&\bullet^2\\\bullet^2}}\right)=\ee{\tf_{ij}^2}\ee{\tf_{kl}^2}=\frac{1}{\beta^2}.
\eeq
There are only two distinct values of $i_1,\ i_2,\ k_1$\ and $k_2$.  As $i_1<k_1$ and $i_2<k_2$, this implies that $i_1=i_2=i$ and $k_1=k_2=k$.  Since both points are degenerate, this means that also $j_1=j_2=j$ and $l_1=l_2=l$.   Now in this case and in all cases that follow, the matrix element is entirely determined by the diagram and the $\Phi$ factors have no $i$ or $k$ dependence, thus the sums over the $i$ and $k$ can be factored out and evaluated separately.  Thus this contribution is
\bea
x_1&=&\frac{\beta^4}{4}\left(\sum_{i<k}^q\right)\frac{1}{\beta^2}\sum_{j,l=1}^q\left(\F{j}{l}+2\pi-4\pi\theta(j-l)\right)^2\nonumber\\
&=&\frac{n^2}{4q^4}\frac{q^2}{2}\sum_{j,l=1}^q\left(\F{j}{l}^2-8\pi\theta(j-l)\F{j}{l}+4\pi^2-16\pi\theta(j-l)+16\pi\theta(j-l)\right)\nonumber\\
&=&n^2\left(\frac{\ee{\Phi^2}}{8}-\pi\ee{\Phi_>}+\frac{\pi^2}{2}\right) \label{x1}
\eea
where we have defined
\beq
\ee{\Phi^2}=\frac{1}{q^2}\sum_{j,l=1}^q\F{j}{l}^2\hsp
\ee{\Phi_>}=\frac{1}{q^2}\sum_{j>l}^q\F{j}{l}.
\eeq

We next consider the diagram
\beq
\left(\vcenter{\xymatrix{&\bullet\ar[d]\\&\bullet\ar[u]\\\bullet^2}}\right)=\ee{\tf_{ij}^2}\ee{\tf_{kl_1}\tf_{kl_2}}=-\frac{1}{q\beta^2}.
\eeq
Here again there are only two values of $i_1,\ i_2,\ k_1$\ and $k_2$ and so again $i_1=i_2=i$ and $k_1=k_2=k$.  One of these is a double point.  If it is $i$ then $j_1=j_2$, but if it is $k$ then $l_1=l_2$.  These two cases give equal contributions, and so we consider the first and multiply by a factor of two.  Altogether
\bea
x_2&=&2\frac{\beta^4}{4}\left(\sum_{i<k}^q\right)\left(-\frac{1}{q\beta^2}\right)\\
&&\times \sum_{j,l_1,l_2=1}^q \left(\F{j}{l_1}+2\pi-4\pi\theta(j-l_1)\right)\left(\F{j}{l_2}+2\pi-4\pi\theta(j-l_2)\right) .\nonumber
\eea
The first line gives $-n^2/(4q^3)$.  To simplify the second line, we can use the binned version of the Bethe equation (\ref{bethebin}) to sum over $l_1$ and $l_2$, leaving
\bea
x_2&=&-\frac{n^2}{4q^3}\sum_{j=1}^q \left(2q K\left(\frac{n}{q}j\right)+4\pi(j-q)+2\pi q - 4\pi j\right)^2\nonumber\\
&&=n^2\left(-\ee{K^2}+\pi^2\right) \label{x2}
\eea
where we have again used the fact that the average value of $K$ is $\pi$.

The third diagram is
\beq
\left(\vcenter{\xymatrix{&\bullet\ar[r]&\bullet\ar[l]\\\bullet^2}}\right)=\ee{\tf_{ij}^2}\ee{\tf_{k_1l}\tf_{k_2l}}=-\frac{1}{q\beta^2}.
\eeq
Now there are three columns, and so there are inequivalent pairings of $i$ and $k$.  One may have $i_1=i_2=i$, $k_1=k_2=k$, $i_1=k_2$ or $i_2=k_1$.  The first two give equal contributions, as there is a symmetry in which $i$ and $k$ are exchanged and the sites are inverted.  Similarly the third and fourth are equal.  More subtly, the third is equal to minus one half of the first.  This is because in the first case the $i$ and $k$ sum is
\beq
\sum_{i=1}^q\sum_{k_1,k_1=i}^q 1=\sum_{i=1}^q(q-i)^2=\frac{q^3}{3}.
\eeq
While in the second it is
\beq
\sum_{i_1=1}^q\sum_{i_2=1}^{i_1-1}\sum_{k_1=i_1+1}^q 1=\sum_{i_1=1}^q i_1(q-i_1)=\frac{q^3}{6}.
\eeq
This explains the factor of two difference.  The signs are different because in the second case one exchanges one pair of $(j,l)$.  Both $\Phi$ and also the zeroed form of $\alpha_3$ are antisymmetric with respect to this interchange.  

Summarizing, we only need to consider the first of the four possibilities, and the contribution of the other diagrams will give a weight factor of $1+1-1/2-1/2=1$.   This is
\beq
x_3=-\frac{\beta^4}{4}\left(\sum_{i<k_1,k_2}^q\right)\frac{1}{q\beta^2}\sum_{j,l=1}^q\left(\F{j}{l}+2\pi-4\pi\theta(j-l)\right)^2.
\eeq
Note that this is equal to our first expression for $x_1$ in Eq.~(\ref{x1}) except for the $\{i,k\}$ integral which is multiplied by a factor of $2q/3$ and the matrix element which is multiplied by $-1/q$.  Therefore
\beq
x_3=-\frac{2x_1}{3}.
\eeq

The next diagram is three by three
\beq
\left(\vcenter{\xymatrix{&&\bullet\ar@{-->}[dl]\\&\bullet\ar@{-->}[ur]&\\\bullet^2}}\right)=\ee{\tf_{ij}^2}\ee{\tf_{k_1l_1}\tf_{k_2l_2}}=\frac{1}{q^2\beta^2}.
\eeq
Again, corresponding to the three columns, there are three possible values of the $i$ and $k$, yielding the same four pairings as above.  The integration factors are the same and so again the weights are $1$, $1$, $-1/2$ and $-1/2$ and so it will suffice to consider the possibility $i_1=i_2=i$.  As $i$ is at a double point, $j_1=j_2$.  However, unlike the previous case, now there are three rows and so $l_1\neq l_2$.  This leaves us with
\bea
x_4&=&\frac{\beta^4}{4}\left(\sum_{i<k_1,k_2}^q\right)\frac{1}{q^2\beta^2}\\ \label{x4}
&\times&\sum_{j,l_1,l_2=1}^q\left(\F{j}{l_1}+2\pi-4\pi\theta(j-l_1)\right)\left(\F{j}{l_2}+2\pi-4\pi\theta(j-l_2)\right).\nonumber\\
\eea
The first line yields $n^2/(12q^3)$.  As in the case of $x_2$, the $l_1$ and $l_2$ may be summed in the last line using the binned Bethe equation,  leaving
\beq
x_4=\frac{n^2}{12q^3}\sum_{j=1}^q \left(2q K\left(\frac{n}{q}j\right)+4\pi(j-q)+2\pi q - 4\pi j\right)^2.
\eeq
Comparing with Eq.~(\ref{x2}) we see that
\beq
x_4=-\frac{x_2}{3}.
\eeq

The final diagram is
\beq
\left(\vcenter{\xymatrix{\bullet\ar[d]&&\\\bullet\ar[u]&&\\&\bullet\ar[r]&\bullet\ar[l]}}\right)=\ee{\tf_{i_1j}\tf_{i_2j}}\ee{\tf_{kl_1}\tf_{kl_1}}=\frac{1}{q^2\beta^2}.
\eeq
Again there are three columns and so the same three values of $i$ and $k$, with the same weights and so we need only consider the first case $i_1=i_2=i$.  Unlike the case of $x_4$, now $i_1=i_2$ implies that $l_1=l_2=l$.  Thus we find 
\bea
x_5&=&\frac{\beta^4}{4}\left(\sum_{i<k_1,k_2}^q\right)\frac{1}{q^2\beta^2}\\
&\times&\sum_{j_1,j_2,l=1}^q\left(\F{j_1}{l}+2\pi-4\pi\theta(j_1-l)\right)\left(\F{j_2}{l}+2\pi-4\pi\theta(j_2-l)\right).\nonumber\\
\eea
The first row is identical to that of $x_4$ in Eq.~(\ref{x4}).  What about the second row?  If one exchanges $j$ with $l$ then the $\Phi$ terms look the same, but with their indices reversed.  Transposing the indices gives a minus sign in each summand.  However $2\pi-4\pi\theta(j-l)$ is also antisymmetric under the exchange of $j$ and $l$, therefore both factors in the second line change sign, leaving the second line invariant as well.  Thus we have found
\beq
x_5=x_4=-\frac{x_2}{3}.
\eeq

Adding all of these terms together we find that the variance of the anchored $\alpha$, at $O(n^2)$, is
\beq
x=\sum_{i=1}^5 x_i=\frac{x_1+x_2}{3}=n^2\left(\frac{\ee{\Phi^2}}{24}-\frac{\pi\ee{\Phi_>}}{3}-\frac{\ee{K^2}}{3}+\frac{\pi^2}{2} \right).
\eeq
Is this zero?  We numerically integrated the continuum expressions for $\Phi$ and $K$ in Eq.~(\ref{kffin}) to obtain
\beq
\ee{\Phi^2}=5.44\hsp\ee{\Phi_>}\sim -1.13\hsp\ee{K^2}=11.48
\eeq
and so
\beq
x\sim (0.23-1.18-3.83+4.93)n^2=0.15n^2.
\eeq
Is this compatible with zero?  It is nearly twice the best fit Gaussian variance found at $n=11$ in Fig.~\ref{n11fig}, but this is not obviously a sign of incompatibility as the $O(n)$ term could easily drive it down, with a coefficient of order unity.




\section{Conclusions}

Our goal is to devise a method to calculate, to arbitrary accuracy, the ground state and first excited state wave functionals of the $\cp^1$ nonlinear sigma model.  We would like to study the behavior of these wave functionals acting on a fixed-time configuration which circumnavigates the target space at each fixed latitude $\theta$, representing a time-slice of an instanton, to learn how the two sides of the equator $\theta=0$ are connected for the various states.  We hope, by analogy with the double well potential in quantum mechanics, that this will teach us how instantons generate the mass gap, and it will shed light on the role of instantons in Yang-Mills theory.

This model is equivalent to a high spin Heisenberg XXX spin chain, for which the states are in principle known, but in a rather unwieldy form which would be difficult to map to the sigma model.  Therefore one needs a prescription to calculate the spin chain matrix elements which is sufficiently simple so that it can be mapped to the sigma model.  

We begin, for sanity's sake, with spin $1/2$.    To cast our problem in a way which is close to continuum field theory, we collected the lattice sites into bins.  We believe that it is the bins, and not individual pairs of sites, which will eventually correspond to points in the continuum field theory.   We then average away all information involving the internal structure of the bins.  In the binning approximation, the richness of this system is smoothed away.  This is the strength of our approach, but we have not shown that this simplified system is in fact equivalent to the unbinned system.  Numerically we can precisely compute quantities for spin chains of length up to $N=22$ sites.  However this means that in the ground state there are at most 11 spin down sites.  We only expect our approximation to work when the bin size $n/q$ and the number of bins $q$ are infinite, but our numerics allow at most $q=n/q=3$.  At these low values of $q$ and $n/q$ we saw no evidence that the intrabin variations are smaller than the interbin variations, and so no evidence that the binning approximation leaves the matrix elements invariant.  

Thus the validity of our binning approach is, for the time being, taken as a postulate.  Once we are able to calculate the matrix elements, we may be able to use them to calculate $N$-point functions.  These are known, and so we can in principle test the consistency of the postulate.   Even if the postulate is true, we expect it to fail at subleading orders in $N$ and $q$.  If these subleading orders contribute to observables, again the postulate fails. 

Assuming this binning postulate, we found that standard combinatorial arguments in terms of partitions describe the behavior of the bins.  Thus instead of complexities which are polynomial in $N$, the chain length $N$ essentially disappears from the problem.  This combinatorial approach partially fixes the behavior of $q$ in the large $N$ limit.

Our strategy is to encode the information about a matrix element in a single function, $\rho(\alpha)$, which is the density of phases $\alpha$ in the CBA.  The Fourier transform of $\rho(\alpha)$ gives a matrix element.  Such an approach would be possible even without binning, but we use the combinatorics of the binning to calculate the moments of $\rho(\alpha)$. 

Our initial hope was that $\rho(\alpha)$ would be a Gaussian, and so this would be straightforward.   However it turned it that the variance was of order $O(N^{3})$.  In the Gaussian approximation this would lead to matrix elements of order $e^{-N^3}$, which is inconsistent with the fact that there are only $2^N$ states.   Our next hope was that $\rho(\alpha)$ is sufficiently close to a Gaussian so that a perturbative approach may be adopted, characterized by a moment expansion whose subleading terms represent the deviation from Gaussianity.  However we found that this Gaussian approximation is quite poor because $\rho(\alpha)$ is rich in substructure which in fact dominates both the moments and the Fourier transform.  

To fix this, we modified $\alpha$ by introducing an anchor which leaves the matrix elements invariant.  This anchor has a number of nice properties.  First, using the binning approximation we were able to show that the variance of the anchored $\alpha$  is only $O(N^2)$.  Numerically we were able to show, at $N\leq 22$, that the anchor reduces the variance by two orders of magnitude.  We have numerically confirmed that the modified $\alpha$ appears to be free of substructure at all even $N\leq 22$, several of which were shown explicitly in the text.   This of course does not guarantee that a moment expansion for $\rho(\alpha)$ will yield a convergent expansion for the matrix elements, but in our opinion it is promising.  Thus our proposal is to calculate the moments of $\alpha$ using the combinatorial methods described and use these to reconstruct $\rho(\alpha)$, whose Fourier transform gives the matrix elements.  We will see if this series converges when we do the calculation.

In general we focused our attention on a single matrix element, that relating the classical and quantum ground states $\langle 0|\Omega\rangle$.  The quantum ground state enters rather superficially in the last step, when one performs a numerical integral, and so it is likely that the generalization to other quantum states is not difficult, although in some cases one must change the number of spin down states $n$.  On the other hand the properties of the classical ground state were used in the motivation of the anchor.  In general, one cannot expect the anchor to possess all of the nice properties described above in the case of matrix elements with other classical states.  However, we checked them numerically in the cases of several classical states and found that $\rho(\alpha)$ appeared to be reasonably well-fit by a Gaussian in all cases except for one designed to be maximally far from the classical ground state.   Our method for calculating matrix elements therefore seems unlikely to work on matrix elements with such high energy states.  That said, it is unclear whether such states survive the continuum limit.  In fact the case considered was not N\'eel ordered and so it does not survive the large $s$ limit.  

What about the Gaussian approximation?  If indeed $\rho(\alpha)$ is a Gaussian, then matrix elements of $O(e^{-N})$ are only obtained if the variance of our anchored $\alpha$ is $O(N)$.  The anchor eliminates the $O(N^3)$ part and we have calculated here the $O(N^2)$ contribution.  We found that the $O(N^2)$ coefficient is quite small and in the last step our approach was numerical.  However it appears to be inconsistent with zero.  If indeed it is nonzero, then what has gone wrong?  Is our method doomed?

If the variance contains a term of $O(N^2)$, then that term will dominate the variance at large $N$, which is the limit of interest.  But the question is whether it will dominate the matrix elements.  If it does, then the matrix elements will be of order $O(e^{-N^2})$ and so cannot be normalized and we will arrive at an inconsistency.  This may indicate, for example, that our binning approximation is invalid.  Whether it dominates the matrix elements depends on the distribution.

Consider the following three distributions $\rho(\alpha)$.  The first is a Gaussian with variance that scales as $O(N^2)$ at large $N$.  The second is the weighted sum of two Gaussians with $N$-independent weights, one with a variance of $O(N)$ and the other with a variance of $O(N^2)$.  The third, which generalizes the second, is of the form
\beq
\rho(\alpha)\sim{\mathrm{exp}}\left[-f(\alpha,N)\right]\hsp
f(\alpha,N)\sim \left\{\begin{tabular}{l}$\alpha^2/N$ \ \ if $\alpha^2<<N$\\ $\alpha^2/N^2$ \ \ if $\alpha^2>>N^2$
\end{tabular} \right. 
\eeq
In the first case, the matrix elements will be $O(e^{-N^2})$ and so we will have an inconsistency.  In the second, at large $N$ the broader Gaussian simply ceases to contribute to the matrix elements, and so the matrix elements are of $O(e^{-N})$ as desired, determined entirely from the thin Gaussian.  These first two cases are rather special and so unlikely.  In the third case, for the first few standard deviations the probability falls rapidly as the distribution seems to be a thin Gaussian.  So long as the cross over to the $O(N^2)$ is at sufficiently high $\alpha^2$ that the area of the thin region does not tend to zero at large $N$, then the matrix elements will again be determined by the thin region and so have the correct behavior.   Of course there is no guarantee that any of these cases is realized.

So which is the case at hand?  Having only calculated the variance, it is too early to say.  The calculation of higher moments can distinguish these cases, although at any finite moment, assumptions about the form of $\rho(\alpha)$ will be necessary to determine the potential.  In other words, a concrete statement of the absence of substructure is needed.  Fig.~\ref{n11fig} shows that, at least at finite $N$, $\rho(\alpha)$ is leptokurtic.  If this persists at infinite $N$, it would be inconsistent with the first case but consistent with the others.   One possible way forward will be to evaluate the full infinite series of moments, which will determine the density function completely.

In fact, it is possible for us to go beyond simply calculating moments.  Eq.~(\ref{box}) is the entire joint probability density function (PDF) for $f_{ij}$, $f_{il}$, $f_{jk}$ and $f_{kl}$.  It can be put in a useful form with the expansion (\ref{tool}) and $c$ can be found by imposing that $\ee{1}=1$ as was done in Subsec.~\ref{punto}.  Summing over $p_2$ and $p_3$ one is left with the joint PDF for $p_1$ and $p_4$ which are $f_{ij}$ and $f_{kl}$.  This is easily converted into a joint PDF for $\tf_{ij}$ and $\tf_{kl}$, which via Eq.~(\ref{aeq}) yields $\rho(\alpha)$, which is the PDF for $\alpha$.  If this can be calculated directly, at some order in $N$, the answer may be inserted into Eq.~(\ref{ft}) to determine the matrix element.  In this way, no assumptions regarding substructure are needed.


Summarizing, we appear to be well along the way to calculating the matrix element of the classical and quantum ground states of the $s=1/2$ model, and the other matrix elements appear to be similar.  It is possible in principle that the anchor that we have adapted does not render $\rho(\alpha)$ sufficiently close to a Gaussian for our moment expansion, but numerical evidence at small $N$ suggests that it does.  This all relies on our binning postulate, which allows us to neglect the internal structure of bins in the limit of a large number $q$ of bins of size $n/q$, which is also taken to be large.  We have not yet needed to specify this limit completely, but it may be that the validity of the binning postulate only allows one limit or it may simply never be valid.   Failure of the postulate need not imply abandoning our program, but it means that we must calculate the corrections resulting from intrabin structure.

And if this all works, how do we get to higher spin?  After all, there is no CBA in these cases\footnote{One proposal has appeared in Ref.~\cite{cbas}.}?  The algebraic Bethe Ansatz provides a much more complicated construction of these states.  However on the bright side they are still constructed from $N$ commuting copies of the creation operators $B(\lambda_i)$, and so there is still a permutation symmetry on the $\lambda_i$.  This lends hope that it may be possible to write a state in some basic form, analogous to a single summand in CBA, which upon symmetrization gives the true state.  Then the technology from $s=1/2$ to handle binnings of permutations could be imported to this more complicated setting.

\section* {Acknowledgement}

\noindent
JE is supported by the CAS Key Research Program of Frontier Sciences grant QYZDY-SSW-SLH006 and the NSFC MianShang grants 11875296 and 11675223.  JE also thanks the Recruitment Program of High-end Foreign Experts for support.

\appendix

\section{The Map Between the Spin Chain and Sigma Model} \label{mapapp}

The $\cp^1$ nonlinear sigma model and the antiferromagnetic XXX spin chain at spin $s$ are equivalent in the limit $s\rightarrow\infty$.  This was shown classically by Haldane in Refs.~\cite{haldane1,haldane2}, where it was seem that classically finite $s$ corresponds to a finite coupling of the sigma model.  At the quantum level, the sigma model coupling runs and so there is no such dimensionless free parameter.  Nonetheless the exact quantum correspondence in the infinite $s$ limit was shown in Ref.~\cite{affleck}.  We will review that argument, following the presentation in Ref.~\cite{fc}.

In Sec.~\ref{xxxsez} we introduced the spin $1/2$ antiferromagnetic XXX spin chain.   The general spin $s$ spin chain, introduced in Ref.~\cite{bab}, is similar.  In this case, the Hilbert space at each lattice site is $\C^{2s+1}$, and the $\mathfrak{su}(2)$ Lie algebra, with generators $S_l^i$ at each lattice site $l$ acts on this Hilbert space in the $(2s+1)$-dimensional representation.   The Hamiltonian must include higher order couplings of neighboring sites if one demands integrability.  However these higher order couplings vanish in the continuum limit.  

Define the following combinations of operators
\beq
n^l_i=\frac{1}{2s}\left(S^i_{2l}-S^i_{2l-1}\right)\hsp p^l_i=S^i_{2l}+S^i_{2l-1}.
\eeq
The intuition for the connection to the $\cp^1$ sigma model is as follows.  In the classical ground state, neighboring spins are antialigned and so $S^i_{2l}=-S^i_{2l-1}$. Classically one may replace the $S^i$ with their eigenvalues and so conclude that $|S_{2l}|^2=s(s+1)$ and so
\beq
|n^l_i|^2=|\frac{1}{s} S^i_{2l}|^2=\frac{s(s+1)}{s^2}
\eeq
which in the large $s$ limit tends to unity.  More nontrivially, in the large $s$ limit this antialignment holds even quantum mechanically, in the sense that the energy required to get a finite fractional difference between the eigenvalues of the spin operators at adjacent sites becomes infinite.  Thus it is plausible that at large $s$, the eigenvalues of $|n^l|$ will be concentrated on unity, and so it is a natural coordinate for the position on $\cp^1$ represented as an $S^2$ in $\R^3$.  In other words, the $\cp^1$ sigma model coordinate corresponds to the N{\'e}el order parameter of the spin chain.

Now for a more rigorous description of the equivalence with the $\cp^1$ model.  The algebra satisfied by these new operators is easily calculated from that of $\mathfrak{su}(2)$ to be
\beq
[p^l_i,p^m_j]=-i\epsilon_{ijk}\delta^{lm} p^l_k\hsp [p^l_i,n^m_j]=-i\epsilon_{ijk}\delta^{lm} n^l_k\hsp 
[n^l_i,n^m_j]=\frac{i}{4s^2} \epsilon_{ijk}\delta^{lm}  p^l. \label{cr}
\eeq
We may recognize the first two of these as the canonical commutation relations of the discretized $\cp^1$ model with coupling $g$ if $n^l_i$ are the coordinates at the $l$th lattice point and $p$ is the canonical momentum
\beq
p=\frac{1}{g}\dot{n}\times n.
\eeq
However the third relation in (\ref{cr}) agrees with the commutation relations of the canonically quantized sigma model only in the limit $s\rightarrow\infty$, where it vanishes.  Therefore the finite $s$ spin chain corresponds to a noncommutative deformation of the $\cp^1$ sigma model.


\end{document}